\documentclass{amsart}%
\usepackage{amsmath,amssymb,amsfonts}
\usepackage{graphicx}
\usepackage{amsmath}
\usepackage{amsfonts}
\usepackage{amssymb}
\usepackage{color}%
\newtheorem{theorem}{Theorem}

\newtheorem{remark}[theorem]{Remark}

\newtheorem{proposition}[theorem]{Proposition}
\newtheorem{corollary}[theorem]{Corollary}
\newtheorem{assumption}{Assumption}

\begin{document}

\author{Milena Radnovi\'c}
\address{Mathematical Institute SANU, Belgrade, Serbia}
\email{milena@mi.sanu.ac.yu}
\author{Vered Rom-Kedar}
\address{The Weizmann Institute of Science, Rehovot, Israel}
\email{vered.rom-kedar@weizmann.ac.il}
\title{Foliations of Isonergy Surfaces and Singularities of Curves}
\maketitle

\begin{abstract}
It is well known that changes in the Liouville foliations of the
isoenergy surfaces of an integrable system imply that the
bifurcation set has singularities at the corresponding energy
level. We formulate certain genericity assumptions for  two
degrees of freedom integrable systems and we prove the opposite
statement: \emph{the essential critical points} of the bifurcation
set appear only if the Liouville foliations of the isoenergy
surfaces change at the corresponding energy levels. Along the
proof, we give full classification of the structure of the isoenergy
surfaces near the critical set under our genericity assumptions and we
give their complete list using Fomenko graphs. This may be viewed
as a step towards completing the Smale program for relating the
energy surfaces foliation structure to singularities of the
momentum mappings for non-degenerate integrable two degrees of
freedom systems.
\end{abstract}

\newpage

\tableofcontents


\section{Introduction}

A decade ago, at his International Congress of Mathematics address
\cite{Mosericm98}, Juergen Moser concluded his lecture with the
following remark:
\begin{quote}
``Most striking to me is the development of integrable systems (over
30 years ago) which did not grow out of any given problem, but out
of phenomenon  which was discovered by numerical experiments in a
problem of fluid dynamics''.
\end{quote}
Indeed, Moser provides examples in these notes showing that the
structure of integrable systems is both surprisingly rich and
surprisingly relevant for the analysis of systems arising in nature.
Moser's mainly concentrated in these notes on the richness of
infinite dimensional integrable systems. It is a great honor for us
to contribute to this issue in his memory, a description of the rich
structure of  integrable two degrees of freedom Hamiltonian systems
which are very much related to the corresponding structure of some
near-integrable infinite dimensional systems \cite{SRK,eshVRKPRL}.

\smallskip

In his well known papers \cite{Sm1,Sm2}, Smale studied the
topology of mechanical systems with symmetry. By Noether theorem
\cite{Ar}, the symmetry group acting on the phase space gives rise
to an integral $J$ of the motion, $J$ being independent of the
total energy $E$. Smale investigated the topology of the mapping
\[
I=J\times E\ :\ \mathcal{M}\to\mathbf{R}^{2},
\]
and pointed out that a main problem in this study is to find the
structure of the bifurcation set of this mapping. In \cite{Sm1},
he considered mechanical systems with $2$ degrees of freedom and
$1$-dimensional symmetry group, and then applied some of the
developed ideas to general mechanical systems with symmetry. In
\cite{Sm2}, he applies this work to describe the topology and
construct the bifurcation set for the Newtonian $n$-body problem
in the plane. The study of the geometry of the level sets of the
momentum map under various symmetries is highly nontrivial (see
\cite{AMM,CB}).

\smallskip

Later on, Fomenko, Bolsinov, Oshemkov, Matveev, Zieschang and
others studied the topological classification of the isoenergy
surfaces of $2$ degrees of freedom integrable systems and their
description by certain invariants, see
\cite{Bol,BF1,BMF,BO,F1,F2,Mat,Osh,FoTs90} and references therein.
They established a beautiful and simple way of representing the
Liouville foliation of an isoenergy surface by a certain graph
with edges and some subgraphs marked with rational and natural
numbers. Roughly, in these graphs, each foliation leaf is shrunk
to a distinct point. Thus, each smooth family of Liouville tori
creates an edge, and edges connect together at vertices that
correspond to the singular leaves (see Figure
\ref{fig:center-center} for a simple example). In Appendix 1, we
give a more precise description of the work of Fomenko and his
school, while all the details can be found in
\cite{Bol,BO,BF1,BMF,F1,F2,Mat} and references therein. In
\cite{LHRK3} higher dimensional analogs to branched surfaces were
developed and constructed for several three degrees of freedom
systems. More recently, Zung studied singularities of integrable
and near-integrable systems as well as the genericity of the
non-degeneracy conditions introduced by the Fomenko school
\cite{BZ,Zung1,Zung2,Zung3} whereas Kalashnikov studied the
behavior of isoenergy surfaces near parabolic circles with
resonances \cite{Kal}. These tools have been applied to describe
various mechanical systems (see \cite{CB,BF2} and references
therein) and even of the motion of a rigid body in a fluid
\cite{OrRy00}. In these works, that typically deal with $n$
degrees of freedom integrable systems with $n>2$, it is seen that
when one fixes all but one of the constants of motion, the graphs
produced for that constant of motion undergo interesting
bifurcations as the energy is changed.

\smallskip

In parallel, Lerman and Umanskii analyzed and completely described
the topological structure of a neighborhood of a singular leaf of
$2$ degrees of freedom systems \cite{LU}. In \cite{L2}, Lerman had
completed the study of singular level sets near fixed points of
integrable $3$-degrees of freedom systems.

\smallskip

In all these works, the main objective is to classify the
integrable systems, thus the study of the topology of the level
sets emerges as the main issue, and the role of the Hamiltonian
and of the integrals of motion may be freely interchanged. In
\cite{LHRK1,LHRK2,LHRK3,SRK}, the implications of these
developments on the near-integrable dynamics were sought. Once the
more typical near-integrable Hamiltonian systems are considered
(see for example
\cite{Ar2,BHS96,GH,Haller,LiLi83,MarMey74,Mbook,SaVe85,verh98,ZSUC91}
and references therein), the Hamiltonian emerges as a special
integral. The Hamiltonian is conserved under perturbations and the
perturbed motion is restricted now to energy surfaces and not to
single level sets. For example, folds of smooth curves belonging
to the bifurcation set, which, in the integrable settings are not
considered as singularities, do correspond to singularities when
the bifurcation curves are viewed as graphs over the Hamiltonian
value. Such folds correspond in the near-integrable setting to the
strongest resonances - to circles of fixed points. These are
expected to break under small perturbations and the perturbed
motion near them can be thus studied
\cite{Haller,LHRK1,LHRK2,LHRK3,SRK}.

\smallskip

With this point of view, the classification of changes in the
foliations of the integrable system amounts to the classification
of the possible behaviors of non-degenerate \emph{near integrable}
$2$ degrees of freedom systems. Indeed, away from the bifurcation
set resonances and KAM tori reign. The normal forms near the
different singularities of the bifurcation set can be now derived
and used to classify the various instabilities that may develop
under small perturbations. The analysis of each of these
near-integrable scenarios is far from being complete; beyond KAM
theory, there is a large body of literature dealing with
persistence results for lower dimensional tori (namely circles in
the $2$ degrees of freedom case) -- see, for example,
\cite{BHY,Elia,Graff,Llave,Posch}, and another large body of
results describing the instabilities arising near singular circles
or fixed points that are not elliptic (see \cite{kov93,Haller,SRK}
and references therein). Our classification reveals several new
generic cases that were not studied in this near-integrable
context.

\smallskip

The article is organized as follows. In Section \ref{sec:setup} we
list and discuss all conditions on the $2$-degrees of freedom systems
for which our results hold. In Section \ref{sec:sing.foliations.thm}
we state the Singularities and Foliations Theorem as the main result
of the present work, and give the proof outline. The detailed proof,
together with the detailed description of the energy surfaces near
singular leaves is in Section \ref{sec:essential} and Appendix 2.
Section \ref{sec:examples} contains a few examples and Section
\ref{sec:conc} contains concluding remarks. Appendix 1 is a brief
overview of the necessary results obtained by Fomenko and his school
on the topological classification of isoenergy surfaces and their
representation by Fomenko graphs. Appendix 2 contains, for
completeness, the description of the isoenergy surfaces for cases
that were fully analyzed previously: the behavior near isolated
fixed points, which is essentially a concise summary of Lerman and
Umanskii \cite{LU} results using Fomenko graphs and is very similar
to the work of Bolsinov \cite{Bol} and the behavior near certain
parabolic circles which follows from \cite{BRF}.

\section{The Generic Structure of the Integrable System}\label{sec:setup}

We consider an integrable Hamiltonian system defined on the
$4$-dimensional symplectic manifold $\mathcal{M}$, with the
Hamiltonian $H$. The manifold $\mathcal{M}$ is the union of
\textit{isoenergy surfaces} $\mathcal{Q}=H^{-1}(h)$,
$h\in\mathbf{R}$.

\smallskip

Our aim is to study how the structure of isoenergy surfaces changes
with the energy. To obtain a finite set of possible behaviors, the
class of integrable Hamiltonian systems with $2$ degrees of freedom
must be restricted so that degenerate cases are excluded. Natural
non-degeneracy assumptions and restrictions on these systems under
which our results hold are listed below. Assumptions
\ref{as:smooth}--\ref{as:non-res} are used in the works of the
Fomenko school \cite{BF2}. Assumptions \ref{as:sing.leaves} and
\ref{as:stability} are new --- these contain the assumptions of
Lerman and Umanskii \cite{LU} and include further restrictions on
the type of singularities we allow.

\subsection{The Structure of Isoenergy Surfaces}

\begin{assumption}\label{as:smooth}
$\mathcal{M}$ is a four-dimensional real-analytic manifold with a
symplectic form $\omega$, and $H$, $K$ are real-analytic functions
on $\mathcal{M}$.
\end{assumption}

\begin{assumption}
\label{as:integrable} $(\mathcal{M},\omega,H)$ is an integrable
system with a first integral $K$.
\end{assumption}

\noindent This assumption means that the functions $H$ {and} $K$
are functionally independent and are commuting with respect to the
symplectic structure on $\mathcal{M}$.

\smallskip

Since the flows generated by the Hamiltonian vector fields $H$ and
$K$ commute, they define the action of the group $\mathbf{R}^{2}$
on $\mathcal{M}$. This action, which we will denote by $\Phi$, is
called the \emph{Poisson action generated by $H$ and $K$} and its
orbits are homeomorphic to a point, line, circle, plane, cylinder
or a $2$-torus \cite{Ar}. The functional independence of the
analytic functions $H$ and $K$ means that the differentials $dH$
and $dK$ may be linearly dependent only on isolated orbits of
$\Phi$.

\begin{assumption}
\label{as:compact} The isoenergy surfaces of the system are
compact.
\end{assumption}

Denote the \emph{momentum map} by $\mu:$
\[
\mu\ :\
\mathcal{M}\rightarrow\mathbf{R}^{2},\quad\mu(p)=(H(p),K(p)).
\]
\textit{Level sets} of the momentum map are $\mu^{-1}(h,k)$,
$(h,k)\in \mathbf{R}^{2}$. The \emph{Liouville foliation} of
$\mathcal{M}$ is its decomposition into connected components of
the level sets. Each orbit of $\Phi$ is completely placed on one
foliation leaf. The rank of $d\mu$ is equal to $2$ at each point
of \emph{a regular leaf}, while \emph{singular leaves} contain
points where gradients of $H$ and $K$ are linearly dependent.

Assumption \ref{as:compact} implies that the leaves are compact.
Thus, by the Arnol'd-Liouville theorem \cite{Ar}, each regular
leaf is diffeomorphic to the $2$-dimensional torus
$\mathbf{T}^{2}$ and the the motion in its neighborhood is
completely described by, for example, the action-angle
coordinates.

\noindent
\begin{assumption}
\label{as:non-res} The Hamiltonian $H$ is \emph{non-resonant},
i.e.\ the Liouville tori in which the trajectories form irrational
windings are everywhere dense in $\mathcal{M}$.
\end{assumption}

\noindent This assumption of non-linearity implies that the
Liouville foliations do not depend on the choice of the integral
$K$.

\subsection{Singular Leaves}

A singular leaf of the system contains points where the rank of
the momentum map is less than maximal, i.e.\ less than $2$ in our
case. \emph{The rank} of a singular leaf is the smallest rank of
its singular points. It is equal to $0$ if there is a fixed point
of the action $\Phi$ on the leaf, or to $1$ if the leaf contains
one-dimensional orbits of $\Phi$ and does not contain any fixed
points. \emph{The complexity} of a singular leaf is the number of
connected components of the set of all points on the leaf with the
minimal rank.

\smallskip

A fixed point of the action $\Phi$ corresponds to an isolated fixed
point of the Hamiltonian flow whereas one-dimensional orbits of
$\Phi$ correspond to invariant circles of the Hamiltonian flow. Next
we list our non-degeneracy assumptions regarding the structure of
the singular leaves.

\smallskip

Let $x\in\mathcal{M}$ be a singular point of the momentum mapping,
$\mathcal{K}$ the kernel of $D\mu(x)$, and $\mathcal{I}$ the space
generated by $X_{H}(x)$, $X_{K}(x)$. Here, $X_{H}(x)$, $X_{K}(x)$
are the Hamiltonian vector fields that are generated by $H$, $K$.
The quotient space $\mathcal{K}/\mathcal{I}$ has a natural
symplectic structure inherited from $\mathcal{K}$. This space is
symplectomorphic to a subspace $R$ of $T_{x}\mathcal{M}$ of
dimension $2k$, $k$ being the corank of the singular point $x$.
The quadratic parts of $H$ and $K$ at $x$ generate a subspace
$\mathcal{F}_{R}^{(2)}(x)$ of the space of quadratic forms on $R$.
We say that $x$ is a \emph{non-degenerate point} if
$\mathcal{F}_{R}^{(2)}(x)$ is a Cartan subalgebra of the algebra
of quadratic forms on $R$ (see \cite{DM,LU,Zung1}).

\smallskip

A singular leaf is \emph{non-degenerate} if all its singular points
are non-degenerate. In the two-degrees of freedom case, there are
families of non-degenerate leaves of rank $1$ --- they contain
non-degenerate invariant circles of the Hamiltonian flow. Such leaves are called  atoms by the Fomenko school. The invariant
circles may typically become degenerate at some isolated values of
the energy --- so to classify how typical systems change with the
energy one must include the classification of some degenerate rank
$1$ orbits. Additionally, non-degenerate leaves of rank $0$ appear
at some isolated values of the energy --- these correspond to
non-degenerate isolated fixed points of the Hamiltonian flow. The
classification of such non-degenerate singular leaves containing one
fixed point of the Poisson action $\Phi$ is done in \cite{LU},
together with the detailed topological description of the
neighborhoods of such leaves.

\smallskip

To formulate precisely what is a typical behavior near the singular
leaves containing rank $1$ orbits, we recall first the reduction
procedure \cite{LU,Nekh}. Consider a closed one-dimensional orbit
$\gamma$ of the action $\Phi$, which we will refer to as \emph{a
circle}. Let $U$ be some neighborhood of a point $m\in\gamma$. Each
level $U\cap\{H=h\}$ is foliated into segments of trajectories of
the Hamiltonian $H$. Identifying each such segment with a point, we
obtain a two-dimensional quotient manifold $D_{h}$ with the induced
$2$-form. The integral $K$ is reduced to a family of functions
$K_{h}$ on $D_{h}$, with $K_{H(m)}$ having a critical point. For
non-degenerate $\gamma$, this family will be equivalent to
$h+x^{2}+y^{2}$ or $h+x^{2}-y^{2}$ depending on whether $\gamma$ is
elliptic or hyperbolic. The obtained normal form, which describes
the local behavior near $\gamma$, does not depend on the point
$m\in\gamma$ and varies smoothly with the energy level $h$.

\smallskip

Typically, the singular leaf containing $\gamma$ is of complexity
$1$, namely it does not contain additional circles: if $\gamma$ is
elliptic the level set consist of $\gamma$ only. If $\gamma$ is
hyperbolic then the singular leaf contains also the separatrices of
$\gamma$, and these, by Assumption \ref{as:compact} are compact.
Usually one expects that these separtrices will close in a
homoclinic loop. However, some exceptions are possible; first, at
isolated energy levels heteroclinic connections may naturally appear
--- then the level set has complexity greater than $1$. Second, when there
are some symmetry constraints, heteroclinic connections may persist
along a curve: for example, such a situation always appears near
non-orientable saddle-saddle fixed points, see Appendix 2. We thus
allow singular leaves with rank $1$ orbits to have complexity of at
most $2$.

\smallskip

Finally, degenerate circles arise when $\gamma$ is parabolic. Then,
generically, the above reduction procedure gives rise to a family of
the form $hx+x^{3}+y^{2}$. If the Hamiltonian has a $\mathbf Z_{2}$
symmetry, parabolic circles of the form $hx^{2}+x^{4}\pm y^{2}$
or two parabolic circles of the generic form that belong to the same level set may appear.

Thus, we impose the following assumptions
on the structure of the singular leaves:

\begin{assumption}\label{as:sing.leaves}
Singular leaves appearing in the Liouville foliation of
$\mathcal{M}$ can be one of the following types:
\begin{itemize}
\item
Non-degenerate leaves of rank $1$ and complexity at most $2$;

\item
Degenerate leaves of rank $1$ and complexity $1$ such that the
reduction procedure leads to the one-parameter family of functions
$K_{h}$ that is equivalent to $hx+x^{3}+y^{2}$;

\item
Non-degenerate leaves of rank $0$ and complexity $1$ that do not
contain any closed one-dimensional orbits of $\Phi$;

\item
Degenerate leaves of rank $1$ and complexity $1$ such that the
reduction procedure leads to the one-parameter family of functions
$K_{h}$ that is equivalent to either $hx^{2}+x^{4}+y^{2}$ or
$hx^{2}+x^{4}-y^{2}$;

\item
Degenerate leaves of rank $1$ and complexity $2$ such that the
reduction procedure near both circles leads to the one-parameter
family of functions $K_{h}$ that is equivalent to $hx+x^{3}+y^{2}$.
\end{itemize}
\end{assumption}

\subsection{The Bifurcation Set}

\emph{The bifurcation set} $\Sigma$ is the set of images of
critical points of the momentum map, i.e.\ it contains all $(h,k)$
values for which $\mu ^{-1}(h,k)$ contains a singular leaf.
$\Sigma$ consists of smooth curves and isolated points \cite{Sm1}.

\smallskip

We define \emph{the critical set} $\Sigma_{c}$ as the set of all
$(h_{c}, k_{c})\in\Sigma$ such that there does not exist a
neighborhood $U$ of $(h_{c},k_{c})$ in $\mathbf{R}^{2}$, for which
$\Sigma\cap U$ is a graph of a smooth function of $h$.

\smallskip

In other words, $\Sigma_{c}$ contains all isolated points of
$\Sigma$ and all the singularities of the curves from $\Sigma$
viewed as graphs over $h$. $\Sigma_{c}$ includes the points
$(h,k)\in\Sigma$ at which these curves intersect, have cusps, or
have folds --- i.e.\ if the line $H=h_{c}$ is tangent to a curve
belonging to $\Sigma$ at $k=k_{c}$ then
$(h_{c},k_{c})\in\Sigma_{c}$.

\smallskip

\begin{figure}[h]
\centering
\begin{minipage}[h]{0.95\textwidth}
\centering
\includegraphics[width=5cm,height=4cm]{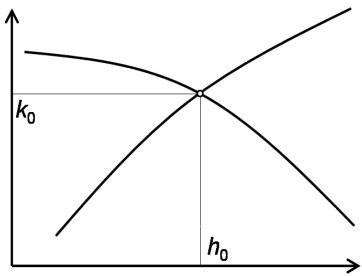}
\parbox{0.95\textwidth}{\caption{Intersection of singularity curves corresponding to
two disconnected singular leaves.}\label{fig:presek}}
\end{minipage}
\end{figure}

\begin{figure}[h]
\centering
\begin{minipage}[h]{0.44\textwidth}
\centering
\includegraphics[width=5cm,height=4cm]{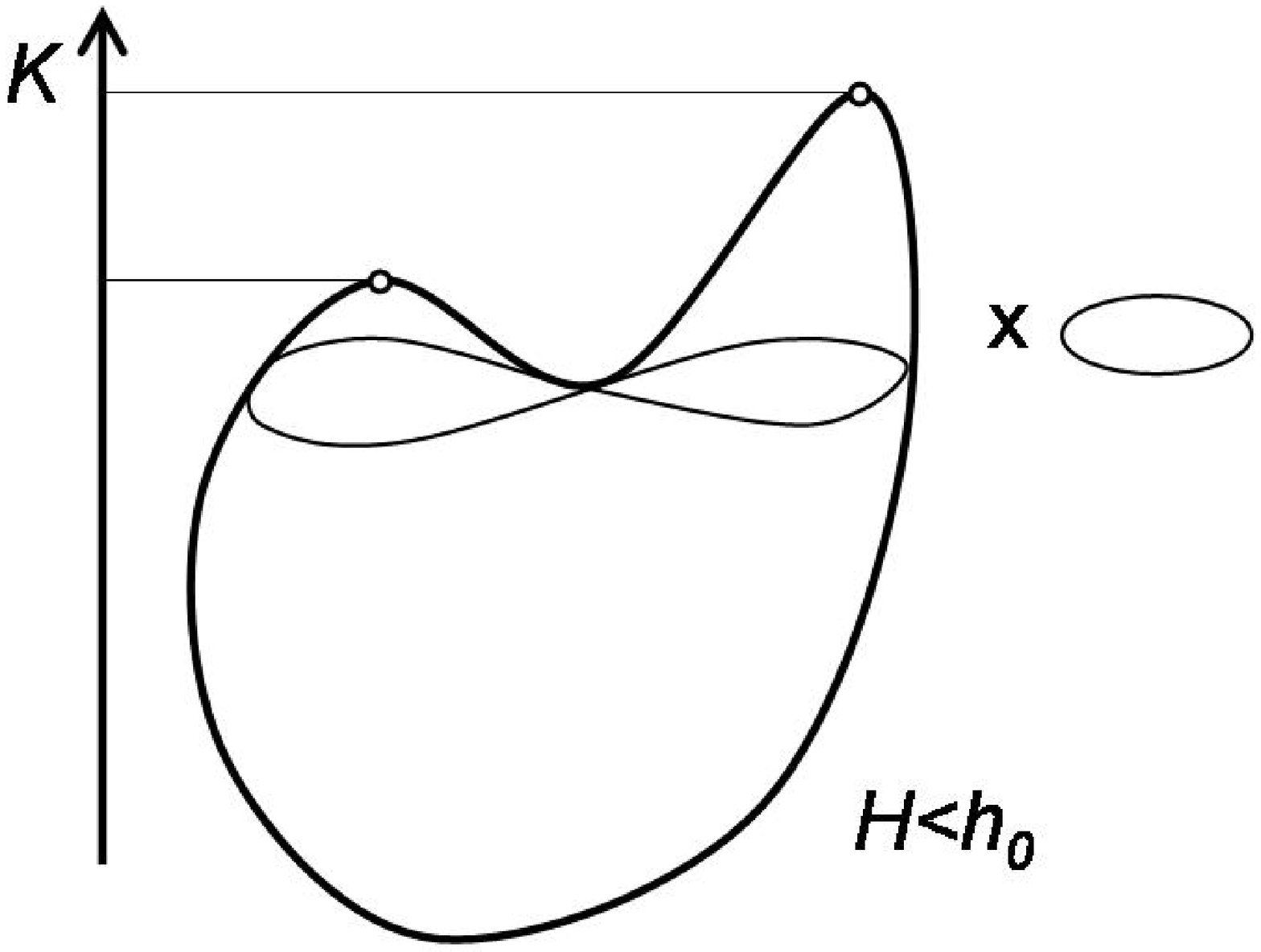}
\end{minipage}
\begin{minipage}[h]{0.44\textwidth}
\centering
\includegraphics[width=5cm,height=4cm]{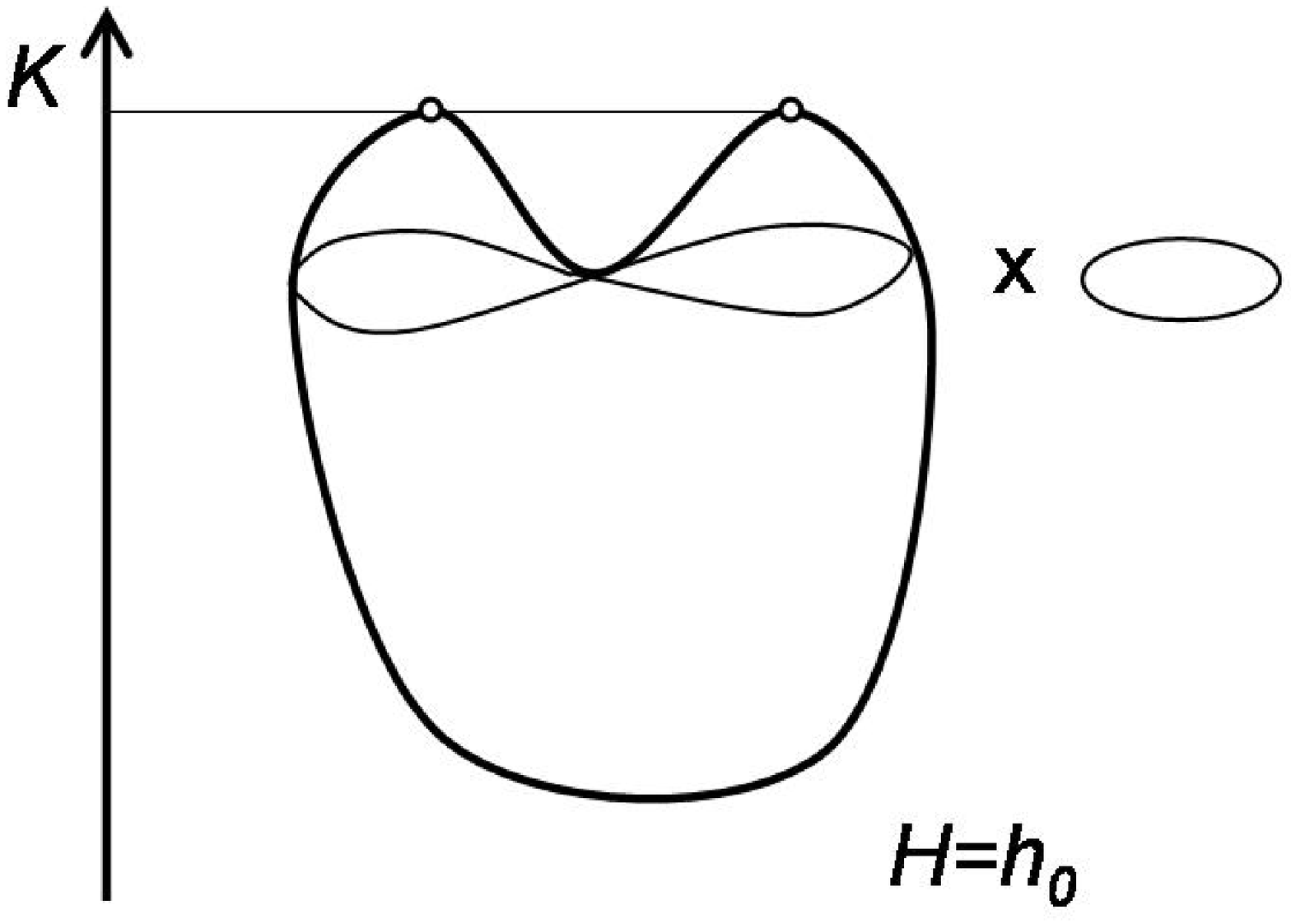}
\end{minipage}
\begin{minipage}[h]{0.44\textwidth}
\centering
\includegraphics[width=5cm,height=4cm]{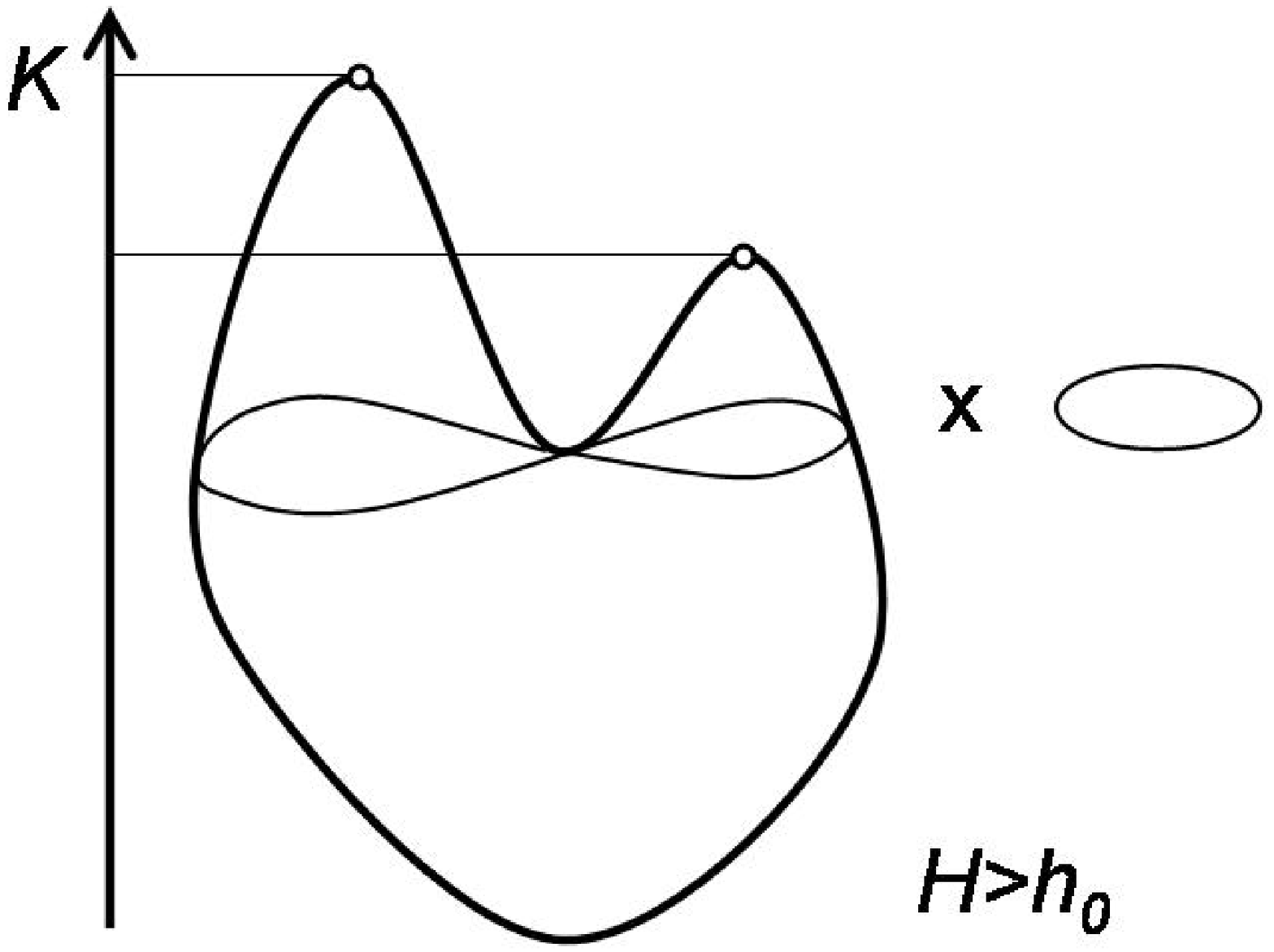}
\end{minipage}
\caption{Isoenergy surfaces for different values of $H$
corresponding to the bifurcation diagram of Figure
\ref{fig:presek}.} \label{fig:mnogostrukosti}
\end{figure}

It is well known that the bifurcations of the Liouville foliations
of the isoenergy surfaces may happen only at critical points of
$\Sigma$ (see, for example, \cite{Bol}). However, the opposite
implication is incorrect: for example, two curves from $\Sigma$
corresponding to two families of singular leaves may intersect at
a point $(h_{0},k_{0})$ (see Figure \ref{fig:presek}) while
$\mu^{-1}(h_{0},k_{0})$ simply contains two disconnected leaves
with no bifurcation happening. Such a situation is illustrated in
Figure \ref{fig:mnogostrukosti}.

Thus, we need to refine the definition of $\Sigma_{c}$ so as to
exclude such trivial cases from our consideration.

\smallskip

Consider a leaf $\mathcal{L}$ of the Liouville foliation. An open
set $V$ which is invariant under the action $\Phi$ and contains
$\mathcal{L}$ will be called an \emph{extended neighborhood of
$\mathcal{L}$}. For such a set $V$, denote by $\Sigma(V)$ the set of
images of critical points of the momentum map restriction to $V$. We
call $\Sigma(V)$ \textit{the local bifurcation set of $V$}. As in
the global case, $\Sigma(V)$ may itself have singularities, e.g.\
folds, common endpoints of curves, intersections or isolated points.

\smallskip

Denote by $\Sigma_{c}(V)\subset\Sigma_{c}$ the singularities of the
local bifurcation set $\Sigma(V)$ as defined above. Clearly, for
sufficiently thin extended neighborhoods, $\Sigma_{c}(V)$ is
independent of $V$, so we may define \emph{a local critical point}
$\Sigma_{c}(\mathcal{L})=\Sigma_{c}(V)$. Notice that
$\Sigma_{c}(\mathcal{L})$ is either empty or equal to $\{\mu
(\mathcal{L})\}$.

\smallskip

We now introduce the \emph{essential critical set} as the union of
all local critical points over all foliation leaves of
$\mathcal{M}$:
\[
\Sigma_{c}^{ess}=\bigcup_{\mathcal{L}}\Sigma_{c}(\mathcal{L}).
\]

\smallskip

Indeed, in the example shown on Figures \ref{fig:presek} and
\ref{fig:mnogostrukosti}, the level set $(h_{0},k_{0})$ consists of
two isolated elliptic circles, and
$\Sigma_{c}(\mathcal{L})=\emptyset$ for both of them. Thus
$(h_{0},k_{0})$ does not belong to the essential critical set.

\smallskip

Clearly, non-degenerate singular leaves of rank $0$ and degenerate
ones allowed by Assumption \ref{as:sing.leaves} will always produce
essential singularities, see \cite{LU,BRF}. However, this is not
true for most non-degenerate singular leaves of rank $1$. Each
non-degenerate one-dimensional orbit of $\Phi$ belongs to a smooth
family of circles that is mapped by $\mu$ to a smooth curve of
$\Sigma$. \emph{Stable singularities} that may appear on such curves
are folds and points of transversal intersection of two curves. We
impose an additional stability condition: that all essential
singularities appear on distinct isoenergy surfaces.

\begin{assumption}\label{as:stability}

\

\begin{itemize}
\item
Let $\mathcal{L}$, $\mathcal{L}^{\prime}$ be two distinct foliation
leaves such that both have non-empty local critical points. Then
$H(\mathcal{L})\neq H(\mathcal{L}^{\prime})$.

\item
Let $\mathcal{L}$ be a non-degenerate singular leaf of rank $1$ such
that $\Sigma_{c}(\mathcal{L})\neq\emptyset$. Then, at the point
$\mu(\mathcal{L})$ one of the following singularities occur:

\begin{itemize}
\item
a point of transversal intersection of two curves of $\Sigma$;

\item
a quadratic fold of a curve of $\Sigma$.
\end{itemize}
\end{itemize}
\end{assumption}

The Singularities and Foliations Theorem is formulated for systems
that satisfy Assumptions \ref{as:smooth}--\ref{as:stability} and
these include many natural mechanical systems (see Section
\ref{sec:examples}).

\section{The Singularities and Foliations Theorem}
\label{sec:sing.foliations.thm}

The main result of this paper is the formulation of the following
theorem and its constructive proof. The main part of the proof is
based on statements appearing in the next section and the
appendices.

\smallskip

We say that isoenergy surfaces are \textit{Liouville equivalent}
if they are topologically conjugate and the homeomorphism
preserves their Liouville foliations \cite{BF1,BF2,BMF}.

\medskip

\noindent\textbf{Singularities and Foliations Theorem.}\ \
\textit{Consider the $2$ degrees of freedom integrable Hamiltonian
system $(\mathcal{M},\omega,H)$ satisfying Assumptions
\ref{as:smooth}--\ref{as:stability}. Then, a change of the Liouville
equivalence class of the isoenergy surfaces occurs at the energy
level $H=h_{c}$ if and only if there exists a value $k_{c}$ such
that $(h_{c},k_{c})\in\Sigma_{c}^{ess}$. Furthermore, there is a
finite number of types of such possible changes. All of them are
listed in Proposition \ref{prop:intersection}, Proposition
\ref{prop:fold}, Corollary \ref{cor:fixed.points}, Corollary
\ref{cor:parabolic}, and represented by the Fomenko graphs of
Figures \ref{fig:intersection.orientable}--\ref{fig:double.fold} and
\ref{fig:center-center}--\ref{fig:parabolic3}.}

\smallskip

\begin{proof}
$\Longrightarrow$ See Proposition 4.3 in \cite{Bol}; We repeat here
the proof for completeness. We prove this first part by
contradiction. Assume that the isoenergy surfaces at $H=h_{1}$ and
$H=h_{2}$ are not Liouville equivalent and that there is no
$(h_{c},k_{c})\in\Sigma_{c}^{ess}$ with $h_{c}\in [h_{1},h_{2}].$
Assumptions \ref{as:smooth}--\ref{as:non-res} imply that with the
exception of isolated values of $H$, the Liouville foliation of
isoenergy surfaces may be completely described with the
corresponding Fomenko invariants \cite{BMF}. Vertices of the graph
joined to the isoenergy surface $H=h$ correspond exactly to
intersections of the line $H=h$ with $\Sigma$.
 Since we assumed there is no
point of the set $\Sigma_{c}^{ess}$ between the lines $H=h_{1}$ and
$H=h_{2}$, the Fomenko graphs, together with all their corresponding
invariants, will change continuously between the values $h_{1}$ and
$h_{2}$. Since the graph with all joined numerical invariants is
given by a finite set of discrete parameters \cite{BMF}, it follows
that all isoenergy surfaces $H=h$, where $h$ is between $h_{1}$ and
$h_{2}$, have the same Fomenko invariants. Thus, the isoenergy
surfaces $H=h_{1}$ and $H=h_{2}$ are Liouville equivalent, which
concludes the proof.

\smallskip

$\Longleftarrow$ Take $(h_{c},k_{c})\in\Sigma_{c}^{ess}$. Then,
there exists a singular leaf $\mathcal{L}$ such that
$\mu(\mathcal{L})=(h_{c},k_{c})$ and
$\Sigma_{c}(\mathcal{L})\neq\emptyset$. According to Assumption
\ref{as:sing.leaves}, there are three types of such singular
leaves: non-degenerate leaves of rank $1$, degenerate leaves of
rank $1$ and non-degenerate leaves of rank $0$.

If $\mathcal{L}$ is a non-degenerate leaf of rank $1$, the statement
follows from Propositions \ref{prop:intersection} and
\ref{prop:fold}. The changes in Liouville foliations are represented
by Figures \ref{fig:intersection.orientable}--\ref{fig:double.fold}.

If $\mathcal{L}$ is a degenerate leaf admitting one of the allowed
normal forms of Assumption \ref{as:sing.leaves}, the statement
follows from the works of \cite{LU,Kal,BRF} as summarized by Proposition \ref{prop:parabolic}. Possible changes in
the Liouville foliations are described in Corollary
\ref{cor:parabolic} (see Appendix 2, Figures
\ref{fig:parabolic1}--\ref{fig:parabolic3}).

Finally, if $\mathcal{L}$ is a non-degenerate leaf of rank $0$,
i.e. it contains a fixed point of the action $\Phi$, the statement
follows from the works of   \cite{LU,Bol} as summarized by   Proposition \ref{prop:fixed.points}. The
corresponding changes in the Liouville foliations are listed in
Corollary \ref{cor:fixed.points} (see Appendix 2, Figures
\ref{fig:center-center}--\ref{fig:focus}).
\end{proof}

Clearly, The Singularities and Foliations Theorem may be applied to
the more restricted class of systems that  do not have any smooth families
of leaves of complexity $2$ (see Section
\ref{sec:examples}). All possible essential singularities of
such systems are included in the list of cases that appear in the
theorem. Yet, for such simple systems, several types of bifurcations that
are listed cannot be realized: the isolated non-orientable
saddle-saddle fixed point, the symmetric degenerate circles
undergoing a symmetric orientable saddle node bifurcation and the
hyperbolic pitchfork bifurcation. Indeed, these are exactly the cases in which curves
corresponding to complexity $2$ leaves appear near the critical set
(see Figures \ref{fig:saddle-saddle.3orientable},
\ref{fig:saddle-saddle.2orientable} and second lines of Figures
\ref{fig:parabolic1} and \ref{fig:parabolic3}).

\smallskip

\section{Changes of Liouville Foliations near Essential Singularities}
\label{sec:essential}
\subsection{Behavior near Non-Degenerate Circles}
\label{sec:non-degenerate.circles}

Consider a non-degenerate leaf $\mathcal{L}$ of rank $1$, such that
$\Sigma_{c}(\mathcal{L})\neq\emptyset$. By Assumption
\ref{as:stability}, the singularity of the bifurcation set appearing
at $\mu(\mathcal{L})$ is a transversal intersection of two smooth
curves or a quadratic fold. We analyze these two possibilities in
Propositions \ref{prop:intersection} and \ref{prop:fold} separately.

\begin{proposition}\label{prop:intersection}
Consider a Hamiltonian system $(\mathcal{M},\omega,H)$ satisfying
Assumptions \ref{as:smooth}--\ref{as:stability}. Suppose that a
non-degenerate leaf $\mathcal{L}$ of rank $1$ is such that
$\Sigma_{c}^{ess}(\mathcal{L})\neq\emptyset$ and $\mu(\mathcal{L})$
is a transversal intersection of two smooth curves from $\Sigma$.
Then, for sufficiently small $|h-H(\mathcal{L})|$, the isoenergy
surfaces for $h<H(\mathcal{L})$ and $h>H(\mathcal{L})$ are not
Liouville equivalent. The Fomenko graphs\footnote{In the Fomenko
graphs, we denote vertices corresponding to different smooth
families of singular circles by different symbols -- circles and
squares.} that describe all the possible structures of the isoenergy
surfaces near $\mathcal{L}$ are given by Figures
\ref{fig:intersection.orientable} and
\ref{fig:intersection.nonorientable}.
\begin{figure}[h]
\centering
\begin{minipage}[h]{0.44\textwidth}
\centering
\includegraphics[width=5cm,height=4cm]{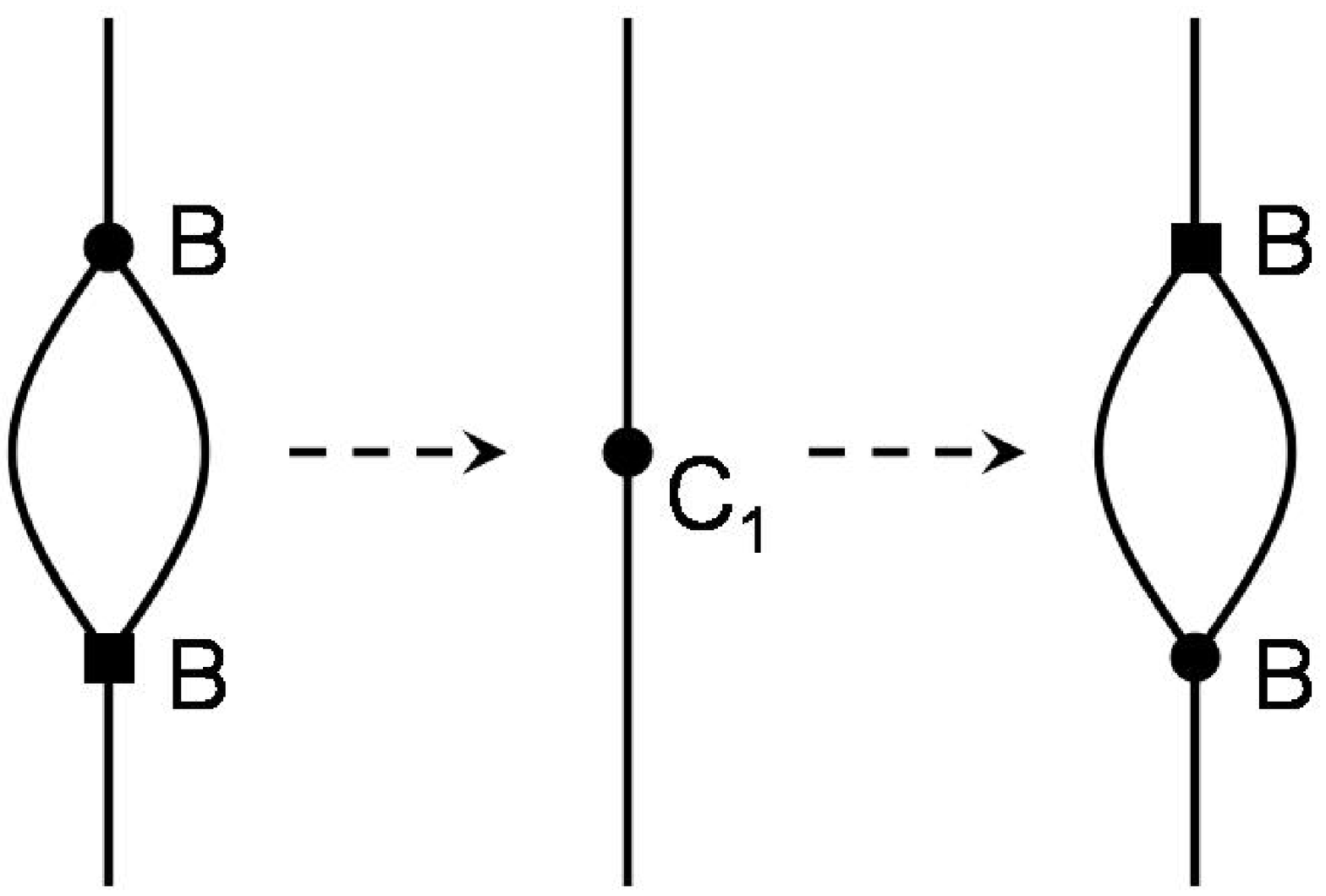}
\end{minipage}
\begin{minipage}[h]{0.44\textwidth}
\centering
\includegraphics[width=5cm,height=4cm]{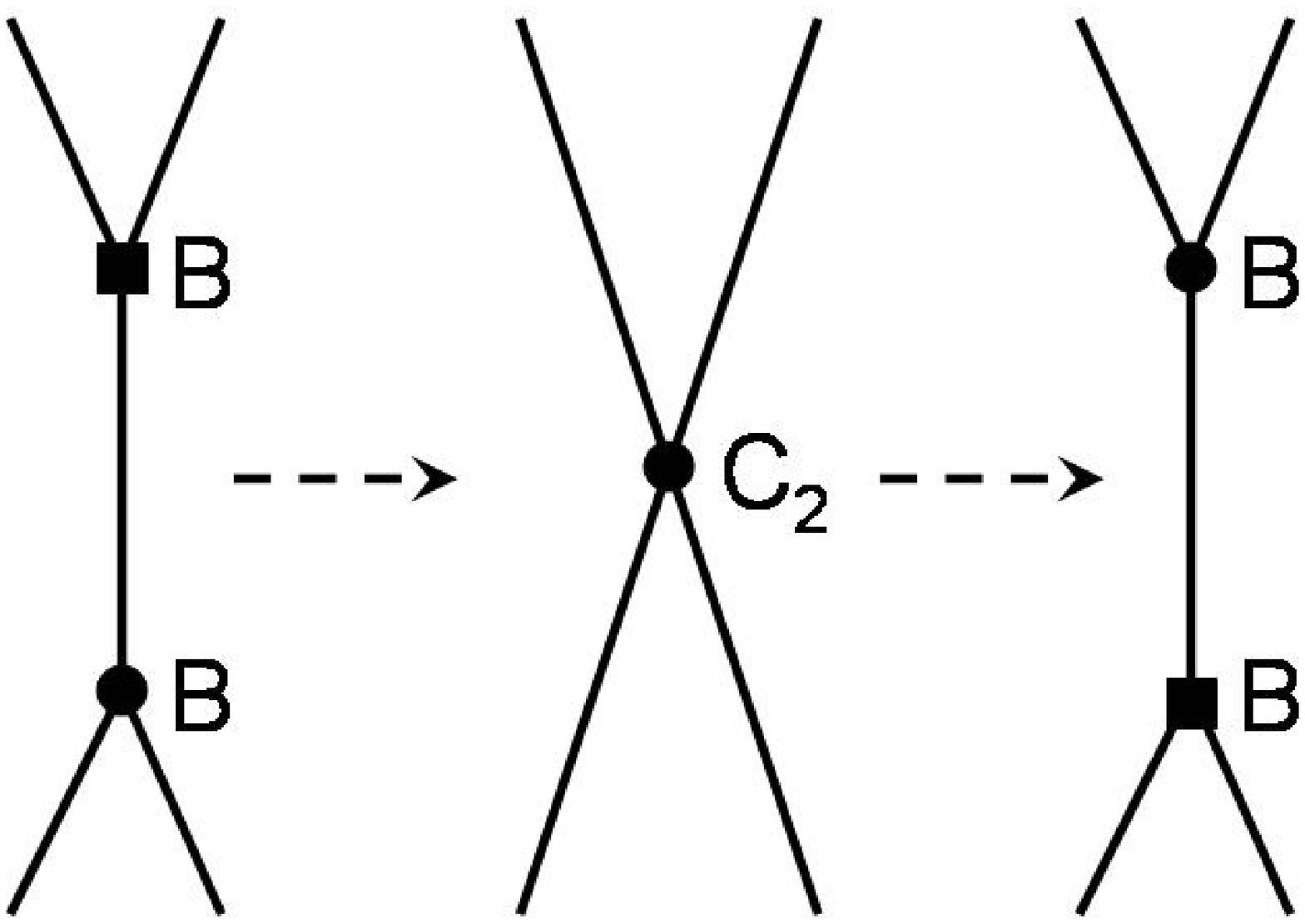}
\end{minipage}
\begin{minipage}[h]{0.44\textwidth}
\centering
\includegraphics[width=5cm,height=4cm]{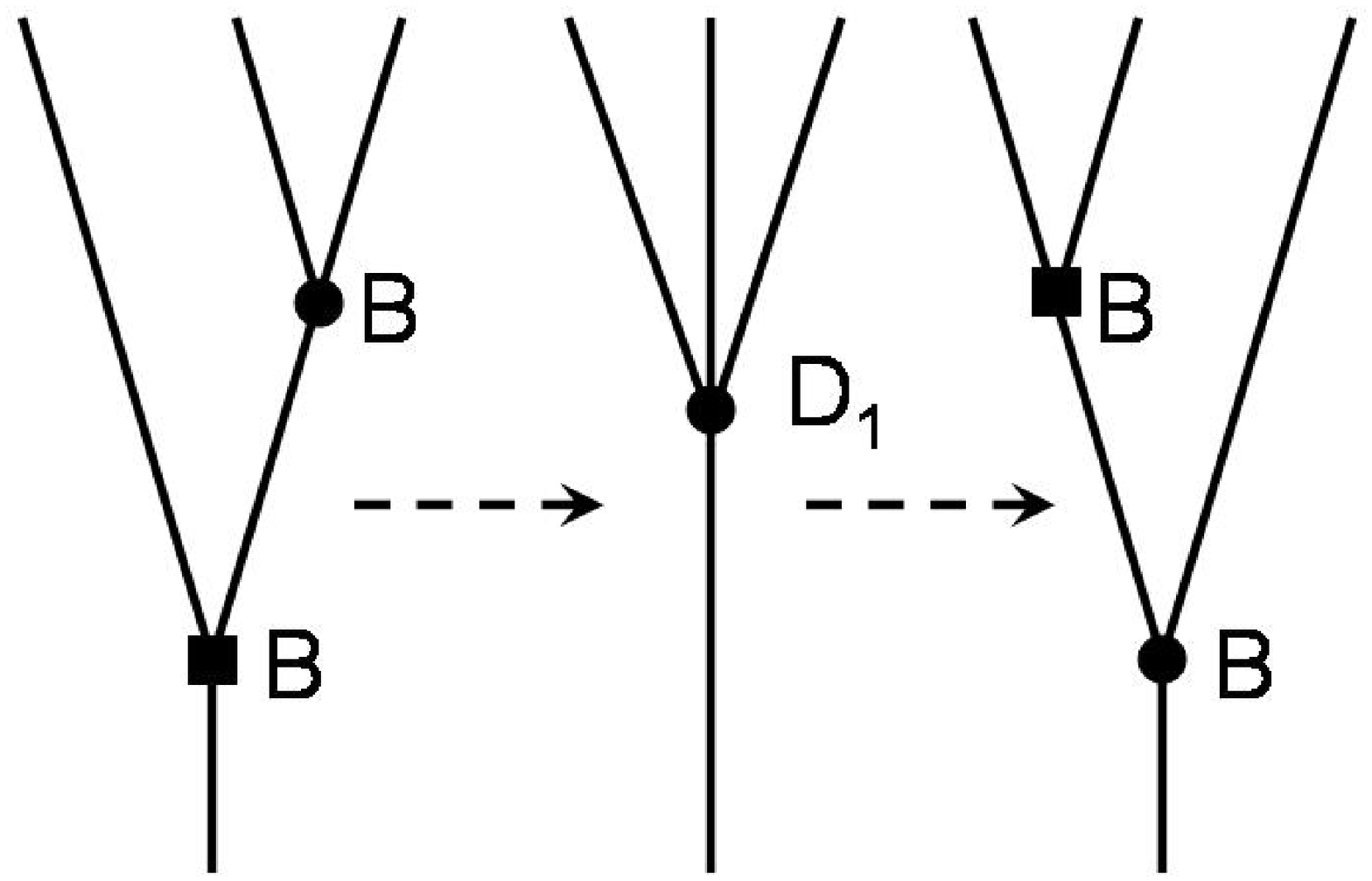}
\end{minipage}\begin{minipage}[h]{0.44\textwidth}
\centering
\includegraphics[width=5cm,height=4cm]{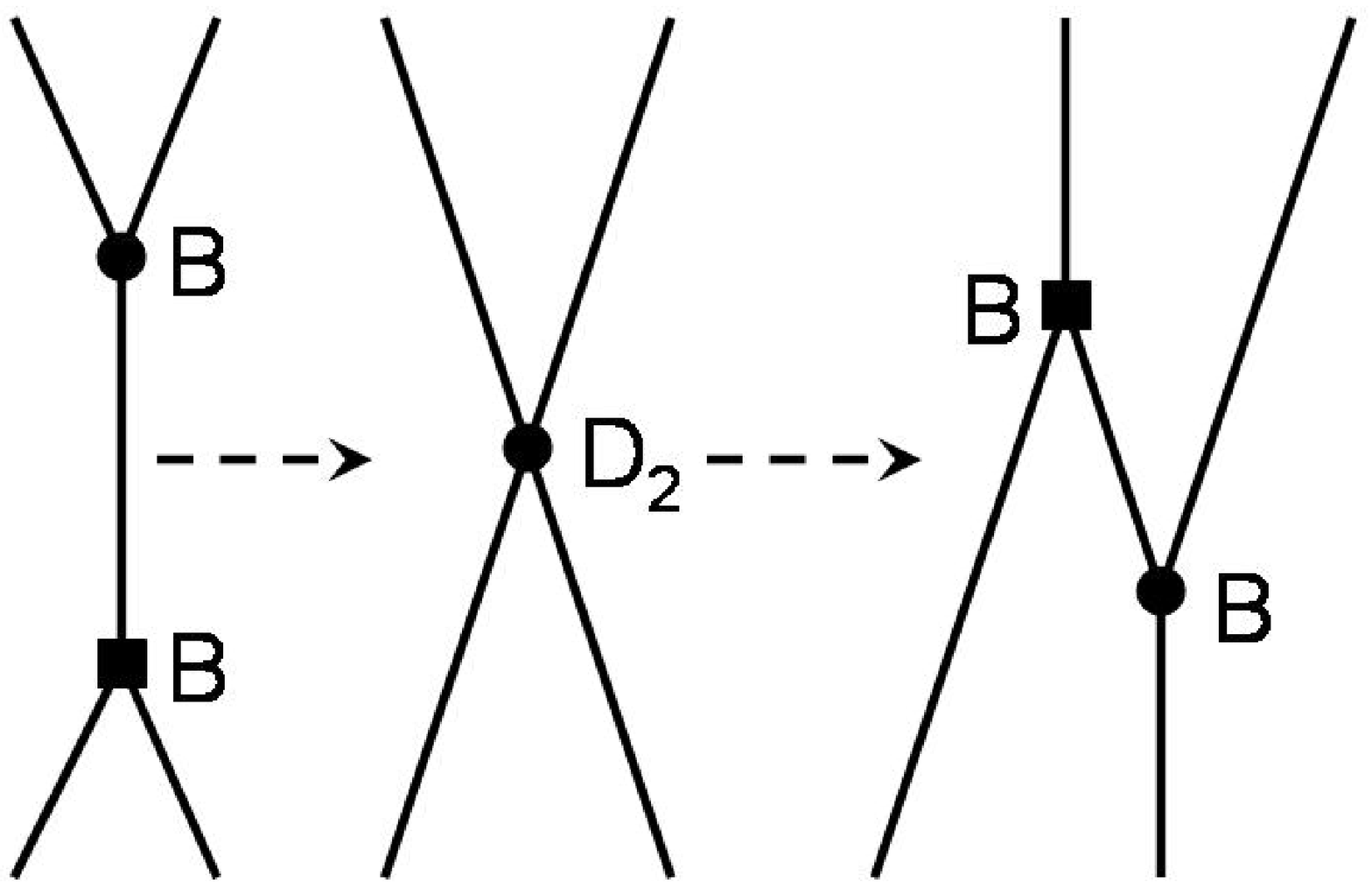}
\end{minipage}
\caption{Global bifurcations in the orientable case. The four
possible changes of the Liouville foliations near an intersection
of two families of orientable hyperbolic circles that creates a
non-degenerate singular leaf of rank $2$ (Proposition
\ref{prop:intersection}).} \label{fig:intersection.orientable}
\end{figure}

\begin{figure}[h]
\centering
\begin{minipage}[h]{0.46\textwidth}
\centering
\includegraphics[width=5cm,height=4cm]{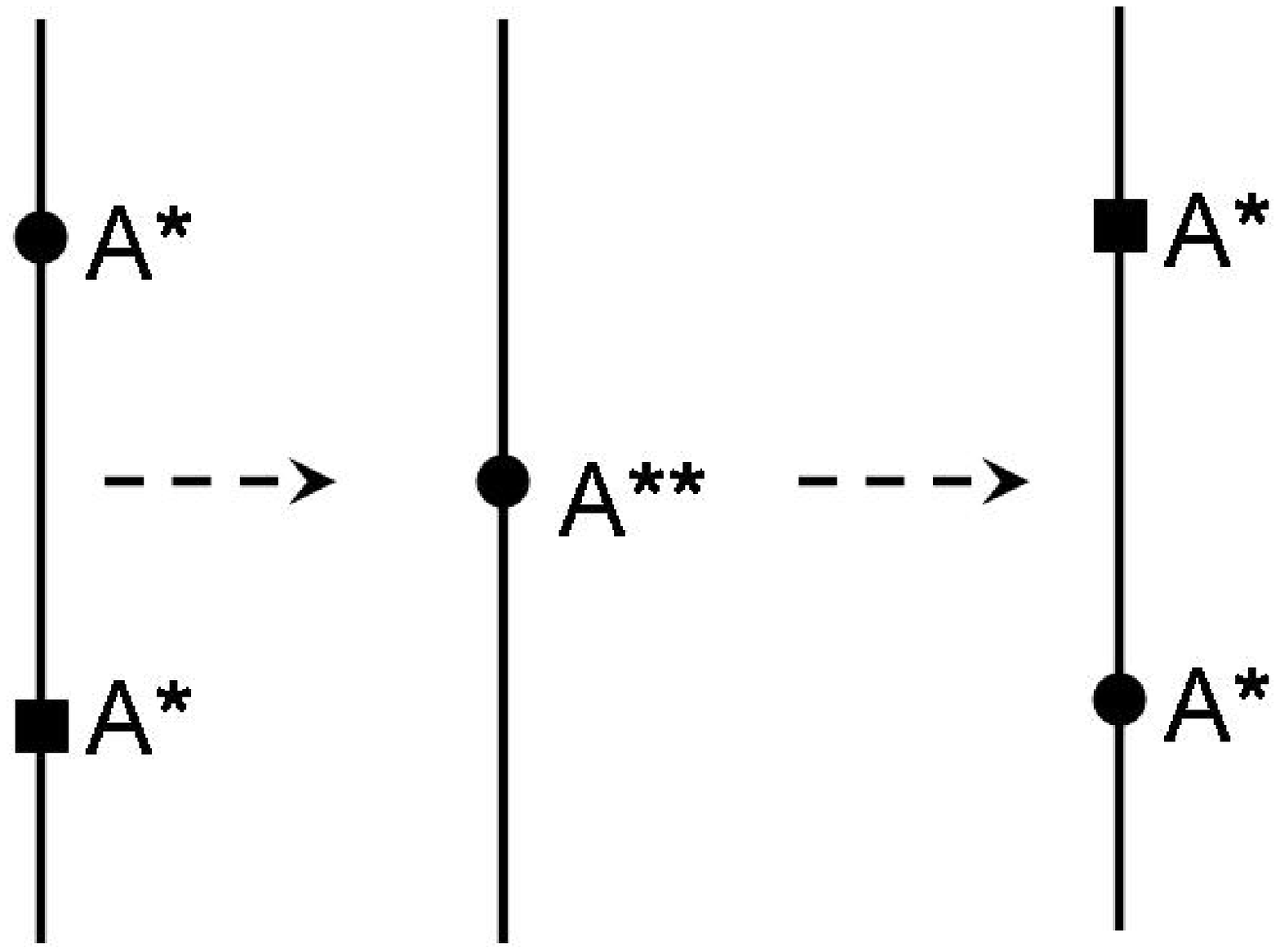}
\end{minipage}
\begin{minipage}[h]{0.46\textwidth}
\centering
\includegraphics[width=5cm,height=4cm]{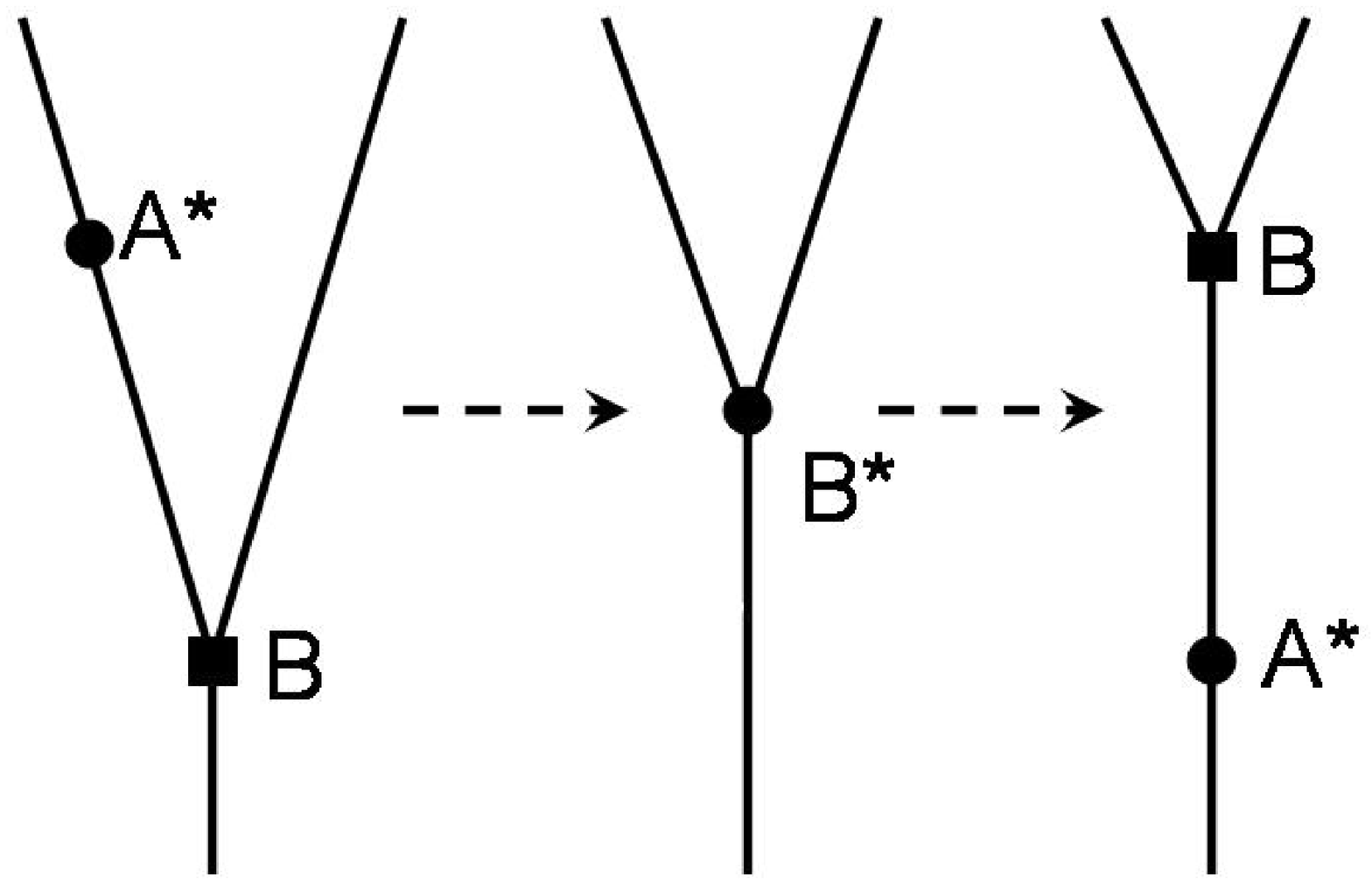}
\end{minipage}
\caption{Global bifurcations in the non-orientable cases. Chan\-ges
of the Liouville foliations near an intersection of two families of
hyperbolic circles that creates a non-degenerate singular leaf of
rank $2$. Left: both families of circles are non-orientable. Right:
one orientable and one non-orientable families. (Proposition
\ref{prop:intersection}).} \label{fig:intersection.nonorientable}
\end{figure}
\end{proposition}

\begin{proof}
Since, by Assumption \ref{as:sing.leaves}, $\mathcal{L}$ may be of
complexity of at most $2$,  each of the curves $c_{1}$ and $c_{2}$
that intersect at $\mu(\mathcal{L})$ corresponds, away from
$\mathcal{L}$, to a family of non-degenerate leaves of complexity
$1$ and rank $1$ and  the complexity of the leaf $\mathcal{L}$ is equal
to $2$. The possible foliations of the isoenergy surface near such a
leaf were completely classified by Fomenko (see \cite{BF2}): there
are exactly six kinds of such foliations, denoted by atoms
$\mathbf{C}_{1}$, $\mathbf{C}_{2}$, $\mathbf{D}_{1}$,
$\mathbf{D}_{2}$, $\mathbf{A}^{\ast\ast}$, $\mathbf{B}^{\ast}$.

\smallskip

Notice that the orientability of the circles along each of the
curves $c_{1}$ and $c_{2}$ is fixed. Consider first the four atoms
of complexity $2$ that have orientable circles: $\mathbf{C}_{1}$,
$\mathbf{C}_{2}$, $\mathbf{D}_{1}$, $\mathbf{D}_{2}$. Each of these
atoms dictates uniquely the form of the foliations in the nearby
isoenergy surfaces as summarized by Figure
\ref{fig:intersection.orientable}.

\smallskip

Suppose first that the Fomenko atom corresponding to the isoenergy
surface of $\gamma$ is $\mathbf{C}_{1}$. Because only one edge of
the Fomenko graph joins the atom $\mathbf{C}_{1}$ from each side,
it follows that there is exactly one Liouville torus projected to
each point placed above or under both the curves $c_{1}$ and
$c_{2}$. This means that two such tori are projected to points
between the curves.

\smallskip

Now consider the case of atom $\mathbf{D}_{1}$. Then, three
families of Liouville tori should appear above the two curves
$c_{1}$ and $c_{2}$, and only one family appears for points below
the two curves. The unique way to achieve this behavior is shown.

\smallskip
The remaining cases, of atoms $\mathbf{C}_{2}$ and
$\mathbf{D}_{2}$ exhibit a similar behavior: both have two
families of tori below and above the two curves $c_{1}$ and
$c_{2}$. To distinguish between these two cases, we note that the
singular level sets corresponding to the atoms $\mathbf{C}_{1}$
and $\mathbf{C}_{2}$ are isomorphic -- both have two oriented
singular periodic trajectories and four heteroclinic separatrices.
Similarly, $\mathbf{D}_{1}$ and $\mathbf{D}_{2}$ are isomorphic:
both have two homoclinic separatrices -- one joined to each circle
from $\mathcal{L}$, and two heteroclinic ones.

\smallskip

Suppose now that $\mathbf{C}_{2}$ is the atom joined to the
isoenergy surface containing $\mathcal{L}$. Now, we will look for
the Fomenko graphs corresponding to the submanifolds
$K=\mathrm{const}$. The atom corresponding to $K=K(\mathcal{L})$
must be one with the singular level set isomorphic to
$\mathbf{C}_{2}$, thus it is $\mathbf{C}_{1}$ or $\mathbf{C}_{2}$.
If it is $\mathbf{C}_{2}$, we would have again two Liouville tori
above any point between the curves $c_{1}$ and $c_{2}$. But, it is
easy to see that this is
not possible since the curves $c_{1}$, $c_{2}$ correspond to $\mathbf{B}%
$-atoms. So, the Fomenko atom for the submanifold
$K=K(\mathcal{L})$ is $\mathbf{C}_{1}$ which was described above.
Hence, there is exactly one family of Liouville tori between the
curves $c_{1}$ and $c_{2}$. Finally, the case of atom
$\mathbf{D}_{2}$ may be similarly analyzed. We conclude that the
atom $\mathbf{D}_{1}$ is joined to the submanifold $K=K(\gamma)$,
and the Fomenko graphs describing the extended neighborhood $V$
are indeed uniquely constructed and can be seen in Figure
\ref{fig:intersection.orientable}.

\smallskip

Cases when at least one of the curves $c_{1}$, $c_{2}$ corresponds
to a family of nonorientable circles are resolved directly from
the structure of the corresponding Fomenko atoms:
$\mathbf{A}^{\ast\ast}$ and $\mathbf{B}^{\ast}$. The graphs are
shown on Figure \ref{fig:intersection.nonorientable}.

\smallskip

Finally, let us remark that the graphs corresponding to the cases
when the atoms $\mathbf{C}_{1}$, $\mathbf{C}_{2}$,
$\mathbf{A^{**}}$ appear, are isomorphic for $h<H(\mathcal{L})$
and $h>H(\mathcal{L})$. Nevertheless, since the $r$-marks
corresponding to the upper and lower edges will be changed, the
Liouville foliations are not equivalent.
\end{proof}

We now list the Liouville foliation structure of the isoenergy
surfaces near the circles of fixed points that appear persistently
when a curve belonging to $\Sigma$ has a fold.

\begin{proposition} \label{prop:fold}
Consider a Hamiltonian system $(\mathcal{M},\omega,H)$ satisfying
Assumptions \ref{as:smooth}-\ref{as:stability}. Suppose that a
non-degenerate leaf $\mathcal{L}$ of rank $1$ is such that
$\mu(\mathcal{L})$ is a fold of a smooth curve from $\Sigma$. Then
$\Sigma_{c}^{ess}(\mathcal{L})\neq\emptyset$. Moreover, for
sufficiently small $|h-H(\mathcal{L})|$, the isoenergy surfaces for
$h<H(\mathcal{L})$ and $h>H(\mathcal{L})$ are not Liouville
equivalent. If $\mathcal L$ is of complexity $1$, then the possible
foliations of isoenergy surfaces near the leaf are completely
described by Figures \ref{fig:fold.elliptic},
\ref{fig:fold.hyp.orientable}, and \ref{fig:fold.hyp.nonorientable}.
For the leaf of complexity $2$, the foliations are shown on Figure
\ref{fig:double.fold}.
\end{proposition}

\begin{proof}
Consider a sufficiently small extended neighborhood $V$ of
$\mathcal{L}$. Clearly, for sufficiently small $\varepsilon$, only
one submanifold $V\cap{H}^{-1}(H(\mathcal{L})+\varepsilon)$,
$V\cap{H}^{-1}(H(\mathcal{L})-\varepsilon)$ contains singular
leaves. Thus, the corresponding isoenergy surfaces are not Liouville
equivalent.

\smallskip

Now, suppose that $\mathcal L$ contains a unique circle $\gamma$.
Denote by $c$ the curve of $\Sigma$ with the fold $\mu(\gamma)$. $c$
divides a small convex neighborhood $N$ of $\mu(\gamma)$ in $\mathbf
R^2$ into a convex and concave part, see Figure
\ref{fig:fold.elliptic}.

\begin{figure}[h]
\centering
\begin{minipage}[h]{0.44\textwidth}
\centering
\includegraphics[width=5cm,height=4cm]{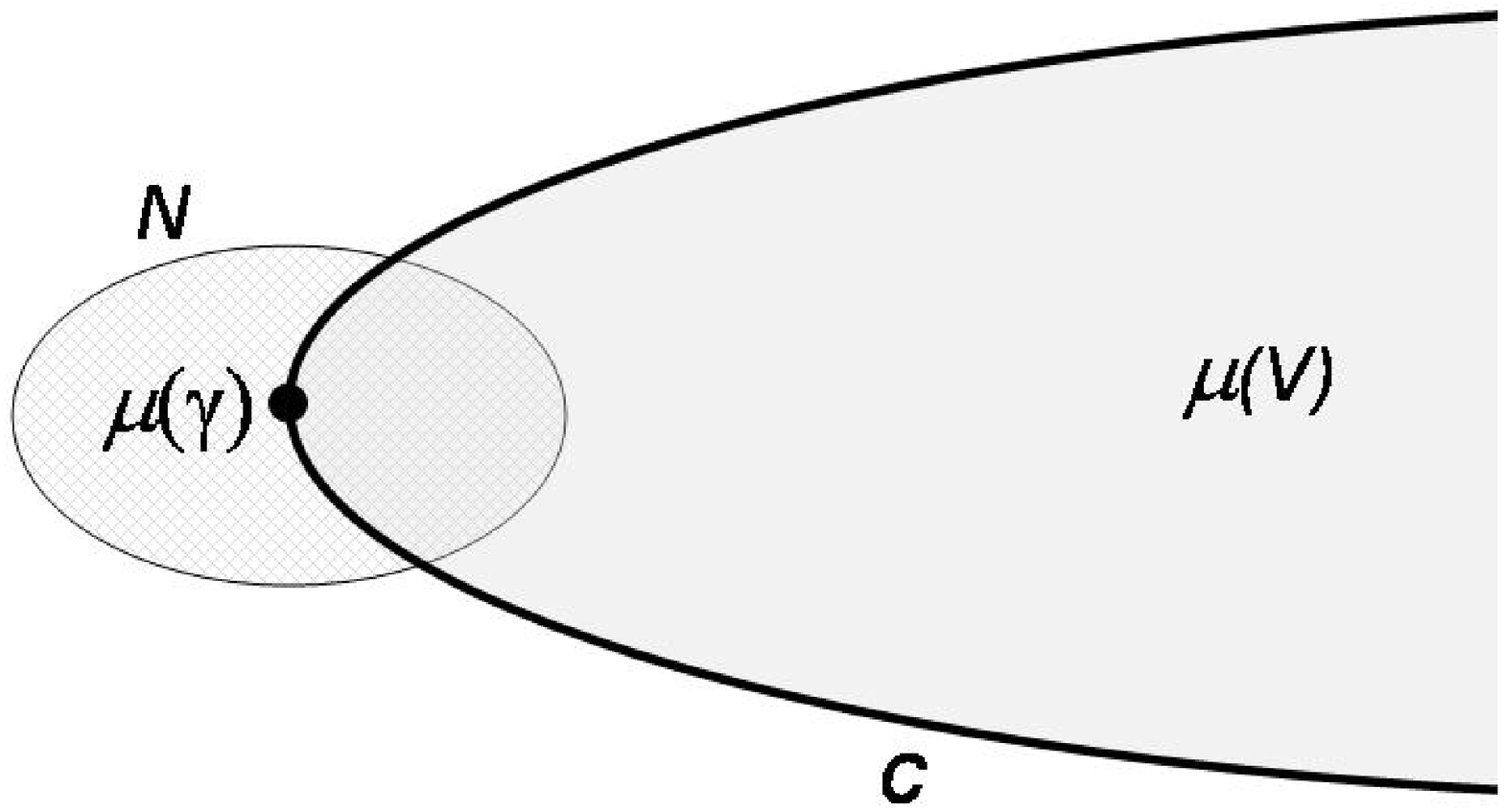}
\end{minipage}
\begin{minipage}[h]{0.44\textwidth}
\centering
\includegraphics[width=5cm,height=4cm]{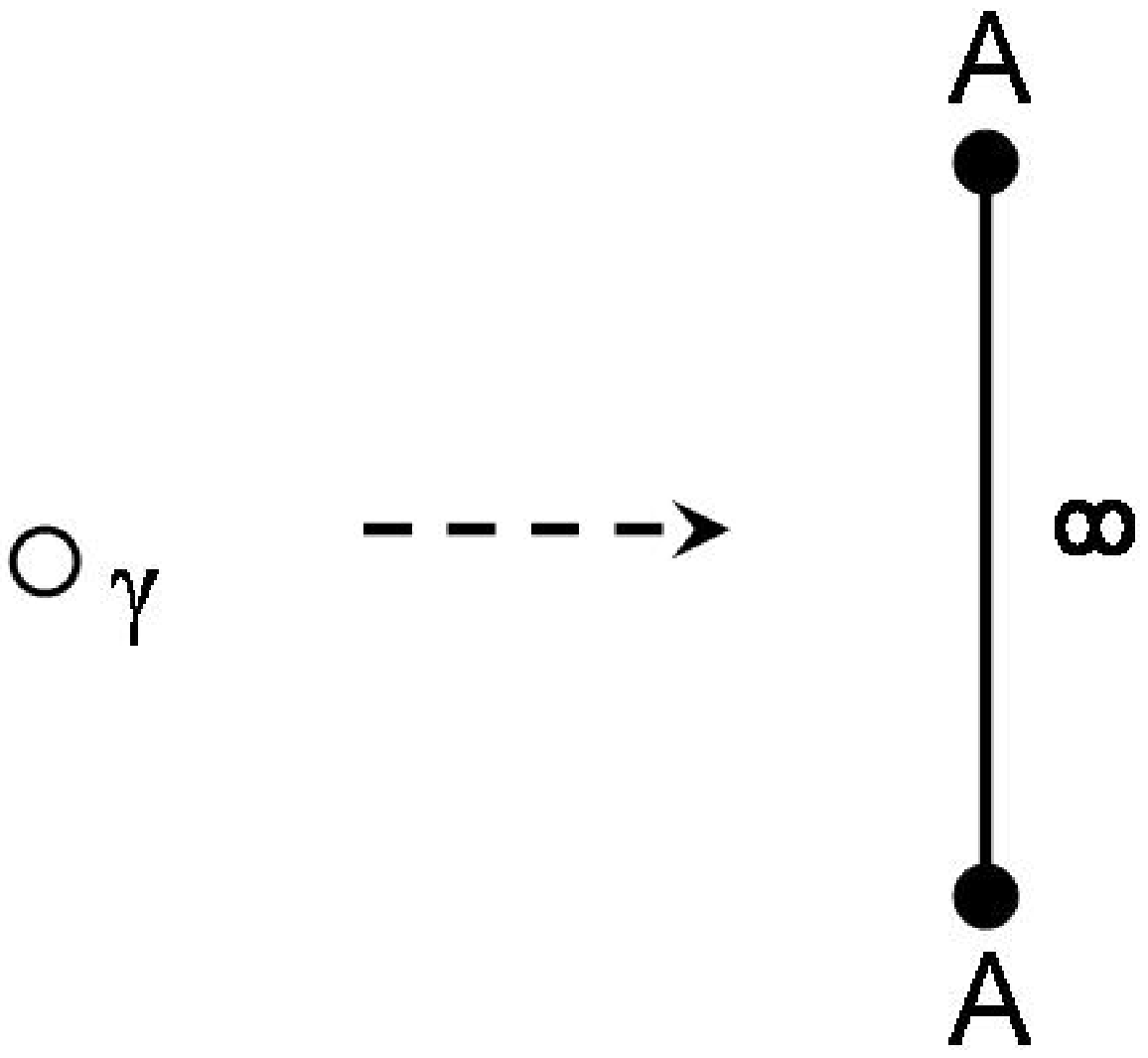}
\end{minipage}
\begin{minipage}[h]{0.44\textwidth}
\centering
\includegraphics[width=5cm,height=4cm]{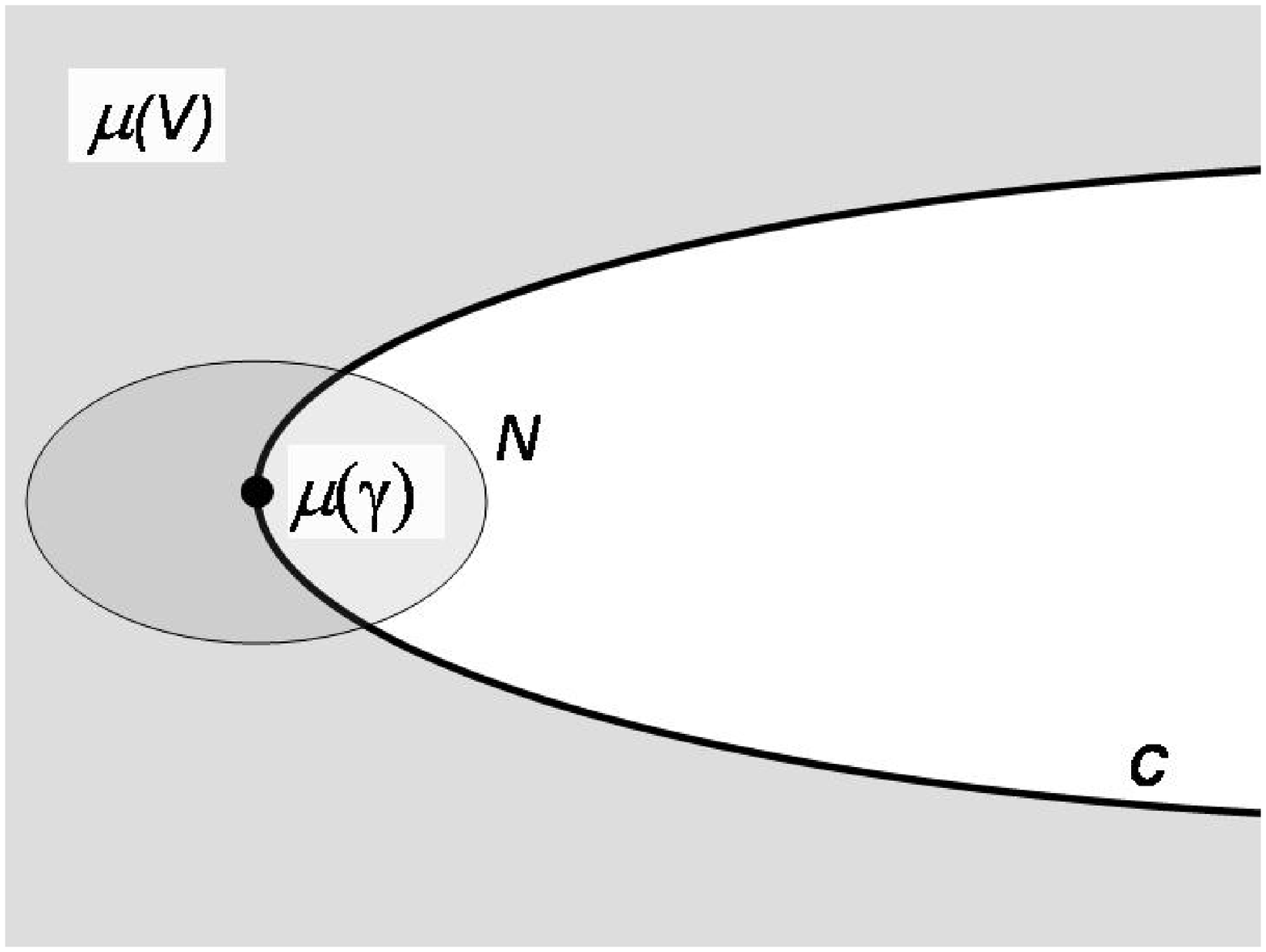}
\end{minipage}
\begin{minipage}[h]{0.44\textwidth}
\centering
\includegraphics[width=5cm,height=4cm]{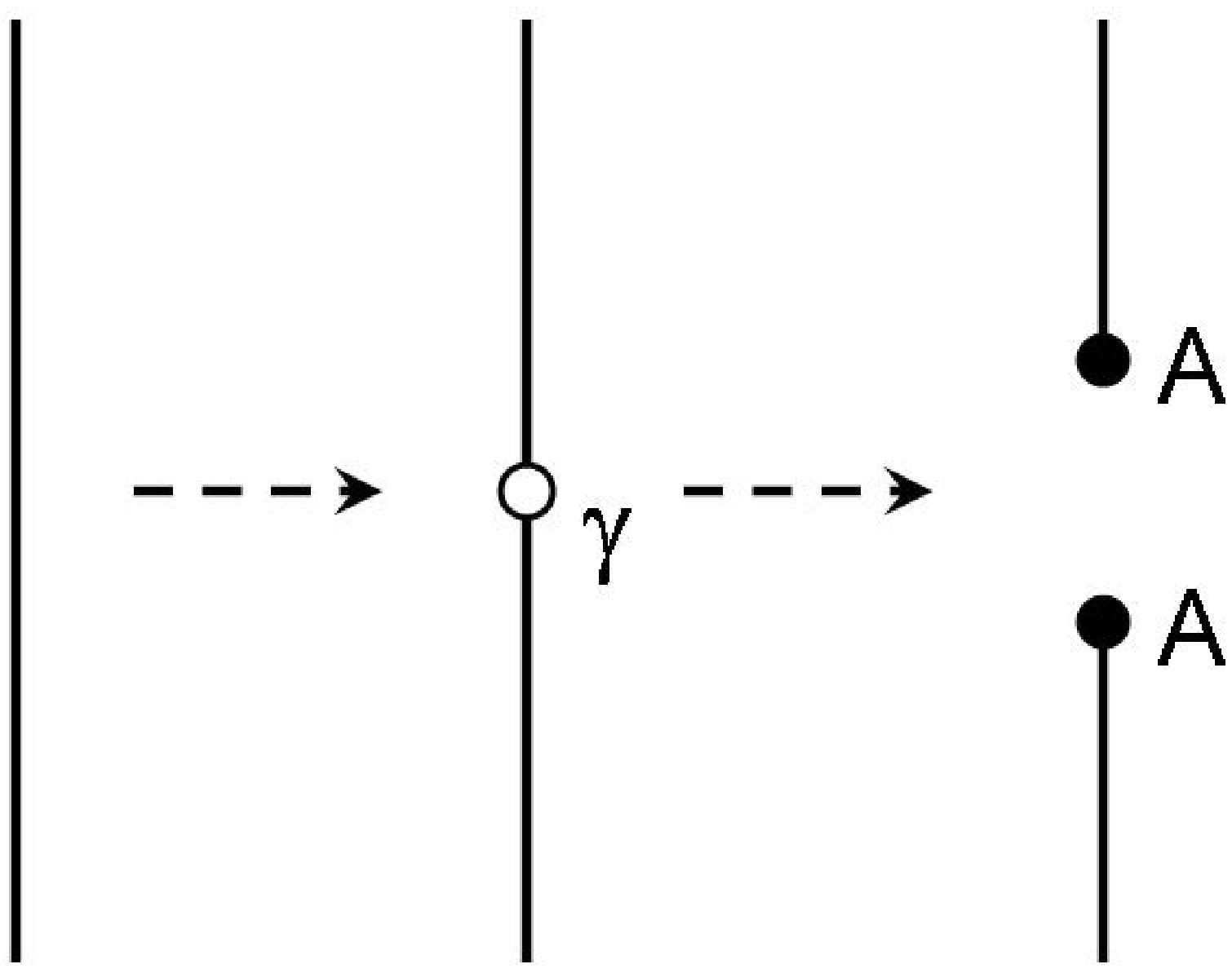}
\end{minipage}
\caption{The two cases of an Elliptic fold; Left: the bifurcation
sets --- the solid curve corresponds to a family of elliptic circles
and the allowed region of motion is shaded. Right: the  changes in
the Liouville foliations as described by the Fomenko graphs
(Proposition \ref{prop:fold}).} \label{fig:fold.elliptic}
\end{figure}

Suppose first that $\gamma$ is a normally elliptic circle, i.e.\
that the curve $c$ corresponds to a family of Fomenko
$\mathbf{A}$-atoms. If $\mu(V)$ has no intersection with the concave
part of $N$ (first line of Figure \ref{fig:fold.elliptic}), the energy surfaces near $\gamma$ are locally
diffeomorphic to $\mathbf{S}^{2}\times\mathbf{S}^{1}$. On the other
hand, if $\mu(V)$ has no intersection with the convex part of $N$ (second line of Figure \ref{fig:fold.elliptic}),
the isoenergy surfaces near $\gamma$ are locally diffeomorphic to
$(0,1)\times\mathbf{T}^{2}$ for $H<H(\gamma)$ and to a disjoint
union of two solid tori for $H>H(\gamma)$, or vice versa.
\begin{figure}[h]
\centering
\begin{minipage}[h]{0.46\textwidth}
\centering
\includegraphics[width=5cm,height=4cm]{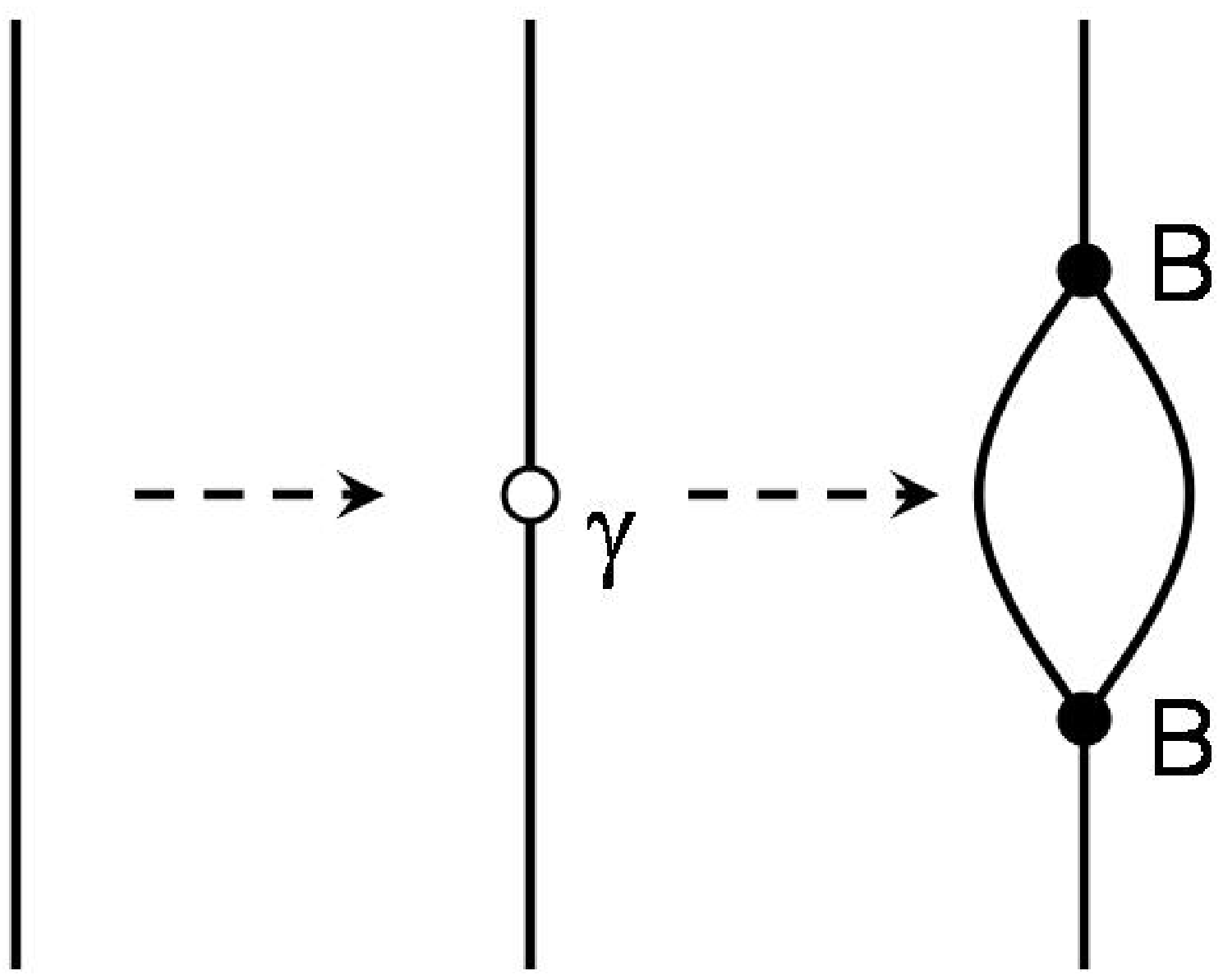}
\end{minipage}
\begin{minipage}[h]{0.46\textwidth}
\centering
\includegraphics[width=5cm,height=4cm]{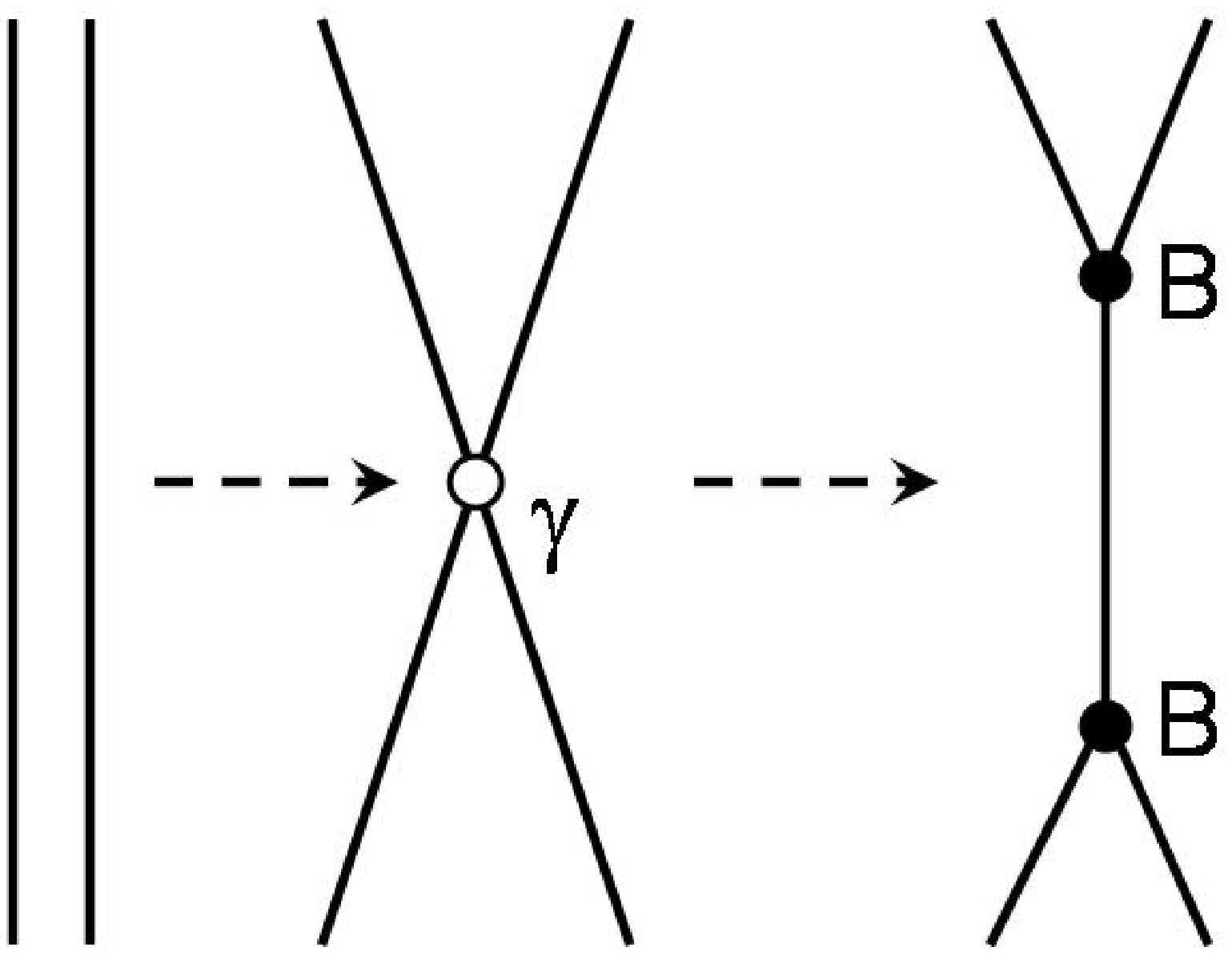}
\end{minipage}
\caption{An orientable hyperbolic fold; The two possible chan\-ges
in the Liouville foliations near a fold of a curve in the
bifurcation set when the curve corresponds to a family of hyperbolic
orientable circles (Proposition \ref{prop:fold}).}
\label{fig:fold.hyp.orientable}
\end{figure}

If $\gamma$ is normally hyperbolic and orientable, there may be one
or two Liouville tori in $V$ over each point in the convex part of
$N$. The corresponding Fomenko graphs are shown on Figure
\ref{fig:fold.hyp.orientable}.

\begin{figure}[h]
\centering
\begin{minipage}[h]{0.44\textwidth}
\centering
\includegraphics[width=5cm,height=4cm]{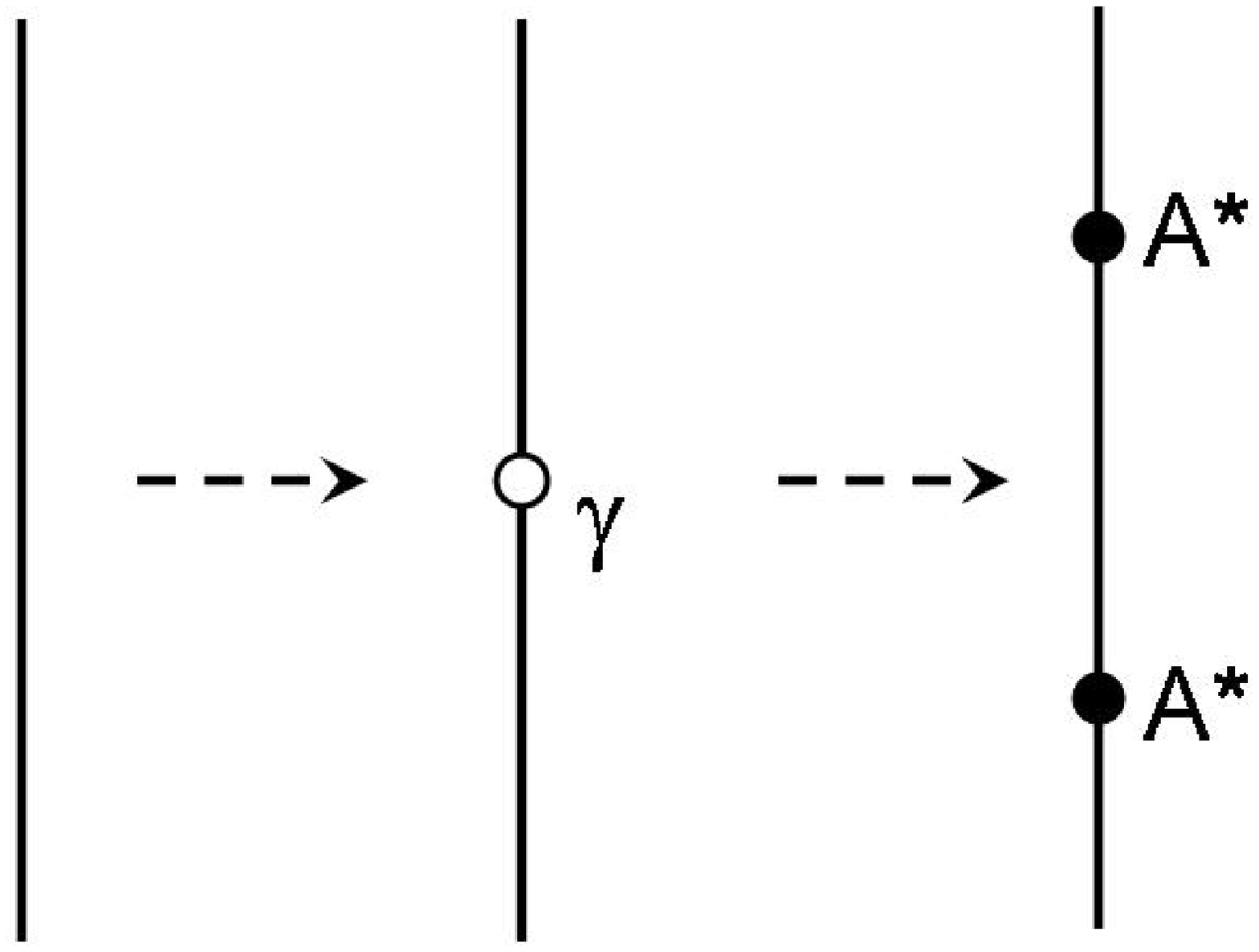}
\end{minipage}
\caption{A non-orientable hyperbolic fold; The change in the
Liouville foliations near a fold of a curve in the bifurcation set
when the curve corresponds to a family of hyperbolic
non-orientable circles (Proposition \ref{prop:fold}).}
\label{fig:fold.hyp.nonorientable}%
\end{figure}

If $\gamma$ is non-orientable normally hyperbolic circle, the
corresponding Fomenko graphs are shown on Figure
\ref{fig:fold.hyp.nonorientable}.

\smallskip

Finally, for $\mathcal L$ of complexity $2$, the Fomenko graphs are
shown on Figure \ref{fig:double.fold}.\end{proof}

\begin{figure}[h]
\centering
\begin{minipage}[h]{0.24\textwidth}
\centering
\includegraphics[width=3cm,height=4.5cm]{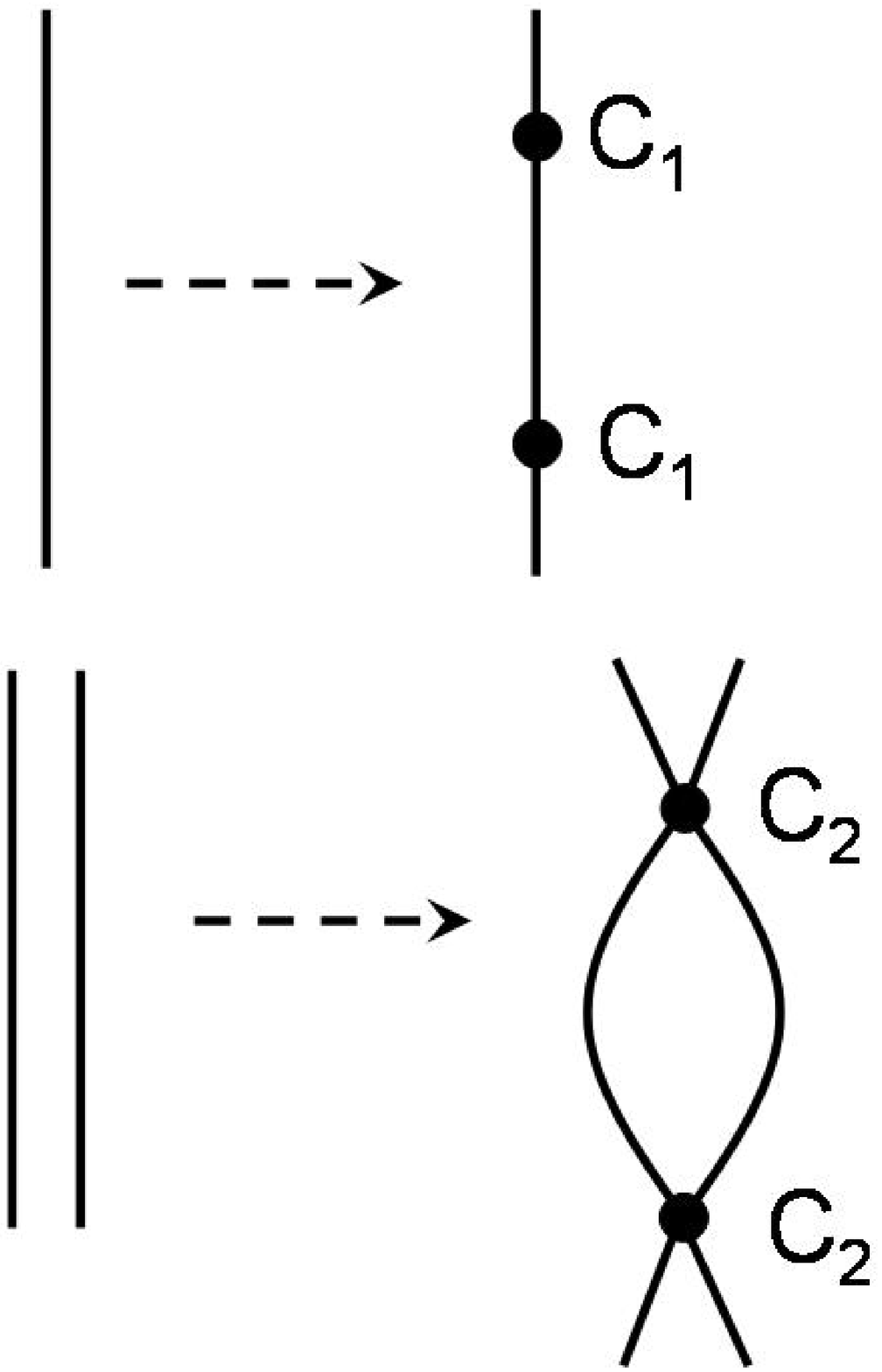}
\end{minipage}
\begin{minipage}[h]{0.24\textwidth}
\centering
\includegraphics[width=3cm,height=4.5cm]{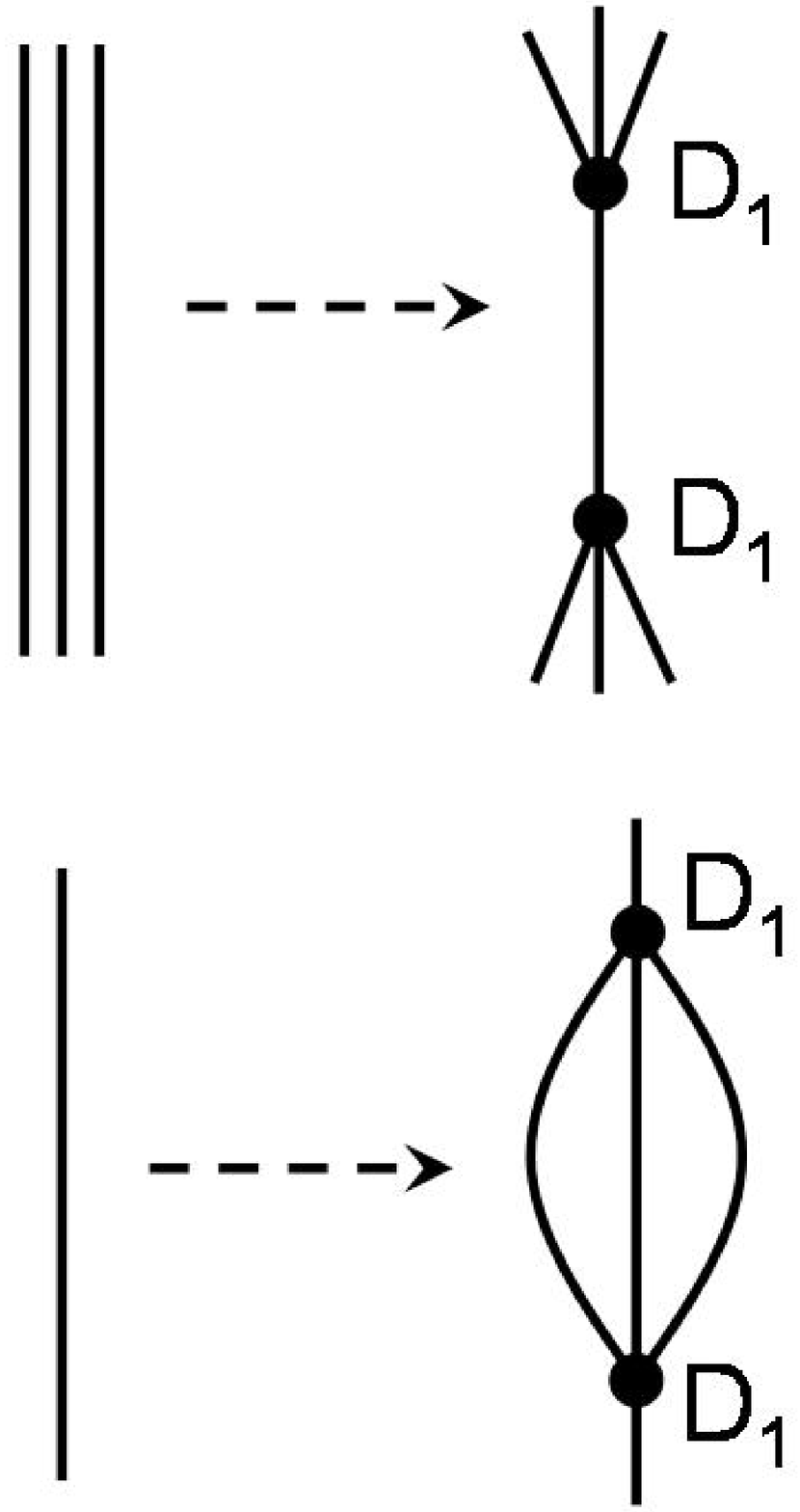}
\end{minipage}
\begin{minipage}[h]{0.24\textwidth}
\centering
\includegraphics[width=3cm,height=4.5cm]{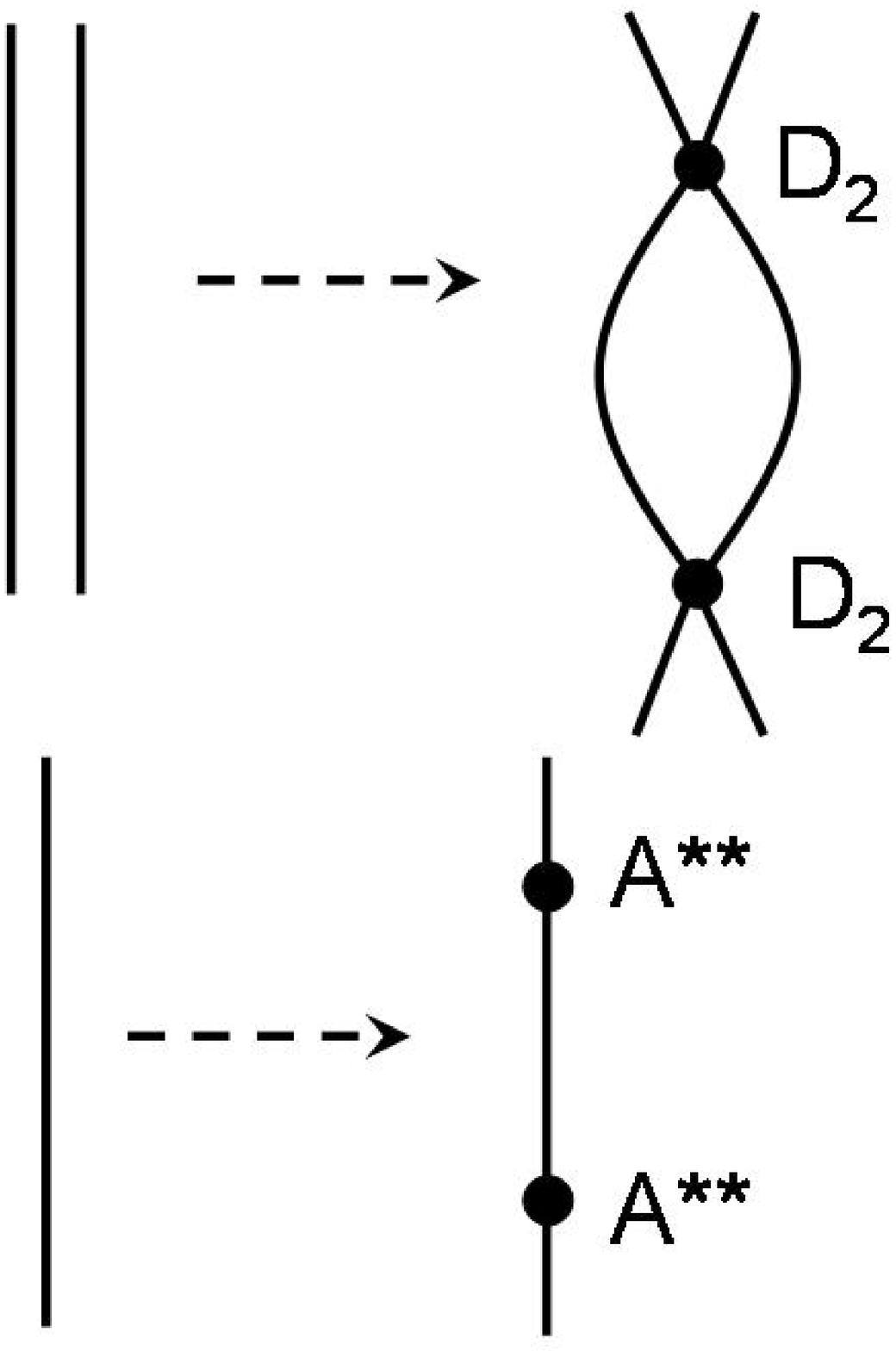}
\end{minipage}
\begin{minipage}[h]{0.24\textwidth}
\centering
\includegraphics[width=3cm,height=4.5cm]{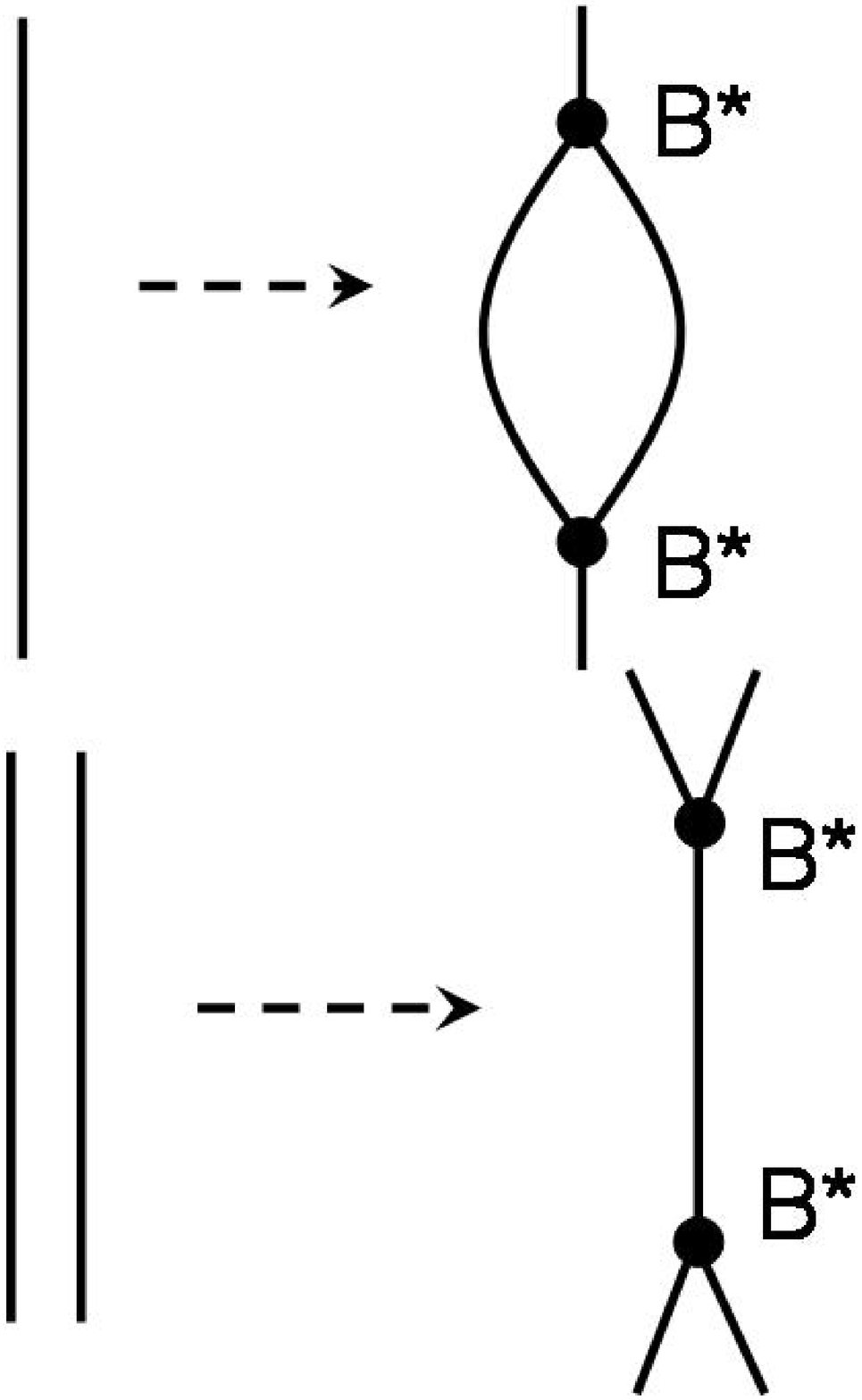}
\end{minipage}
\caption{A fold of complexity $2$; The possible changes in the
Liouville foliations near a fold of a curve in the bifurcation set
when the curve corresponds to a family of non-degenerate complexity
$2$ leaves (Proposition \ref{prop:fold}).}
\label{fig:double.fold}%
\end{figure}

\subsection{Behavior near Parabolic Circles}
\label{sec:parabolic.circles}

The behavior of the level sets near degenerate singular leaves that
are allowed by Assumption \ref{as:sing.leaves}, namely near
parabolic circles, was essentially described in the generic,
non-symmetric case in \cite{LU} and \cite{Kal}, and by \cite{BRF} under some additional symmetry assumptions. From these works we
can immediately conclude:

\begin{proposition}\label{prop:parabolic}
Consider a Hamiltonian system $(\mathcal{M},\omega,H)$ satisfying
Assumptions \ref{as:smooth}--\ref{as:sing.leaves}, and its
degenerate leaf $\mathcal{L}$ of rank $1$. Then
 $\mu(\mathcal L)\in\Sigma_c^{ess}$.
Moreover, for sufficiently small $|h-H(\mathcal{L})|$, the isoenergy
surfaces for $h<H(\mathcal{L})$ and $h>H(\mathcal{L})$ are not
Liouville equivalent. The bifurcation sets and foliations of
isoenergy surfaces near $\mathcal{L}$ are represented by Figures
\ref{fig:parabolic1}, \ref{fig:parabolic2}, and \ref{fig:parabolic3}
in Appendix 2.
\end{proposition}

\begin{proof}
The local bifurcation sets near parabolic circles satisfying
Assumption \ref{as:sing.leaves} and some more general cases, are
constructed in \cite{BRF}, see also \cite{LU} and \cite{Kal}. The
bifurcation sets are represented by Figures
\ref{fig:parabolic1}--\ref{fig:parabolic3} in Appendix 2. Clearly,
in all cases several curves meet at $\mu(\mathcal{L})$ so it is an
essential singularity of the bifurcation set.

\smallskip

In addition, since the number of singular leaves on the isoenergy
surfaces near $\mathcal{L}$ is not the same for $h<H(\mathcal{L})$
and $h>H(\mathcal{L})$, they cannot be Liouville equivalent.

\smallskip

The corresponding Fomenko graphs are represented by Figures
\ref{fig:parabolic1}--\ref{fig:parabolic3} in Appendix 2 --- these
graphs are easily inferred from the circular graphs of \cite{BRF}.
Notice that curves of leafs of complexity 2 are created in some of
the symmetric cases.
\end{proof}

\subsection{Behavior near Fixed Points}\label{sec:fixed.points}

The book of Lerman and Umanskii \cite{LU} provides the analysis and
complete description of the structure of the level sets near
non-degenerate fixed points, under somewhat weaker conditions than
Assumptions \ref{as:smooth}--\ref{as:sing.leaves}. In Appendix 2, we
succinctly summarize this description of the foliation near the
fixed points in Corollary \ref{cor:fixed.points} by using the
bifurcation diagrams and Fomenko graphs (see also \cite{Bol,Mat} for
a very similar description). We remark that although the resulting
graphs are simple, the topology associated with each of these cases
may be highly non-trivial as is described in details in \cite{LU}.
The following proposition is needed to prove our main singularity
and bifurcation theorem.

\begin{proposition}\label{prop:fixed.points}
Let $(\mathcal{M},\omega,H)$ be a Hamiltonian system satisfying
Assumptions \ref{as:smooth}-\ref{as:sing.leaves} and
$p\in\mathcal{M}$ a fixed point of the Poisson action. Then
$\mu(p)\in\Sigma_{c}^{ess}$.

Moreover, for sufficiently small $|h-H(p)|$, the isoenergy
surfaces for $h<H(p)$ and $h>H(p)$ are not Liouville equivalent.
The bifurcation sets and the possible structures of the isoenergy
surfaces near $p$ are shown on Figures
\ref{fig:center-center}--\ref{fig:focus} in Appendix 2.
\end{proposition}

\begin{proof}
The local bifurcation sets near different types of fixed points
are constructed in \cite{LU}. Here, they are represented by
Figures \ref{fig:center-center}--\ref{fig:focus} in Appendix 2,
and it is clear that in each case $\mu(p)$ is an essential
singularity of the bifurcation set as several curves meet at
$\mu(p)$.

Relying on \cite{LU} and \cite{Bol}, we list in Appendix 2 the
marked Fomenko graphs for the isoenergy surfaces close to $p$ for
all the non-degenerate cases. When those graphs are different for
$h<H(p)$ and $h>H(p)$, the proposition immediately follows.
However, in the few cases that are listed below, the corresponding
graphs are the same, so additional arguments are needed.

The first case appears when $p$ is of the center-center type and
$\mu(p)$ is not a local extremum point for the Hamiltonian $H$
(second line of Figure \ref{fig:center-center}). Yet, the
$r$-marks corresponding to the edges of the Fomenko graphs
represented by Figure \ref{fig:center-center} are different. Thus,
the isoenergy surfaces are not Liouville equivalent.

The second case is when $p$ is of the saddle-saddle type, with all
four one-dimensional orbits of action in its leaf being orientable,
when all near-by isoenergy surfaces have the same number of singular
leaves, as shown on the second line of Figure
\ref{fig:saddle-saddle.4orientable}. However, similarly as before
the $r$-marks corresponding to the upper and lower edges of Fomenko
graphs are not the same on for $h<H(p)$ and $h>H(p)$.

The last case is when $p$ is of focus-focus type, as shown by
Figure \ref{fig:focus}. Then the leaf of $p$ is the only singular
leaf in a sufficiently small extended neighborhood. As it is noted
in Remark \ref{rem:monodromy} of Appendix 2, the $r$-marks on the
corresponding edges of Fomenko graphs are changed, which concludes
the proof.
\end{proof}

\section{Examples}
\label{sec:examples}

\subsection{Examples of Systems with only Isolated Leaves of Complexity $2$}

Integrable Hamiltonian systems of the form:
\[
H(x,y,I,\theta) = H_{0}(x,y,I)
 \quad
(x,y,I,\theta)\in\mathbf{R}^{3}\times\mathbf{S}^{1}
\]
with non-degenerate dependence on $(x,y,I)$ (and in particular no
$\mathbf Z_2$--symmetry in the $(x,y)$--plane) satisfy Assumptions
\ref{as:smooth}-\ref{as:stability}. For example, for an open dense
set of $\varepsilon,\omega>0$, the following system
\begin{equation}\label{eq:ham0}
H(x,y,I,\theta;\varepsilon,\omega)
 =
\frac{y^{2}}{2} + I\,\frac{x^{2}}{2} + \varepsilon x^{3} + \frac{x^{4}}{4} + (I+\omega)^{2},
 \quad
(x,y,I,\theta)\in \mathbf{R}^{3}\times\mathbf{S}^{1}
\end{equation}
satisfies these conditions: it has compact level sets and compact
isoenergy surfaces and is non-degenerate. Since it has no fixed
points, and the only rank $1$ orbits correspond to $(x(I),y(I))$
values at which $\nabla_{(x,y)}H_{0}=0$, Assumptions
\ref{as:sing.leaves}--\ref{as:stability} may be easily verified.

\smallskip

To produce non-degenerate fixed points in the above construction we
may take:
\[
I=u^{2}+v^{2}-c
\]
where $c\neq0$ is yet another non-degeneracy parameter, so that
equation (\ref{eq:ham0}) can now have center-center or center-saddle
non-degenerate fixed points, depending on the value of $c$. Then the
system is defined for $I\geq-c$.

\subsection*{The Truncated Forced Nonlinear Schroedinger Equation}

\begin{figure}[h]
\centering
\begin{minipage}[h]{\textwidth}
\centering
\includegraphics[width=12.5cm,height=2.8cm]{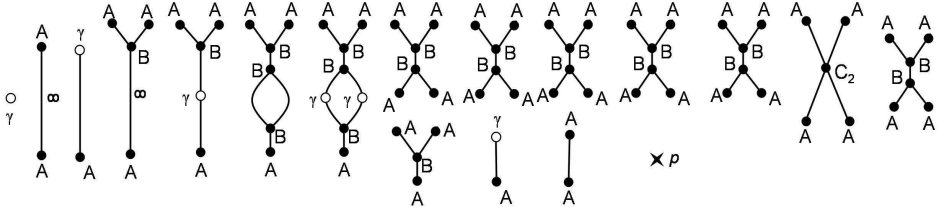}
\parbox{0.95\textwidth}{\caption{Fomenko graphs for the truncated NLS.}\label{fig:nls}}
\end{minipage}
\end{figure}

The trunca\-ted for\-ced nonlinear Schroedinger equation corresponds
to a two-mode Galerkin truncation of the corresponding forced
integrable partial differential equation, see
\cite{BMFO88,BFMO86,BFLKWT83,BL86,CMK02,SRK} and references therein.
The corresponding truncated Hamiltonian is:
\begin{equation*}
H(c,c^{\ast},b,b^{\ast};\varepsilon)=H_{0}(c,c^{\ast},b,b^{\ast})+\varepsilon
H_{1}(c,c^{\ast},b,b^{\ast}),\label{eq:hccb}%
\end{equation*}
with  the Poisson brackets $\{f,g\}=-2i\left(  \left\langle \frac{\partial
}{\partial c},\frac{\partial}{\partial c^{\ast}}\right\rangle +\left\langle
\frac{\partial}{\partial b},\frac{\partial}{\partial b^{\ast}}\right\rangle
\right)  $, where
\begin{align*}
H_{0}  & =\frac{1}{8}|c|^{4}+\frac{1}{2}|b|^{2}|c|^{2}+\frac{3}{16}%
|b|^{4}-\frac{1}{2}(\Omega^{2}+k^{2})|b|^{2}-\frac{\Omega^{2}}{2}|c|^{2}%
+\frac{1}{8}(b^{2}c^{\ast2}+b^{\ast2}c^{2})\label{eq:h0h1}\\
H_{1}  & =\frac{-i}{\sqrt{2}}(c-c^{\ast}).\nonumber
\end{align*}
At $\varepsilon=0$ the Hamiltonian flow possesses an additional
integral of motion:%

\begin{equation*}
I=\frac{1}{2}(|c|^{2}+|b|^{2})\label{eq:idef}%
\end{equation*}
and is thus integrable, see
\cite{BMFO88,BFMO86,BFLKWT83,BL86,CMK02}. Using this integral of
motion one may bring this system to the form (\ref{eq:ham0}), and
thus the bifurcation diagrams and the corresponding Fomenko graphs
may be constructed for any concrete numerical values of the
parameters (see Figures 1 and 2 from \cite{SRK}). A representative
sequence of graphs from \cite{SRK} (here we add  the labels
of the Fomenko atoms and do not provide a different symbol to each of the four different families of circles that appear in this problem), is shown in Figure \ref{fig:nls}.

We see that in this example, the following bifurcations appear:
\begin{itemize}
\item Folds of elliptic and orientable hyperbolic circles
(Proposition \ref{prop:fold}).
\item Center-Center point (Corollary \ref{cor:fixed.points}).
\item Symmetric orientable parabolic circles with the elliptic pitchfork normal form $hx^2+x^4+ y^2$(Corollary
\ref{cor:parabolic}).
\item Global bifurcation of orientable circles creating a $\mathbf{C}_{2}$-atom (Proposition \ref{prop:intersection}).
\end{itemize}
 We see that
rank-1 orbits of complexity 2 appear only at isolated values (at an
orientable global bifurcation).

\subsection{Examples of Integrable Rigid Body Motion}

The topological structure of integrable cases of the rigid body
motion had been extensively studied and analyzed by the Fomenko
school and others \cite{BRF,Osh}, producing examples of many of the
bifurcations that are listed in Corollary \ref{cor:fixed.points} and
Corollary \ref{cor:parabolic}.

\smallskip

For example, the topology of the Kovalevskaya top dynamics is
studied in detail in \cite{BRF}. In the bifurcation diagram, there
appears a line corresponding to a non-orientable family of circles
($\mathbf{A}^{*}$--atoms). It follows that for a perturbation which
is resonant with one circle of this non-orientable family ---
namely, for the appropriate choice of integrals --- we will have a
bifurcation diagram corresponding to a fold of a non-orientable
family of circles, as in Proposition \ref{prop:fold}. We also note
that in this case there is a saddle-saddle fixed point with $2$
non-orientable orbits in its leaf, which creates in the bifurcation
set a curve along which the complexity $2$ atom $\mathbf{C}_{2}$
appears.

\section{Conclusions}\label{sec:conc}

The main result of this paper is the Singularities and
Foliations Theorem, which, together with Propositions
\ref{prop:intersection}, \ref{prop:fold} and Corollaries
\ref{cor:fixed.points}, \ref{cor:parabolic} from Appendix 2 gives
a classification of the changes in the isoenergy surfaces for a
class of two degrees of freedom Hamiltonian systems.

\smallskip

Surprisingly, the study of the level sets structure near singular
circles had lead to the discovery of several new phenomena that
may be of future interest in the near-integrable context; first,
the persistent appearance of folds of curves in the
energy-momentum bifurcation diagrams and its implications had been
fully analyzed. Notably, the appearance of a circle of hyperbolic
fixed points with non-orientable separatrices (type
$\mathbf{A}^{\ast}$) had emerged as a "generic" scenario which had
not been studied yet; under small perturbations we expect to
obtain various multi-pulse orbits to some resonance zone which
will replace the circle of fixed points. While the tools developed
by Haller \cite{HaWi93,Haller} and Kova\v{c}i\v{c} \cite{kov93} to
study the orientable case should apply, the nature of the chaotic
set may be quite different. Second, the complete listing of the
structure of the isoenergy surfaces near all non-degenerate global
bifurcations may lead to a systematic study of the emerging
chaotic sets of the various cases under small perturbations. Let
us note that such a listing follows from \cite{BF2}, but is
usually presented in the different context of the completely
integrable dynamics.

\smallskip

Beyond the classification result, the present paper together with
\cite{LHRK1,LHRK2, LHRK3, SRK} supply a tool for studying
near-integrable systems; In \cite{SRK} we proposed that the
bifurcations of the level sets in such systems may be studied by a
three level hierarchal structure. The first level corresponds to
finding singular level sets on any given energy surface -- namely
of finding the bifurcation set and the structure of the
corresponding isoenergy surfaces as described by the Fomenko
graphs. The second level consists of finding the singularities of
the bifurcation set of the system, and finding which ones are
essential -- namely identifying the essential critical set -- the
set of all values of the energy where the foliation of the
isoenergy surfaces changes. The third level consists of
investigating singularities in the critical set as parameters are
varied.

\smallskip

Combining Fomenko school works and Lerman and Umanskii works leads
to a complete classification of the first level for the
non-degenerate systems considered here. We supply a complete
classification of the second level in this hierarchy for the systems
satisfying Assumptions \ref{as:smooth}--\ref{as:stability}. A
classification of the third level and of  more complex systems that
may arise under symmetric settings is yet to be constructed.

\smallskip

Having this classification of the second level means that given the
structure of an isoenergy surface at some fixed energy, we can now
develop the global structure of all isoenergy surfaces by
continuation, using our classification of the energy surface
structure near the critical set. Namely, a topological continuation
scheme may be implemented , and it will be interesting to relate it
to \cite{DW} where a continuation scheme for finding the action
angle coordinates is found.

\subsection*{Acknowledgements}


We thank Prof.~L.~Lerman, A.~Bolsinov, V.~Dragovi\'c, and
H.~Hanssman for important comments and discussion. The research is
supported by the Minerva foundation, by the Russian-Israeli joint
grant and by the Israel Science Foundation (grant no 273/07). One of
the authors (M.R.) is supported by the Serbian Ministry of Science
and Technology, Project no.\ 144014: \textit{Geometry and Topology
of Manifolds and Integrable Dynamical Systems}.

\section*{Appendix 1: Topological Classification of Iso\-energy Surfaces}

In this section, we describe the representation of isoenergy
surfaces and some of their topological invariants by Fomenko graphs.

\smallskip

Consider a Hamiltonian system defined on the manifold $\mathcal{M}$,
satisfying Assumptions \ref{as:smooth}--\ref{as:non-res}.

\smallskip

Two isoenergy surfaces $\mathcal{Q}$ and $\mathcal{Q}^{\prime}$
are \textit{Liouville equivalent} if there exists a homeomorphism
between them preserving their Liouville foliations.

\smallskip

The set of topological invariants that describe completely isoenergy
manifolds containing regular and non-degenerate leaves of rank $1$,
up to their Liouville equivalence, consists of:

\begin{itemize}
\item[$\bullet$] \textit{The oriented graph} $G$, whose vertices correspond to
the singular connected components of the level sets of $K$, and edges to
one-parameter families of Liouville tori;

\item[$\bullet$] \textit{The collection of Fomenko atoms}, such that each atom
marks exactly one vertex of the graph $G$;

\item[$\bullet$] \textit{The collection of pairs of numbers} $(r_{i}%
,\varepsilon_{i})$, with $r_{i}\in([0,1)\cap\mathbf{Q})\cup\{\infty\}$,
$\varepsilon_{i}\in\{-1,1\}$, $1\le i\le n$. Here, $n$ is the number of edges
of the graph $G$ and each pair $(r_{i},\varepsilon_{i})$ marks an edge of the graph;

\item[$\bullet$] \textit{The collection of integers} $n_{1},n_{2},\dots,n_{s}%
$. The numbers $n_{k}$ correspond to certain connected components of the
subgraph $G^{0}$ of $G$. $G^{0}$ consists of all vertices of $G$ and the edges
marked with $r_{i}=\infty$. The connected components marked with integers
$n_{k}$ are those that do not contain a vertex corresponding to an isolated
critical circle (an $\mathbf{A}$ atom) on the manifold $\mathcal{Q}$.
\end{itemize}

Let us clarify the meaning of some of these invariants.

\smallskip

First, we are going to describe the construction of the graph $G$
from the manifold $\mathcal{Q}$. Each singular leaf of the Liouville
foliation corresponds to exactly one vertex of the graph. If we cut
$\mathcal{Q}$ along such leaves, the manifold will fall apart into
connected sets, each one consisting of one-parameter family of
Liouville tori. Each of these families is represented by an edge of
the graph $G$. The vertex of $G$ which corresponds to a singular
leaf $\mathcal{L}$ is incident to the edge corresponding to the
family $\mathcal{T}$ of tori if and only if
$\partial\mathcal{T}\cap\mathcal{L}$ is non empty.

\smallskip

Note that $\partial\mathcal{T}$ has two connected components, each
corresponding to a singular leaf. If the two singular leaves
coincide, then the edge creates a loop connecting one vertex to
itself.

\smallskip

Now, when the graph is constructed, one need to add the orientation
to each edge. This may be done arbitrarily, but, once determined,
the orientation must stay fixed because the values of numerical
Fomenko invariants depend on it.

\smallskip

The Fomenko atom which corresponds to a singular leaf $\mathcal{L}$
of a singular level set is determined by the topological type of the
set $\mathcal{L}_{\varepsilon}$. The set
$\mathcal{L}_{\varepsilon}\supset{\mathcal{L}}$ is the connected
component of
 $\{\ p\in\mathcal{Q}\ |\ c-\varepsilon<K(p)<c+\varepsilon\ \}$,
where $c=K(\mathcal{L})$, and $\varepsilon>0$ is such that $c$ is
the only critical value of the function $K$ on $\mathcal{Q}$ in
interval $(c-\varepsilon,c+\varepsilon)$.

\smallskip

Let
 $\mathcal{L}_{\varepsilon}^{+} = \{\ p\in\mathcal{L}_{\varepsilon}\ |\ K(p)>c\ \}$,
 $\mathcal{L}_{\varepsilon}^{-} = \{\ p\in\mathcal{L}_{\varepsilon}\ |\ K(p)<c\ \}$.
Each of the sets $\mathcal{L}_{\varepsilon }^{+} $,
$\mathcal{L}_{\varepsilon}^{-}$ is a union of several connected
components, each component being a one-parameter family of Liouville
tori. Each of these families corresponds to the beginning of an edge
of the graph $G$ incident to the vertex corresponding to
$\mathcal{L}$.

\smallskip

Let us say a few words on the topological structure of the set
$\mathcal{L}$. This set consists of at least one fixed point or
closed one-dimensional orbit of the Poisson action $\Phi$ on
$\mathcal{M}$ and several (possibly none) two-dimensional orbits of
the action. The set of all zero-dimensional and one-dimensional
orbits in $\mathcal{L}$ is called the \textit{garland} (see
\cite{LU}) while two-dimensional orbits are called
\textit{separatrices}.

\smallskip

The trajectories on each of these two-dimensional separatrices is
homoclinically or heteroclinically tending to the lower-dimensional
orbits. The Liouville tori of each of the families in
 $\mathcal{L}_{\varepsilon}\setminus\mathcal{L}
=
 \mathcal{L}_{\varepsilon}^{+}\cup\mathcal{L}_{\varepsilon}^{-}$
tend, as the integral $K$ approaches $c$, to a closure of a subset
of the separatrix set.

\

\textbf{Fomenko Atoms of Complexity 1 and 2}

\medskip

Fomenko and his school completely described and classified
non-degenerate leaves of rank $1$.

\smallskip

If $\mathcal{L}$ is not an isolated critical circle, then a
sufficiently small neighborhood of each $1$-dimensional orbit in
$\mathcal{L}$ is isomorphic to either two cylinders intersecting
along the base circle, and then the orbit is \textit{orientable}, or
to two Moebius bands intersecting each other along the joint base
circle, then the $1$-dimensional orbit is \textit{non-orientable}.

\smallskip

The number of closed one-dimensional orbits in $\mathcal{L}$ is
called \textit{the complexity} of the corresponding atom.

\smallskip

There are exactly three Fomenko atoms of complexity $1$.

\subsubsection*{The atom $\mathbf{A}$.}

This atom corresponds to a normally elliptic singular circle, which
is isolated on the isoenergy surface $\mathcal{Q}$. A small
neighborhood of such a circle in $\mathcal{Q}$ is diffeomorphic to a
solid torus. One of the sets $\mathcal{L}_{\varepsilon}^{+}$,
$\mathcal{L}_{\varepsilon}^{-}$ is empty, the other one is
connected. Thus only one edge of the graph $G$ is incident with the
vertex marked with the letter atom $\mathbf{A}$.

\subsubsection*{The atom $\mathbf{B}$.}

In this case, $\mathcal{L}$ consists of one orientable normally
hyperbolic circle and two $2$-dimensional separatrices -- it is
diffeomorphic to a direct product of the circle $\mathbf{S}^{1}$ and
the plane curve given by the equation $y^{2}=x^{2}-x^{4}$. Because
of its shape, we will refer to this curve as the \textit{`figure
eight'}. The set $\mathcal{L}_{\varepsilon }\setminus\mathcal{L}$
has $3$ connected components, two of them being placed in
$\mathcal{L}_{\varepsilon}^{+}$ and one in
$\mathcal{L}_{\varepsilon}^{-}$, or vice versa. Let us fix that two
of them are in $\mathcal{L}_{\varepsilon }^{+}$. Each of these two
families of Liouville tori limits as $K$ approaches $c$ to only one
of the separatrices. The tori in $\mathcal{L}_{\varepsilon }^{-}$
tend to the union of the separatrices.

\subsubsection*{The atom $\mathbf{A}^{*}$.}

$\mathcal{L}$ consists of one non-orientable hyperbolic circle and
one $2$-di\-men\-sional separatrix. It is homeomorphic to the smooth
bundle over $\mathbf{S}^{1}$ with the `figure eight' as fiber and
the structural group consisting of the identity mapping and the
central symmetry of the `figure eight'. Both
$\mathcal{L}_{\varepsilon}^{+}$ and $\mathcal{L}_{\varepsilon }^{-}$
are $1$-parameter families of Liouville tori, one limiting to the
separatrix from outside the `figure eight' and the other from the
interior part of the `figure eight'.

\smallskip

The atoms of complexity $2$ appear in the non-degenerate integrable
two degrees of freedom systems considered here on isoenergy surfaces
close to fixed points of the system, and, for isolated values of the
energy $H$, at level sets corresponding to global bifurcations of
the system. Fomenko showed that there are exactly six different
atoms of complexity $2$, and these are described next. It is
instructive to use the Fomenko graphs depicted in the figures of
propositions 1 and 2 to understand the structure of the level sets
near these complexity 2 atoms.

\subsubsection*{The atom $\mathbf{C}_{1}$.}

$\mathcal{L}$ consists of two orientable circles $\gamma_{1}$,
$\gamma_{2}$ and four heteroclinic $2$-dimensional separatrices
$\mathcal{S}_{1}$, $\mathcal{S}_{2}$, $\mathcal{S}_{3}$,
$\mathcal{S}_{4}$. Trajectories on $\mathcal{S}_{1}$,
$\mathcal{S}_{3}$ are approaching $\gamma_{1}$ as time tend to
$\infty$, and $\gamma_{2}$ as time tend to $-\infty$, while those
placed on $\mathcal{S}_{2}$, $\mathcal{S}_{4}$ approach $\gamma_{2}$
as time tend to $\infty$, and $\gamma_{1}$ as time tend to
$-\infty$. Each of the sets $\mathcal{L}_{\varepsilon}^{+}$,
$\mathcal{L}_{\varepsilon}^{-}$ is connected and contains only one
family of Liouville tori. Both families of tori deform to the whole
separatrix set, as the value of $K$ approaches $c$.

\subsubsection*{The atom $\mathbf{C}_{2}$.}

The level set $\mathcal{L}$ is the same as in the atom
$\mathbf{C}_{1}$. Each of the sets $\mathcal{L}_{\varepsilon}^{+}$,
$\mathcal{L}_{\varepsilon}^{-}$ contains two families of Liouville
tori. As $K$ approaches to $c$, the tori
from one family in $\mathcal{L}_{\varepsilon}^{-}$ deform to $\mathcal{S}%
_{1}\cup\mathcal{S}_{2}$, and from the other one to $\mathcal{S}_{3}%
\cup\mathcal{S}_{4}$. The tori from one family in
$\mathcal{L}_{\varepsilon }^{+}$ is deformed to
$\mathcal{S}_{1}\cup\mathcal{S}_{4}$, and from the other to
$\mathcal{S}_{2}\cup\mathcal{S}_{3}$.

\subsubsection*{The atom $\mathbf{D}_{1}$.}

$\mathcal{L}$ consists of two orientable circles $\gamma_{1}$,
$\gamma_{2}$ and four $2$-dimensional separatrices
$\mathcal{S}_{1}$, $\mathcal{S}_{2}$, $\mathcal{S}_{3}$,
$\mathcal{S}_{4}$. Trajectories of $\mathcal{S}_{1}$,
$\mathcal{S}_{2}$ homoclinically tend to $\gamma_{1}$, $\gamma_{2}$
respectively. Trajectories on $\mathcal{S}_{3}$ are approaching
$\gamma_{1}$ as time tend to $\infty$, and $\gamma_{2}$ as time tend
to $-\infty$, while those placed on $\mathcal{S}_{4}$, approach
$\gamma_{2}$ as time tend to $\infty$, and $\gamma_{1}$ as time tend
to $-\infty$. One of the sets $\mathcal{L}_{\varepsilon}^{+}$,
$\mathcal{L}_{\varepsilon}^{-}$ contains three, and the other one
family of Liouville tori. Lets say that
$\mathcal{L}_{\varepsilon}^{+}$ contains three families. As $K$
approaches to $c$, these families deform to $\mathcal{S}_{1}$,
$\mathcal{S}_{2}$ and $\mathcal{S}_{3}\cap\mathcal{S}_{4}$
respectively, while the family contained in
$\mathcal{L}_{\varepsilon}^{-}$ deform to the whole separatrix set.

\subsubsection*{The atom $\mathbf{D}_{2}$.}

The level set $\mathcal{L}$ is the same as in the atom
$\mathbf{D}_{1}$. Each of the sets $\mathcal{L}_{\varepsilon}^{+}$,
$\mathcal{L}_{\varepsilon}^{-}$ contains two families of Liouville
tori. As $K$ approaches to $c$, the tori
from one family in $\mathcal{L}_{\varepsilon}^{-}$ deform to $\mathcal{S}_{1}%
$, and from the other one to $\mathcal{S}_{2}\cup\mathcal{S}_{3}%
\cup\mathcal{S}_{4}$. The tori from one family in
$\mathcal{L}_{\varepsilon
}^{+}$ is deformed to $\mathcal{S}_{2}$, and from the other to $\mathcal{S}%
_{1}\cup\mathcal{S}_{3}\cup\mathcal{S}_{4}$.

\subsubsection*{The atom $\mathbf{B}^{*}$.}

$\mathcal{L}$ consists of the orientable circle $\gamma_{1}$, the
non-orientable circle $\gamma_{2}$ and three separatrices
$\mathcal{S}_{1}$, $\mathcal{S}_{2}$, $\mathcal{S}_{3}$.
Trajectories on $\mathcal{S}_{1}$ homoclinically tend to
$\gamma_{1}$. Trajectories on $\mathcal{S}_{2}$ are approaching
$\gamma_{1}$ as time tend to $\infty$, and $\gamma_{2}$ as time tend
to $-\infty$, while those placed on $\mathcal{S}_{3}$, approach
$\gamma_{2}$ as time tend to $\infty$, and $\gamma_{1}$ as time tend
to
$-\infty$. One of the sets $\mathcal{L}_{\varepsilon}^{+}$, $\mathcal{L}%
_{\varepsilon}^{-}$ contains two, and the other one family of
Liouville tori. Lets say that $\mathcal{L}_{\varepsilon}^{+}$
contains two families. Then tori
from these two families deform to $\mathcal{S}_{1}$ and $\mathcal{S}_{2}%
\cup\mathcal{S}_{3}$ respectively, as $K$ approaches to $c$. The
tori from the $\mathcal{L}_{\varepsilon}^{-}$ deform to the union of
the whole separatrix set.

\subsubsection*{The atom $\mathbf{A}^{**}$.}

$\mathcal{L}$ consists of two non-orientable circles $\gamma_{1}$,
$\gamma _{2}$ and two separatrices $\mathcal{S}_{1}$,
$\mathcal{S}_{2}$. Trajectories on $\mathcal{S}_{1}$ are approaching
$\gamma_{1}$ as time tend to $\infty$, and $\gamma_{2}$ as time tend
to $-\infty$, while those placed on $\mathcal{S}_{2}$, approach
$\gamma_{2}$ as time tend to $\infty$, and
$\gamma_{1}$ as time tend to $-\infty$. Each of the sets $\mathcal{L}%
_{\varepsilon}^{+}$, $\mathcal{L}_{\varepsilon}^{-}$ consists of one
family of Liouville tori. As $K$ approaches to $c$, the tori from
both families deform to the whole separatrix set.

\

This concludes the complete list of all atoms of complexities $1$
and $2$, i.e.\ all the possible singular level sets which involve
one or two circles and their separatrices. The appearance of
higher complexities violates our non-degeneracy assumptions. These
are expected to appear when additional symmetries, additional
parameters, or higher dimensional systems are considered.

\

\textbf{Numerical Fomenko Invariants}

\medskip

The meaning of the numerical invariant $r_{i}$ is roughly explained next.

\smallskip

Each edge of the Fomenko graph corresponds to a one-parameter family of
Liouville tori. Let us cut each of these families along one Liouville torus.
The manifold $\mathcal{Q}$ will disintegrate into pieces, each corresponding
to the singular level set, i.e. to a part of the Fomenko graph containing only
one vertex and the initial segments of the edges incident to this vertex. To
reconstruct $\mathcal{Q}$ from these pieces, we need to identify the
corresponding boundary tori. This can be done in different ways, implicating
different differential-topological structure of the obtained manifold, and the
number $r_{i}$ contains the information on the rule of the gluing on each edge.

\smallskip

Two important values of $r_{i}$ are $0$ and $\infty$. Consider one edge of the
graph and the corresponding smooth one-parameter family $\mathcal{L}$ of
Liouville tori. After cutting, the family falls apart into two families of
tori diffeomorphic to the product $(0,1]\times\mathbf{T}^{2}\simeq
(0,1]\times\mathbf{S}^{1}\times\mathbf{S}^{1}$. Such a family can be trivially
embedded to the full torus $\mathbf{D}^{2}\times\mathbf{S}^{1}$ by the mapping
$\phi\times\mathrm{id}_{\mathbf{S}^{1}}$, with
\[
\phi\ :\ (0,1]\times\mathbf{S}^{1}\ \rightarrow\ \mathbf{D}^{2},\quad
\phi(r,\alpha)=(r\cos\alpha,r\sin\alpha),
\]
where we take $\mathbf{S}^{1}=\mathbf{R}/2\pi\mathbf{Z}$, $\mathbf{D}%
^{2}=\{\ (x,y)\in\mathbf{R}^{2}\ |\ x^{2}+y^{2}\leq1\ \}$.

\smallskip

Now, via the gluing, the family $\mathcal{L}$ will be embedded to a the
three-di\-men\-sional manifold $\mathcal{T}$ obtained from two solid tori by
identifying their boundaries. The value $r_{i}=0$ corresponds to the case when
the central circles of the two solid tori are linked in $\mathcal{T}$ with the
linking coefficient $1$. $\mathcal{T}$ is then diffeomorphic to the sphere
$\mathbf{S}^{3}$. The value $r_{i}=\infty$ corresponds to the case when the
central circles are not linked in $\mathcal{T}$. In this case, $\mathcal{T}%
\simeq\mathbf{S}^{2}\times\mathbf{S}^{1}$.

\smallskip

For more detailed description and precise methods for determining the
invariant $r_{i}$, see \cite{BF2, MF}.

\smallskip

The invariants $\varepsilon_{i}$, $n_{k}$ are not important for our
exposition; their detailed description with the examples can be found in
\cite{BF2}.

\section*{Appendix 2: Isonergy Surfaces near Isolated Fixed Points and Parabolic Circles}

Here we provide for completeness the description of isoenergy
surfaces for cases that were fully analyzed previously: the
behavior near isolated fixed points, which is essentially a
concise summary of Lerman and Umanskii \cite{LU} results using
Fomenko graphs \cite{Bol} and the behavior near certain parabolic
circles which follows from \cite{BRF}.

 The bifurcation
diagrams that we draw for all these cases correspond to a slight
modification of the standard energy-momentum diagrams and include
the following elements:
\begin{itemize}
\item Solid curves correspond to families of normally elliptic circles,
i.e.\ to Fomenko $\mathbf{A}$-atoms.
\item Dashed curves correspond to families of normally hyperbolic
circles.
\item Double lines are used to indicate that two families of circles
appear on the same curve in the bifurcation set.
\item Grey area indicates the local region of allowed motion.
\end{itemize}

The topology of the level sets and of the extended neighborhoods of
non-de\-ge\-ne\-ra\-te simple fixed points is described in detail in
\cite{LU}. In the following corollary, we give a concise summary of
these results with an emphasize on listing the possible changes in
the Liouville foliations of the isoenergy surfaces. This is achieved
by constructing the Fomenko graphs that correspond to these
isoenergy surfaces close to the fixed point. We should note that in
\cite{Bol}, circular Fomenko graphs corresponding to certain closed
3-dimensional submanifolds of $\mathcal{M}$ that are contained in a
neighborhood of the fixed point are constructed. More precisely, the
considered $3$-dimensional manifold is the inverse image by the
momentum mapping $\mu$ of a sufficiently small circle in the
$(H,K)$-plane, such that $\mu(p)$ is the center of the circle. Thus,
using both the detailed exposition of \cite{LU} and the results of
\cite{Bol} and \cite{Mat}, the list is completed.

\begin{corollary}\label{cor:fixed.points}
Consider a system $(\mathcal{M},\omega,H)$ satisfying Assumptions
\ref{as:smooth}--\ref{as:non-res} and its simple isolated
non-degenerate fixed point $p$. Then, for a sufficiently small
extended neighborhood $V$ of $p$, all possible cases of the
bifurcation set $\Sigma(V)$ and the corresponding Fomenko graphs
are listed in Figures \ref{fig:center-center}--\ref{fig:focus}.

\begin{figure}[h]
\centering
\begin{minipage}[h]{0.44\textwidth}
\centering
\includegraphics[width=4.5cm,height=3.6cm]{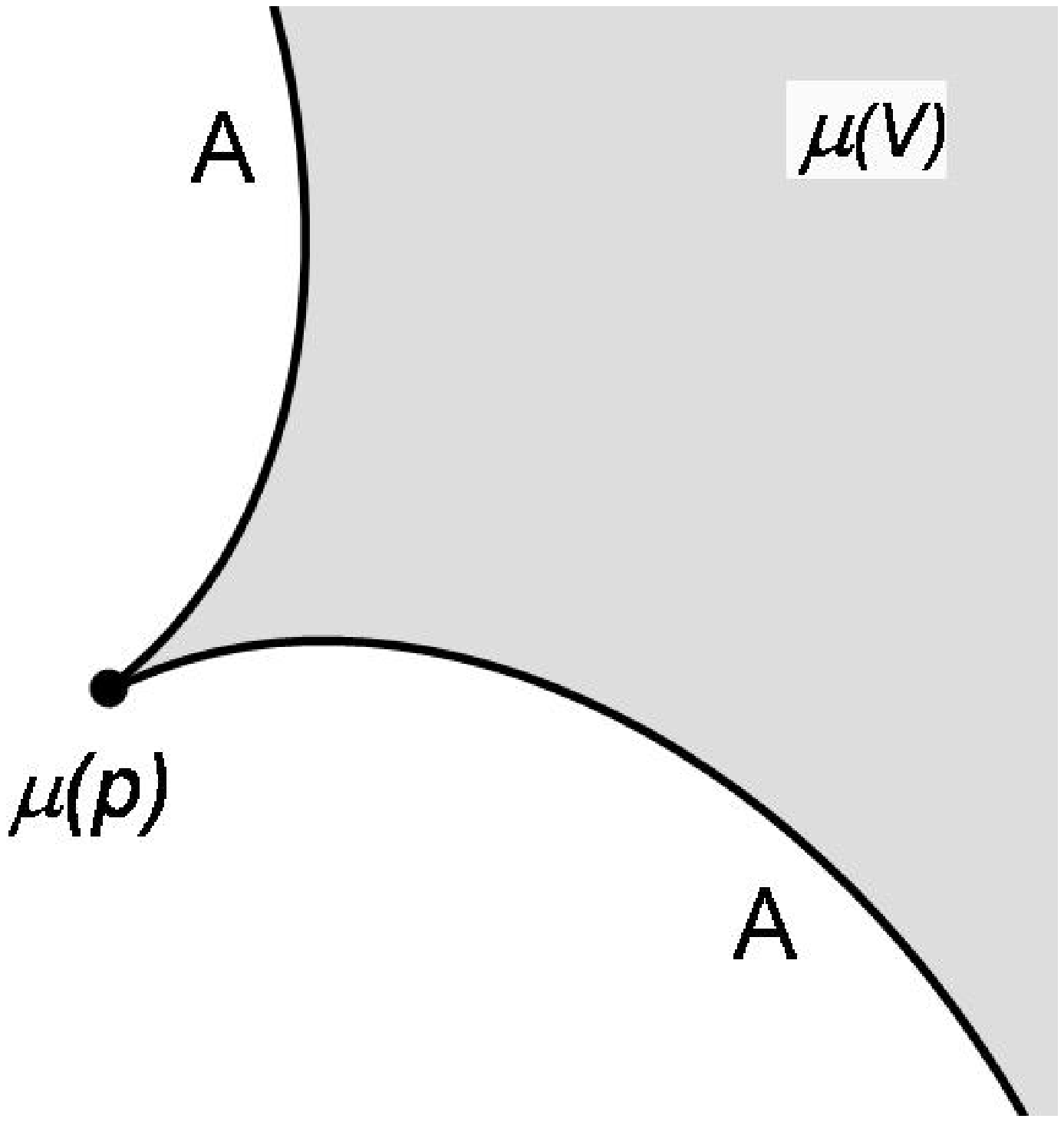}
\end{minipage}
\begin{minipage}[h]{0.44\textwidth}
\centering
\includegraphics[width=4.5cm,height=3.6cm]{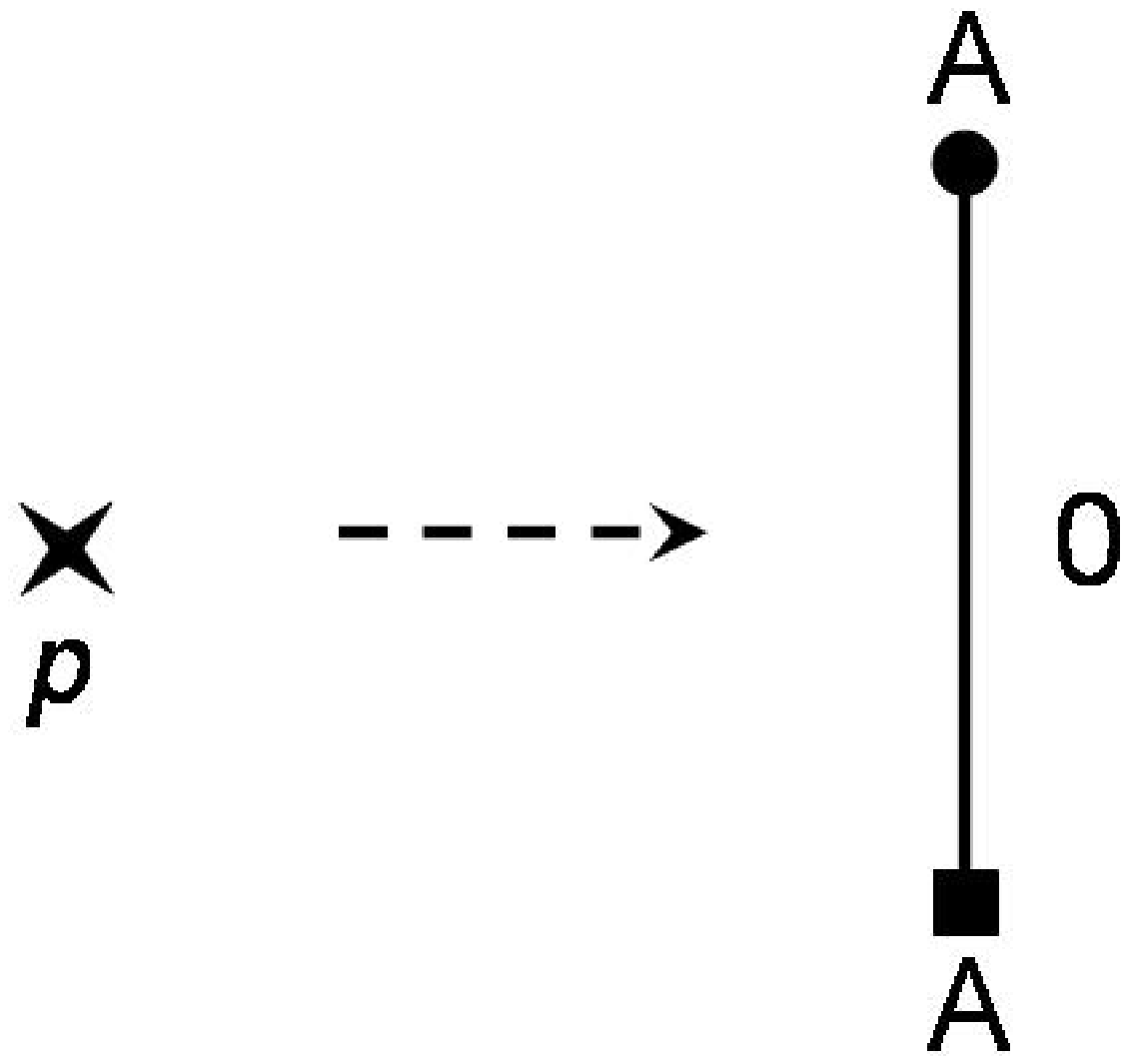}
\end{minipage}
\begin{minipage}[h]{0.44\textwidth}
\centering
\includegraphics[width=4.5cm,height=3.6cm]{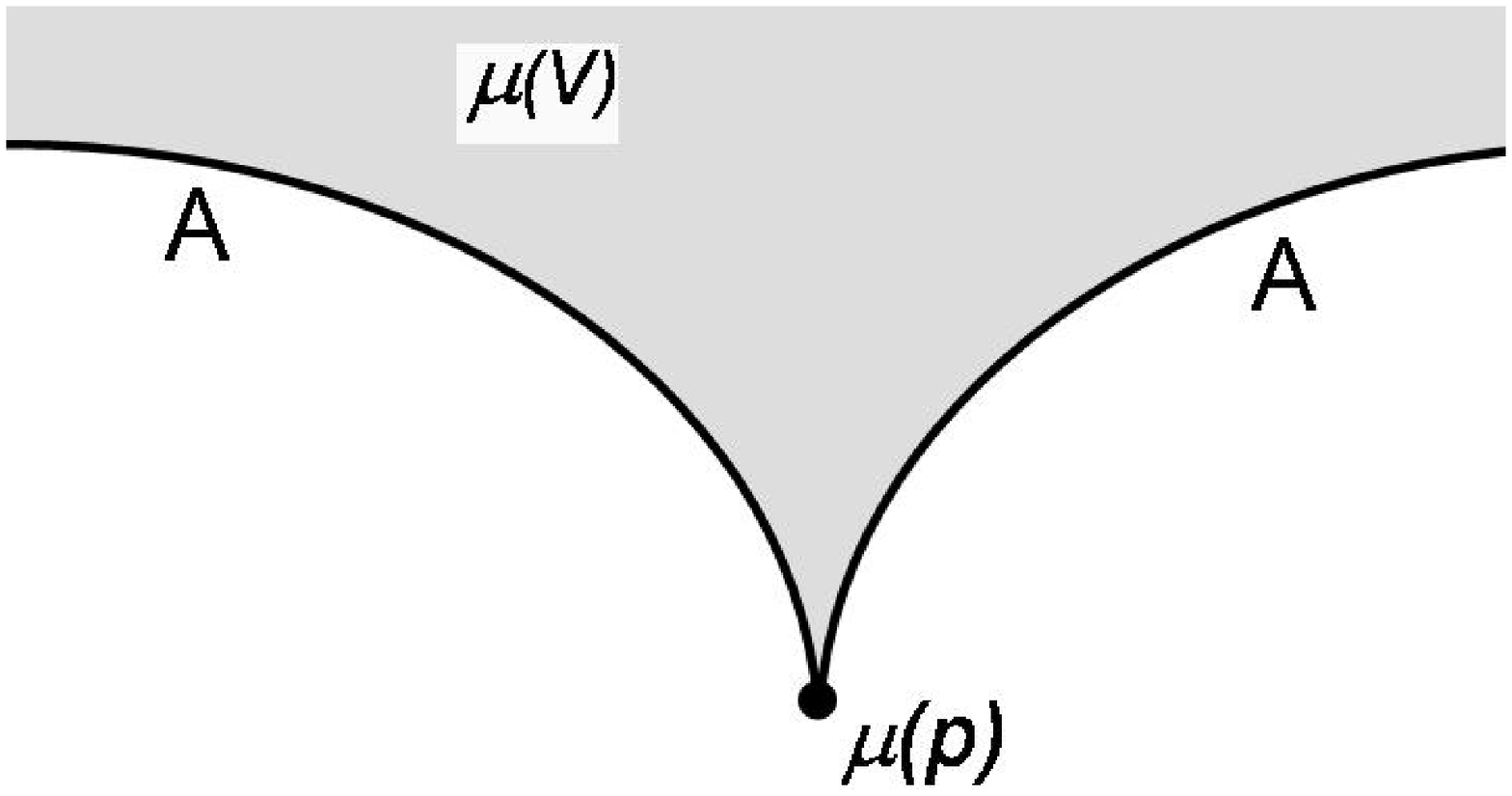}
\end{minipage}
\begin{minipage}[h]{0.44\textwidth}
\centering
\includegraphics[width=4.5cm,height=3.6cm]{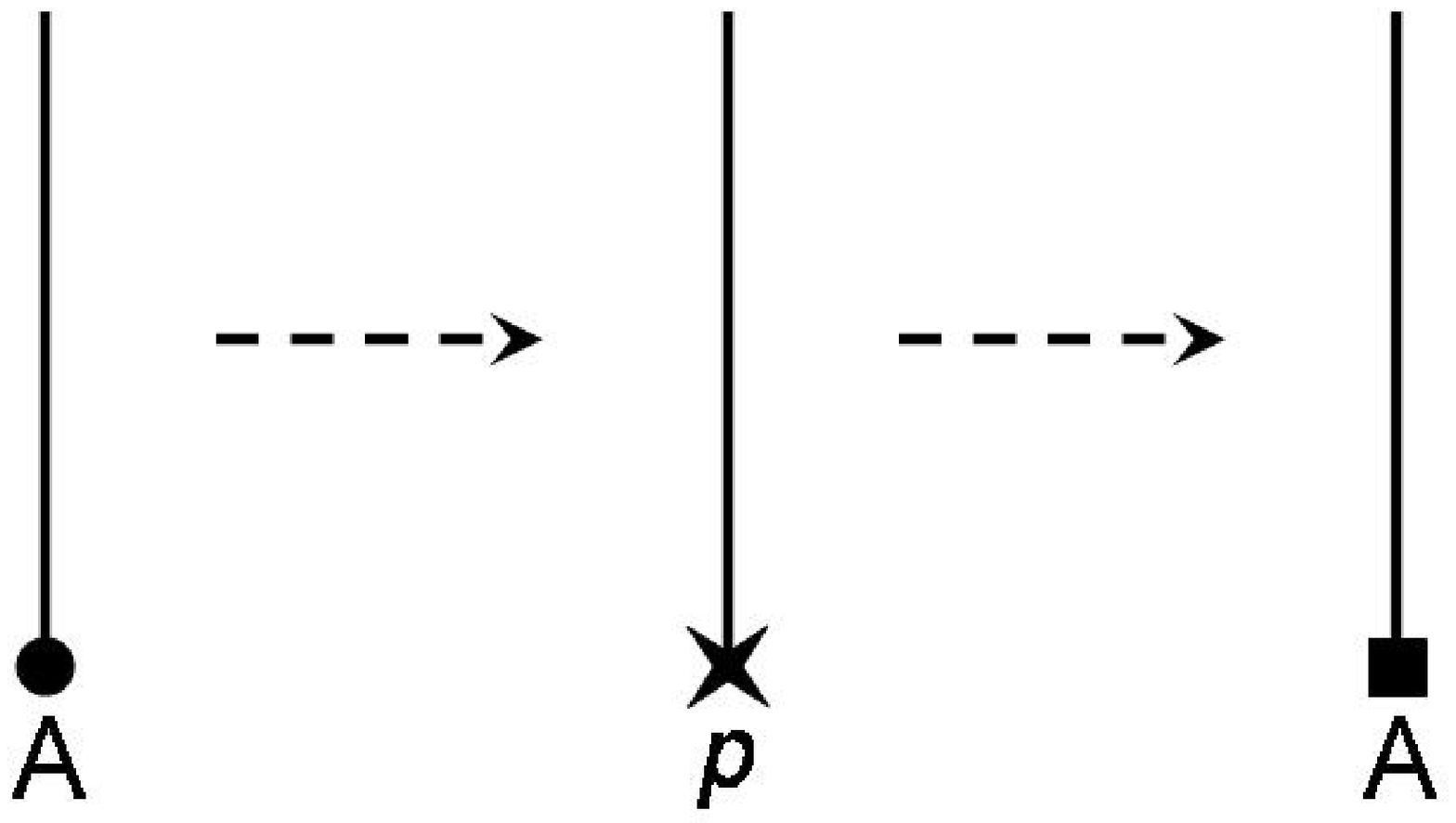}
\end{minipage}
\caption{The center-center fixed point. The two possible
bifurcation sets and Fomenko graphs near a fixed point of the
\emph{center-center type}.} \label{fig:center-center}
\end{figure}

\begin{figure}[h]
\centering
\begin{minipage}[h]{0.44\textwidth}
\centering
\includegraphics[width=4.5cm,height=3.6cm]{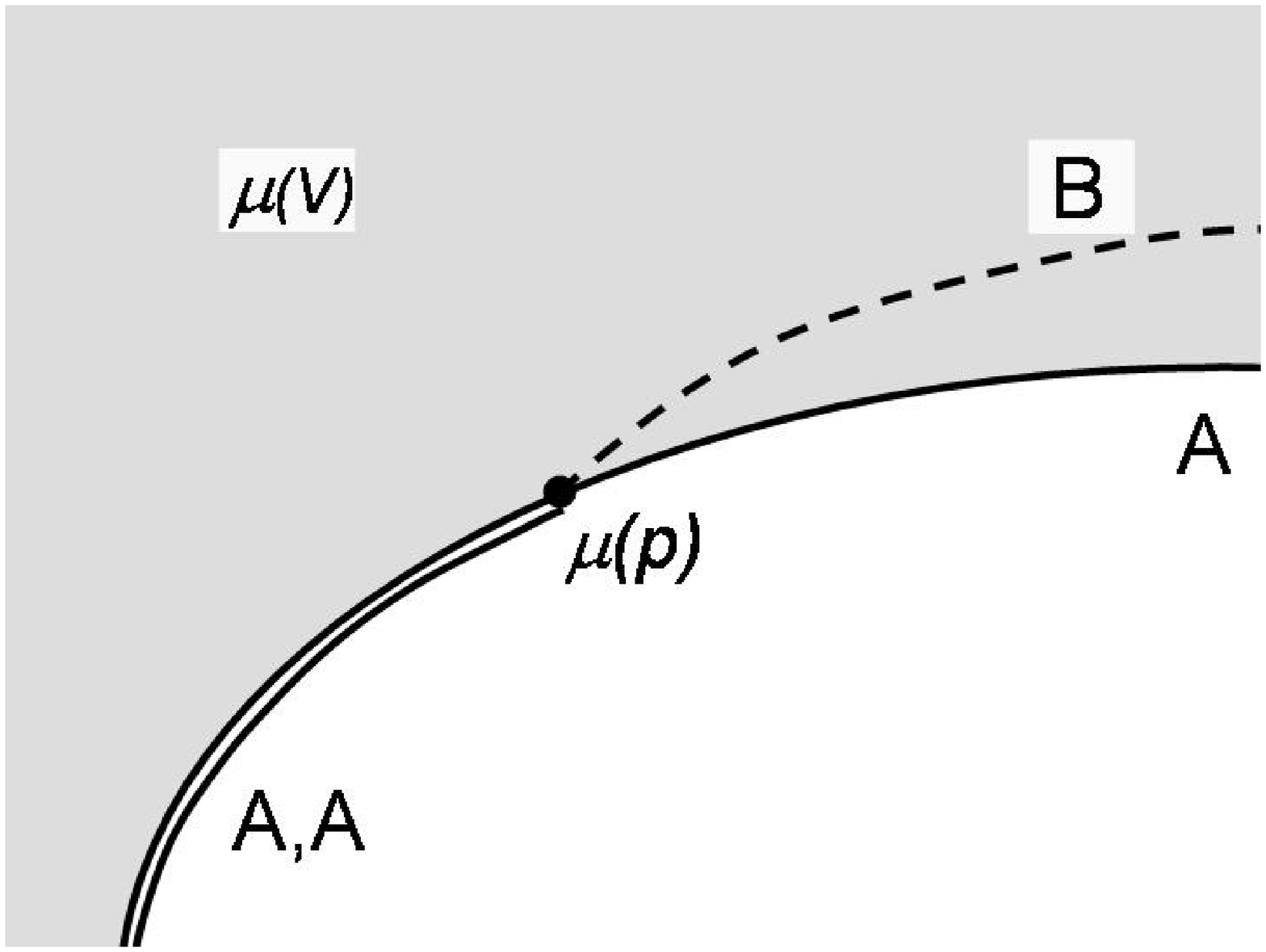}
\end{minipage}
\begin{minipage}[h]{0.44\textwidth}
\centering
\includegraphics[width=4.5cm,height=3.6cm]{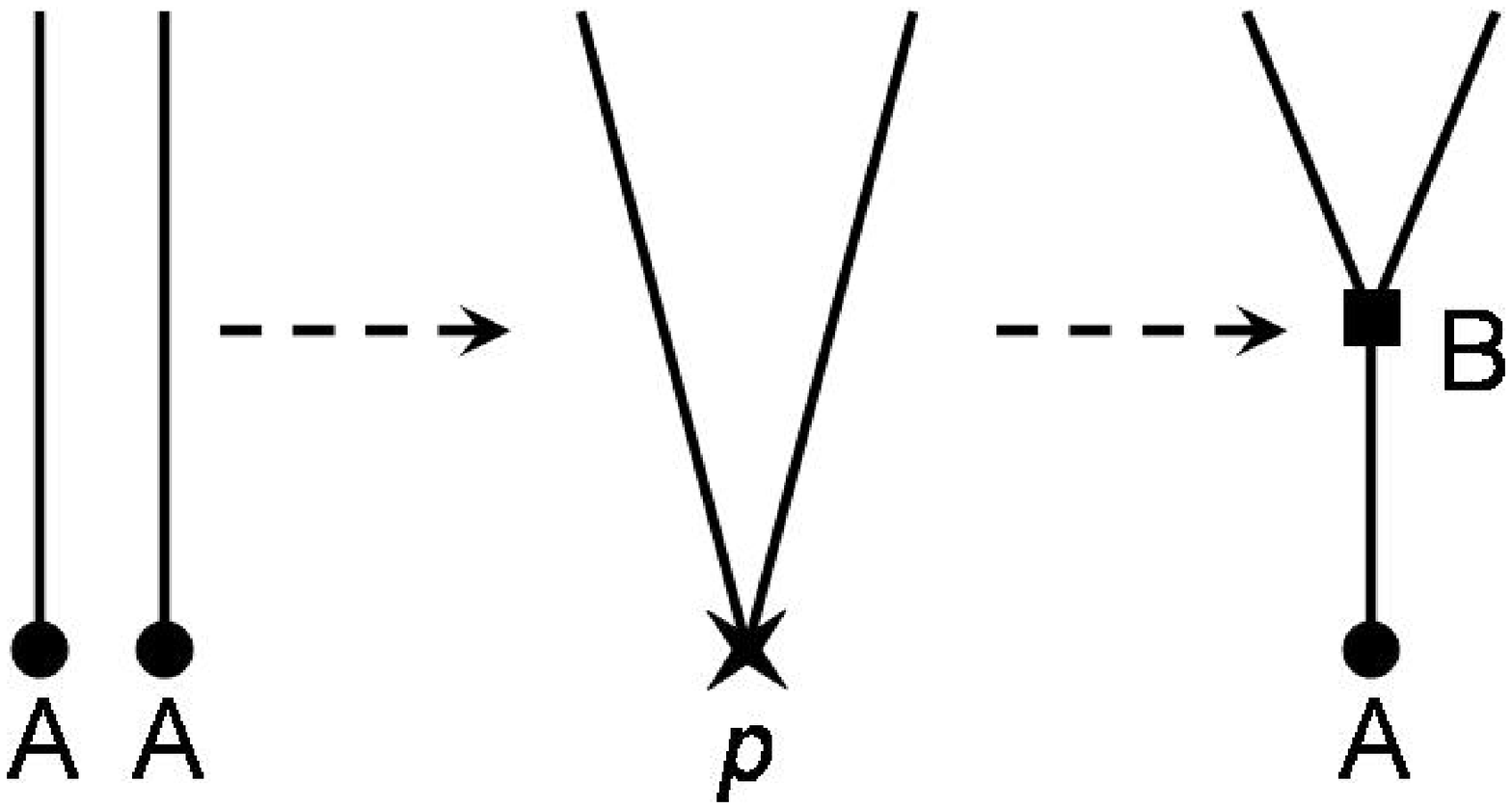}
\end{minipage}
\begin{minipage}[h]{0.44\textwidth}
\centering
\includegraphics[width=4.5cm,height=3.6cm]{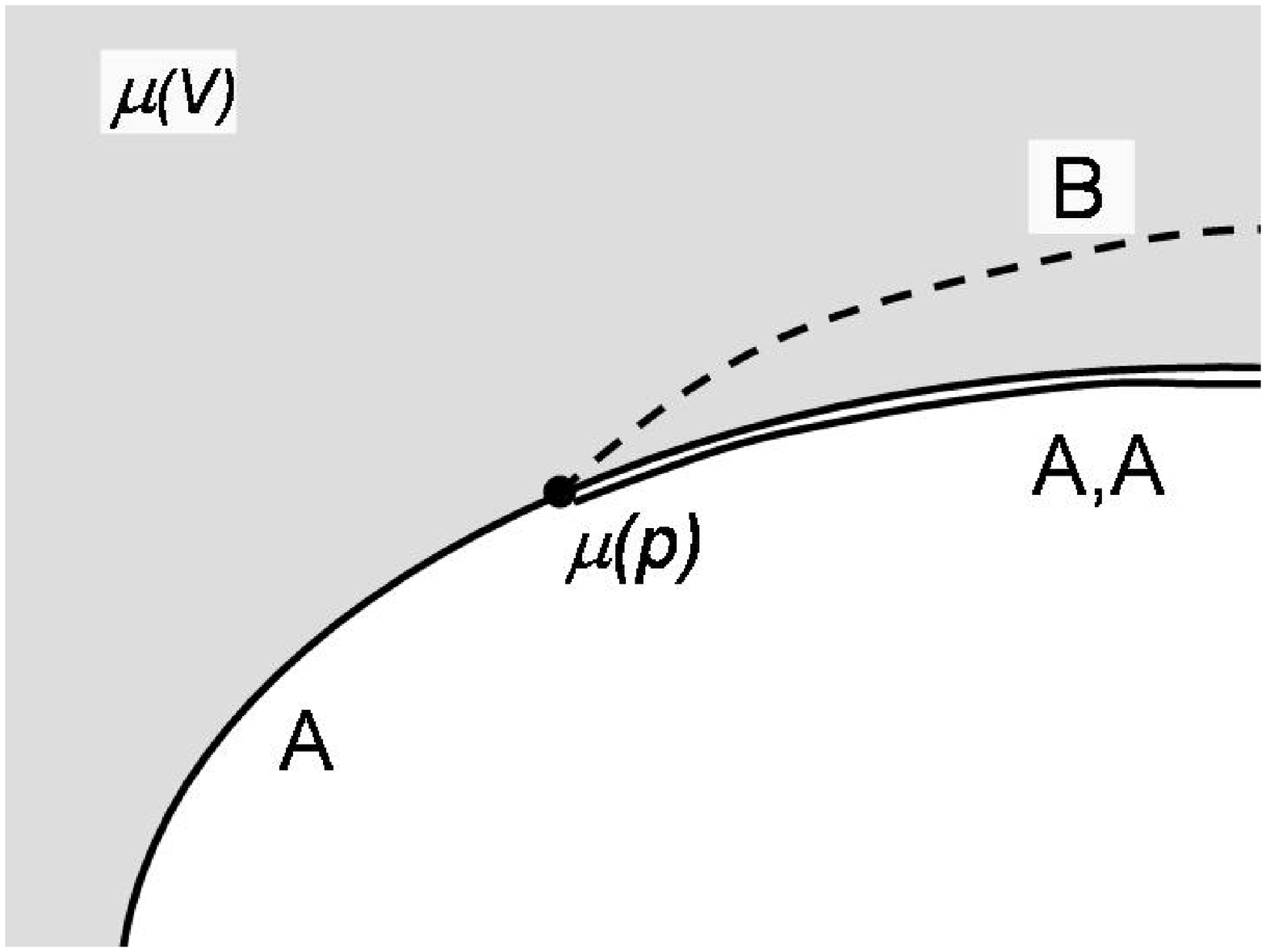}
\end{minipage}
\begin{minipage}[h]{0.44\textwidth}
\centering
\includegraphics[width=4.5cm,height=3.6cm]{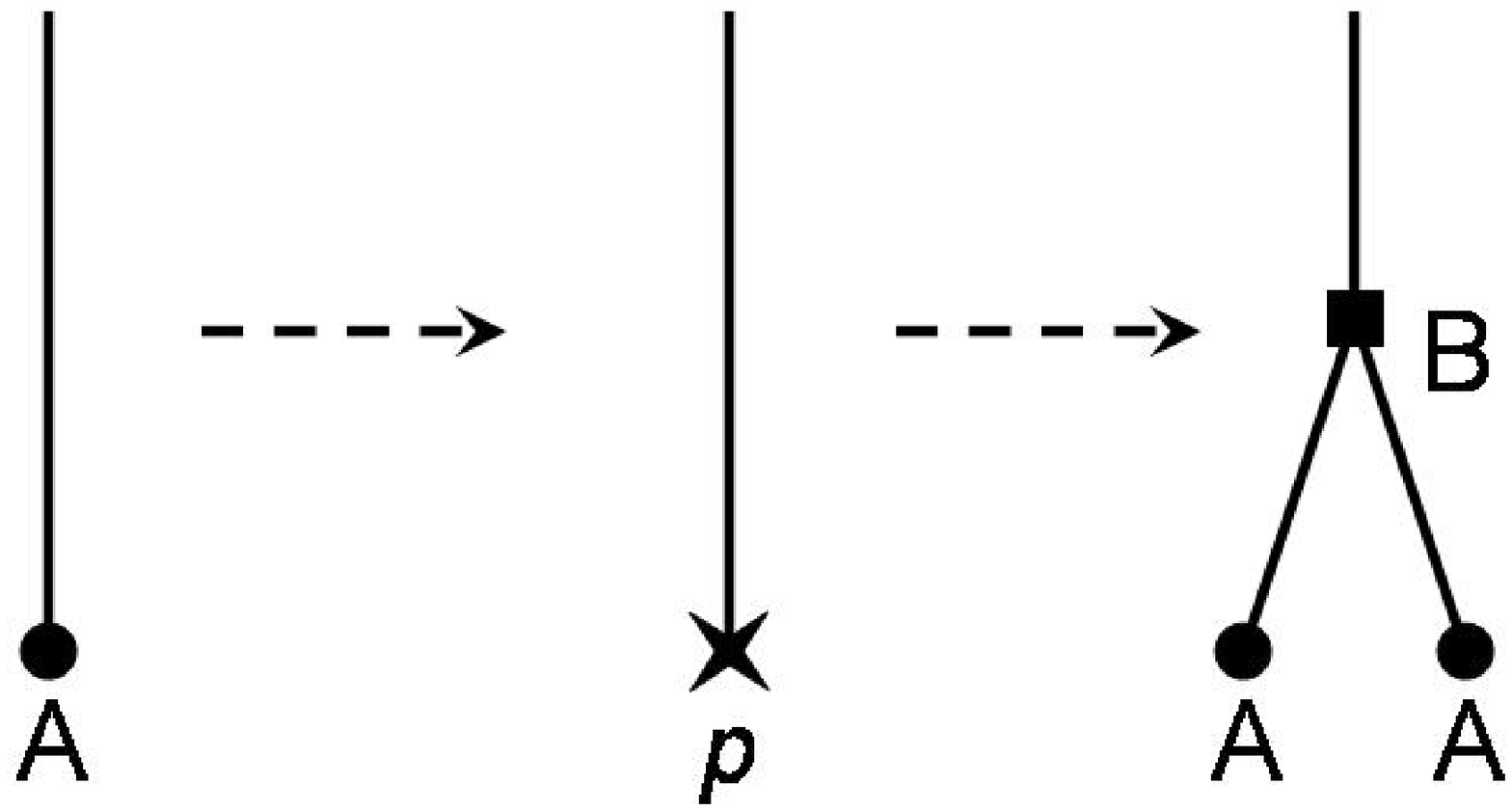}
\end{minipage}
\caption{The saddle-center fixed point. The two possible
bifurcation sets and Fomenko graphs near fixed point of
the \emph{saddle-center type}.}%
\label{fig:saddle-center}%
\end{figure}

\begin{figure}[h]
\centering
\begin{minipage}[h]{0.44\textwidth}
\centering
\includegraphics[width=4.5cm,height=3.6cm]{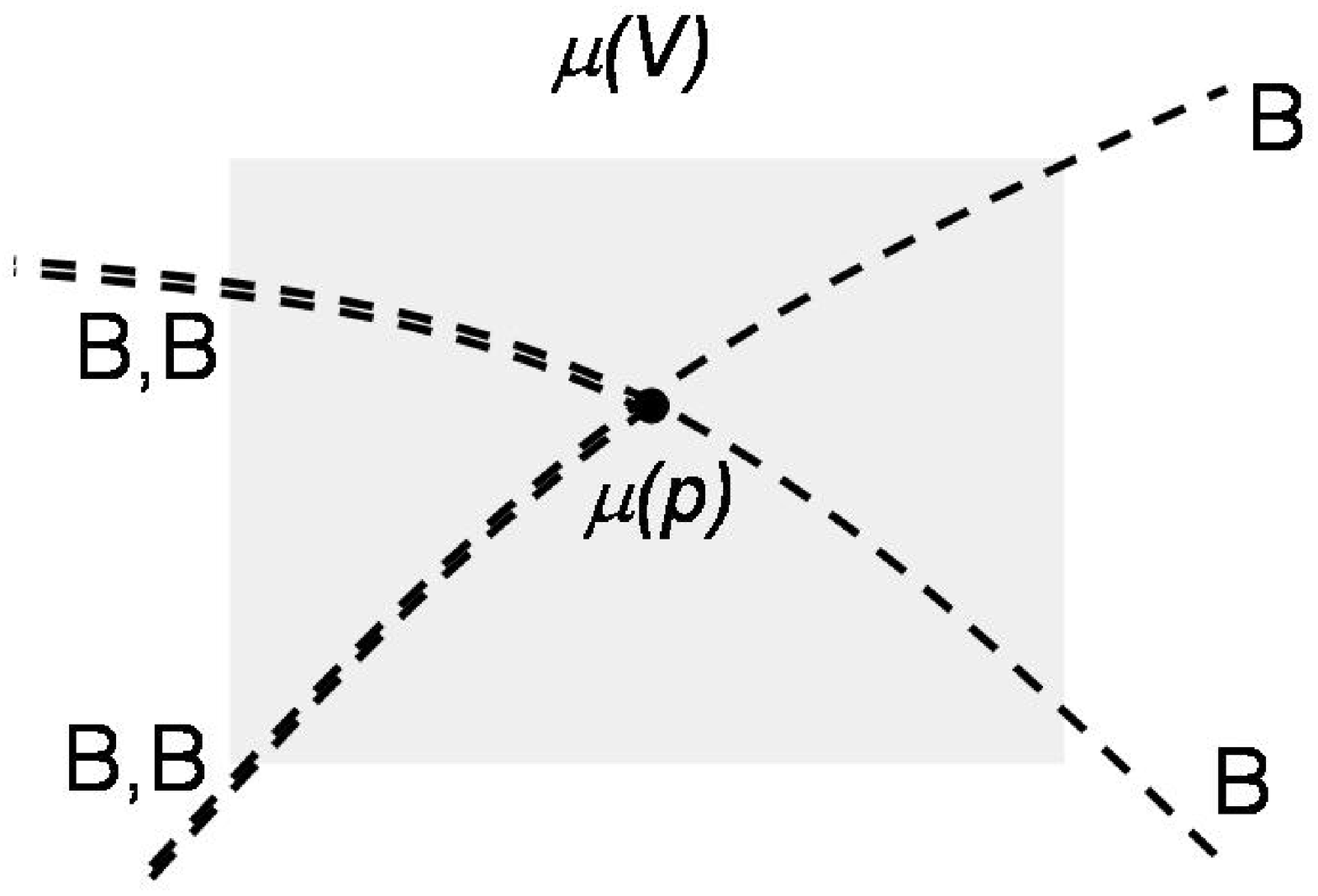}
\end{minipage}
\begin{minipage}[h]{0.44\textwidth}
\centering
\includegraphics[width=4.5cm,height=3.6cm]{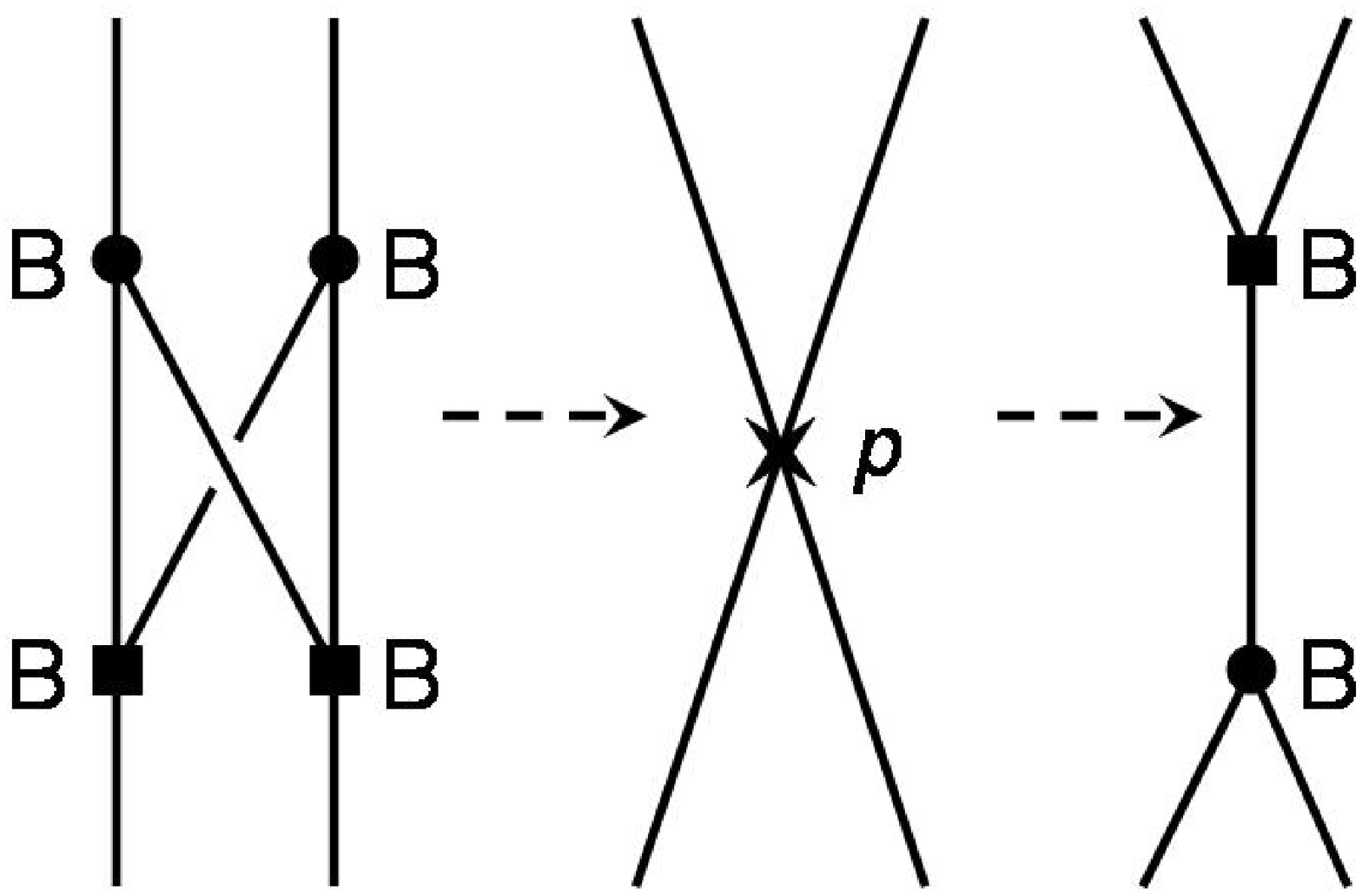}
\end{minipage}
\begin{minipage}[h]{0.44\textwidth}
\centering
\includegraphics[width=4.5cm,height=3.6cm]{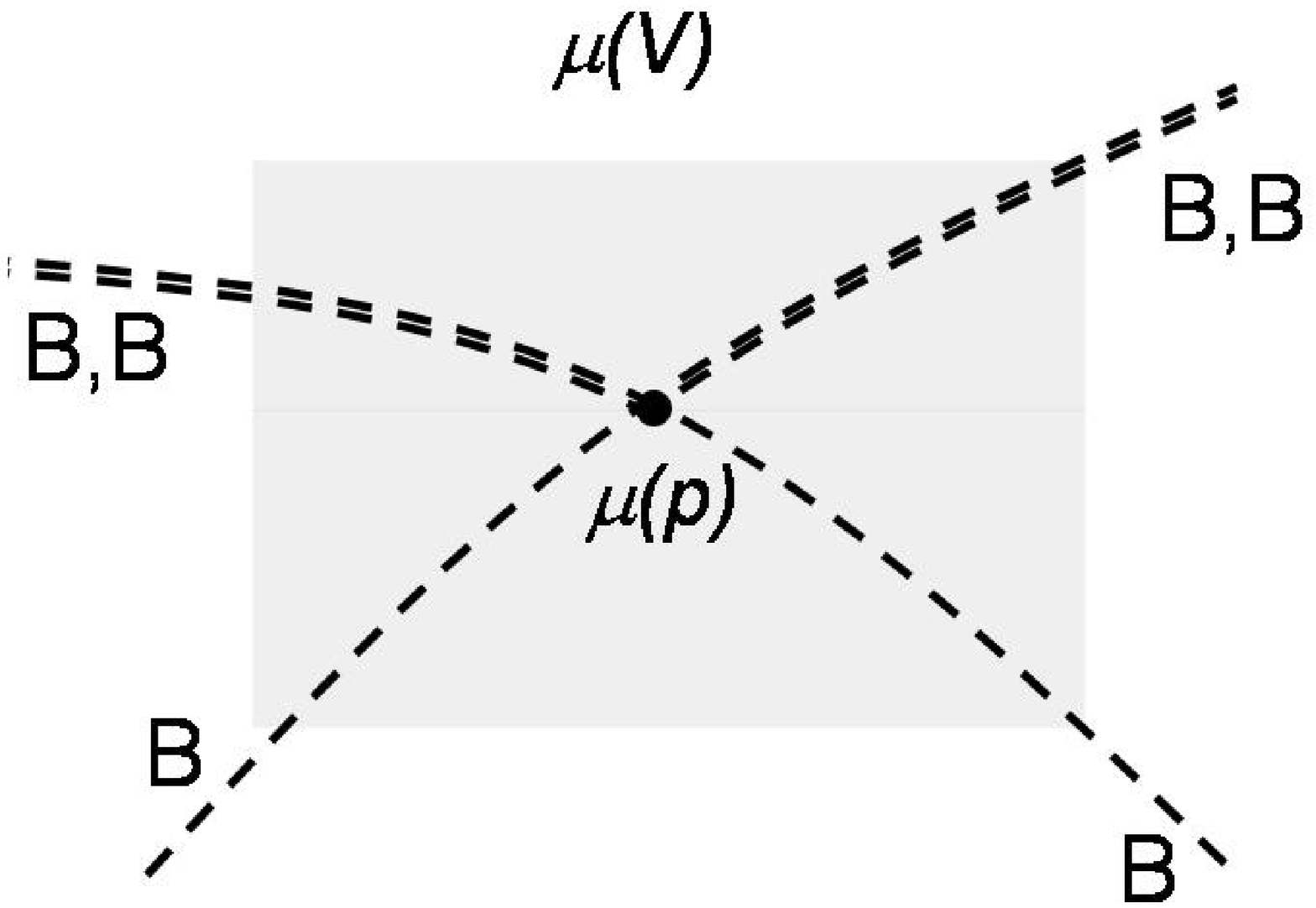}
\end{minipage}
\begin{minipage}[h]{0.44\textwidth}
\centering
\includegraphics[width=4.5cm,height=3.6cm]{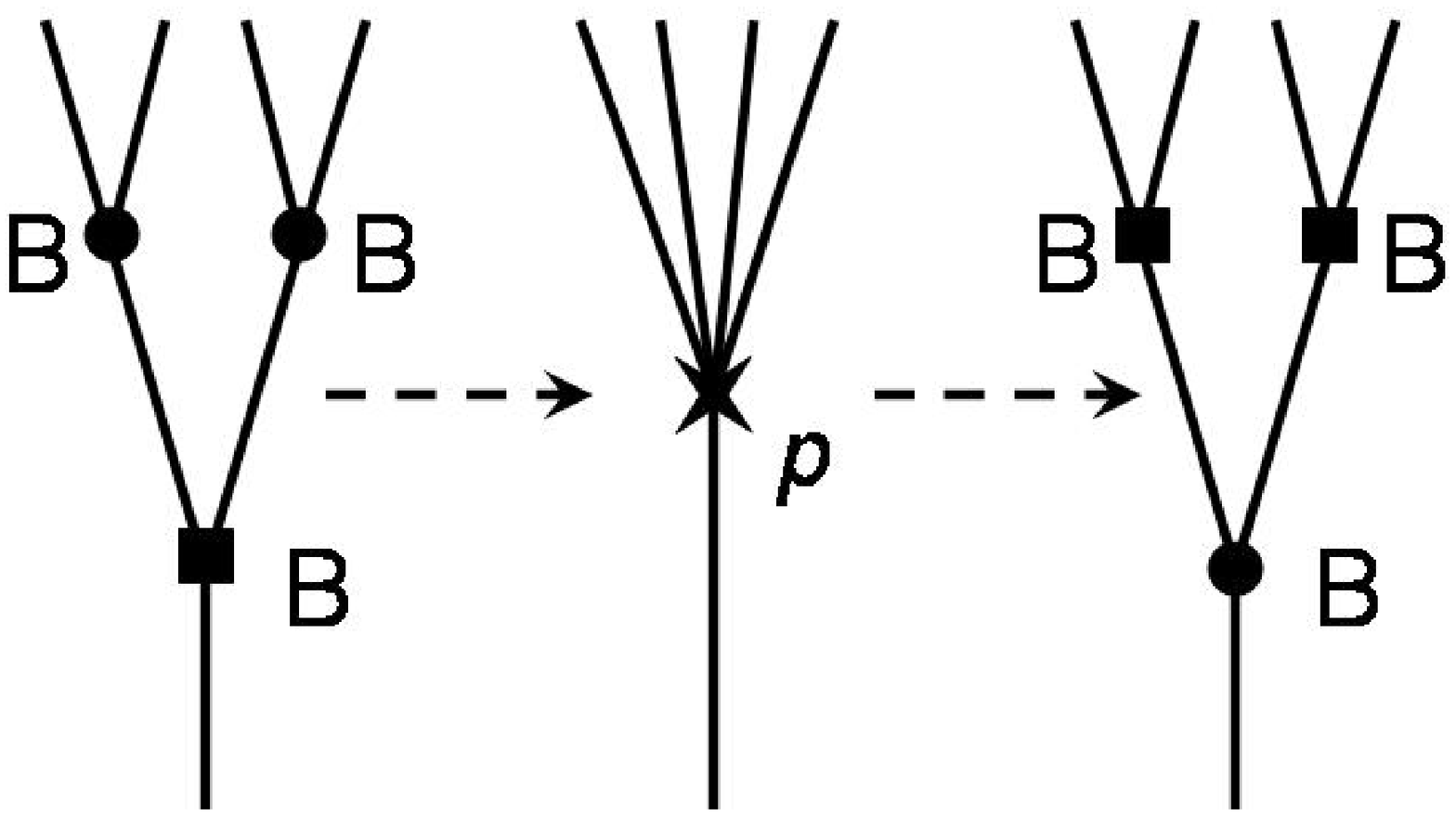}
\end{minipage}
\caption{The orientable saddle-saddle fixed point. The two possible
bifurcation sets and Fomenko graphs near a fixed point of the
\emph{saddle-saddle type}, when all one-dimensional orbits in its
leaf are orientable.}
\label{fig:saddle-saddle.4orientable}%
\end{figure}

\begin{figure}[h]
\centering
\begin{minipage}[h]{0.44\textwidth}
\centering
\includegraphics[width=4.5cm,height=3.6cm]{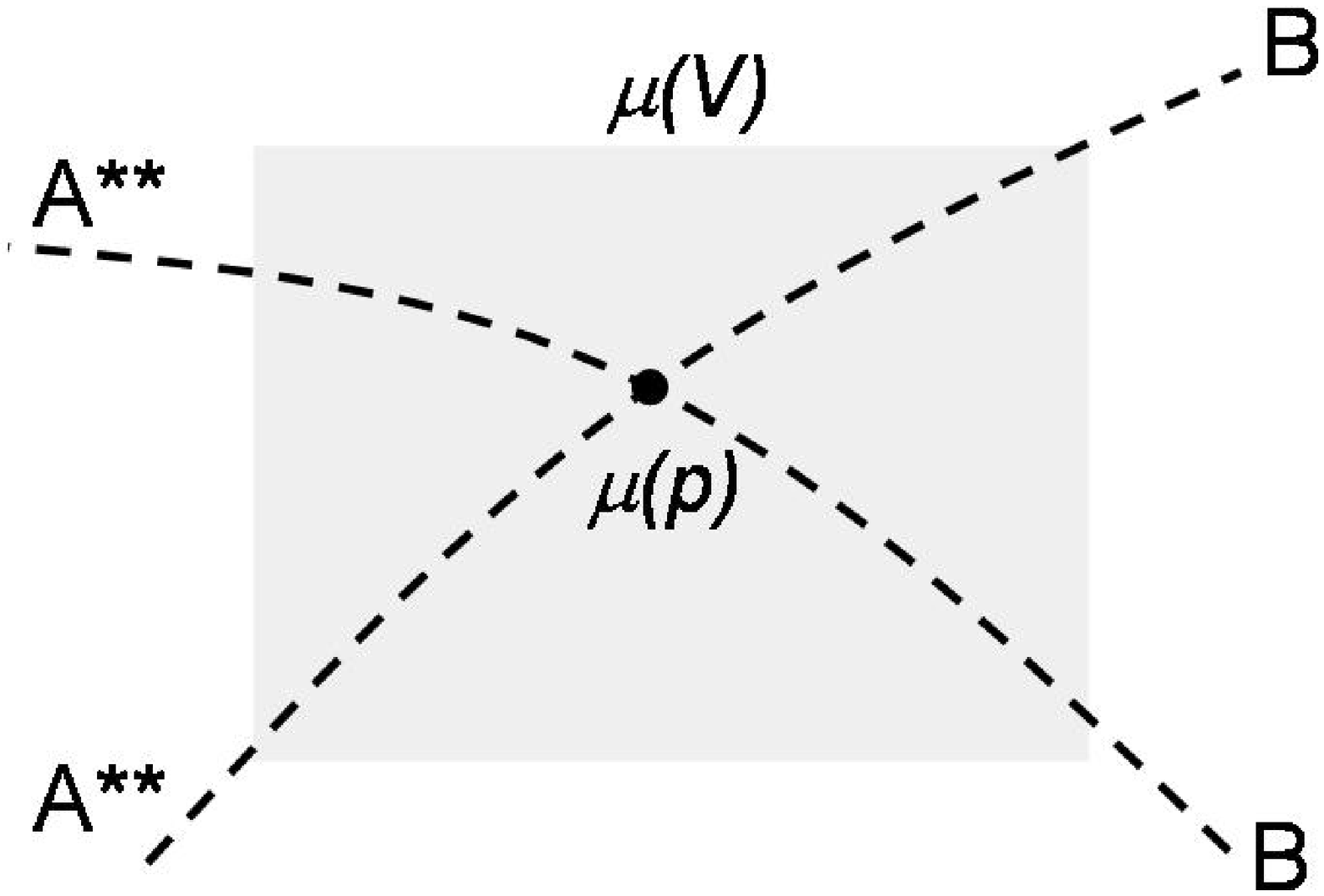}
\end{minipage}
\begin{minipage}[h]{0.44\textwidth}
\centering
\includegraphics[width=4.5cm,height=3.6cm]{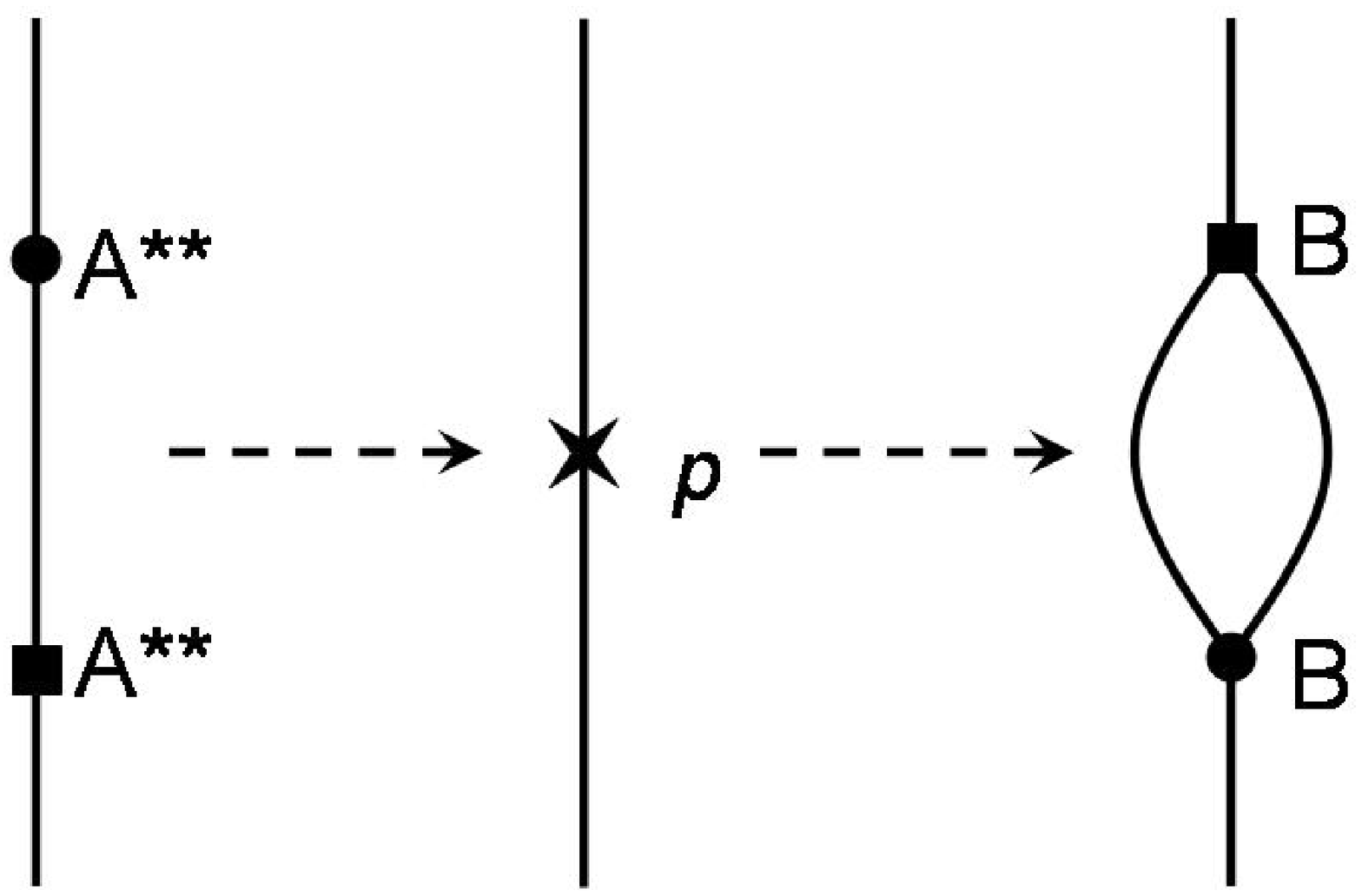}
\end{minipage}
\begin{minipage}[h]{0.44\textwidth}
\centering
\includegraphics[width=4.5cm,height=3.6cm]{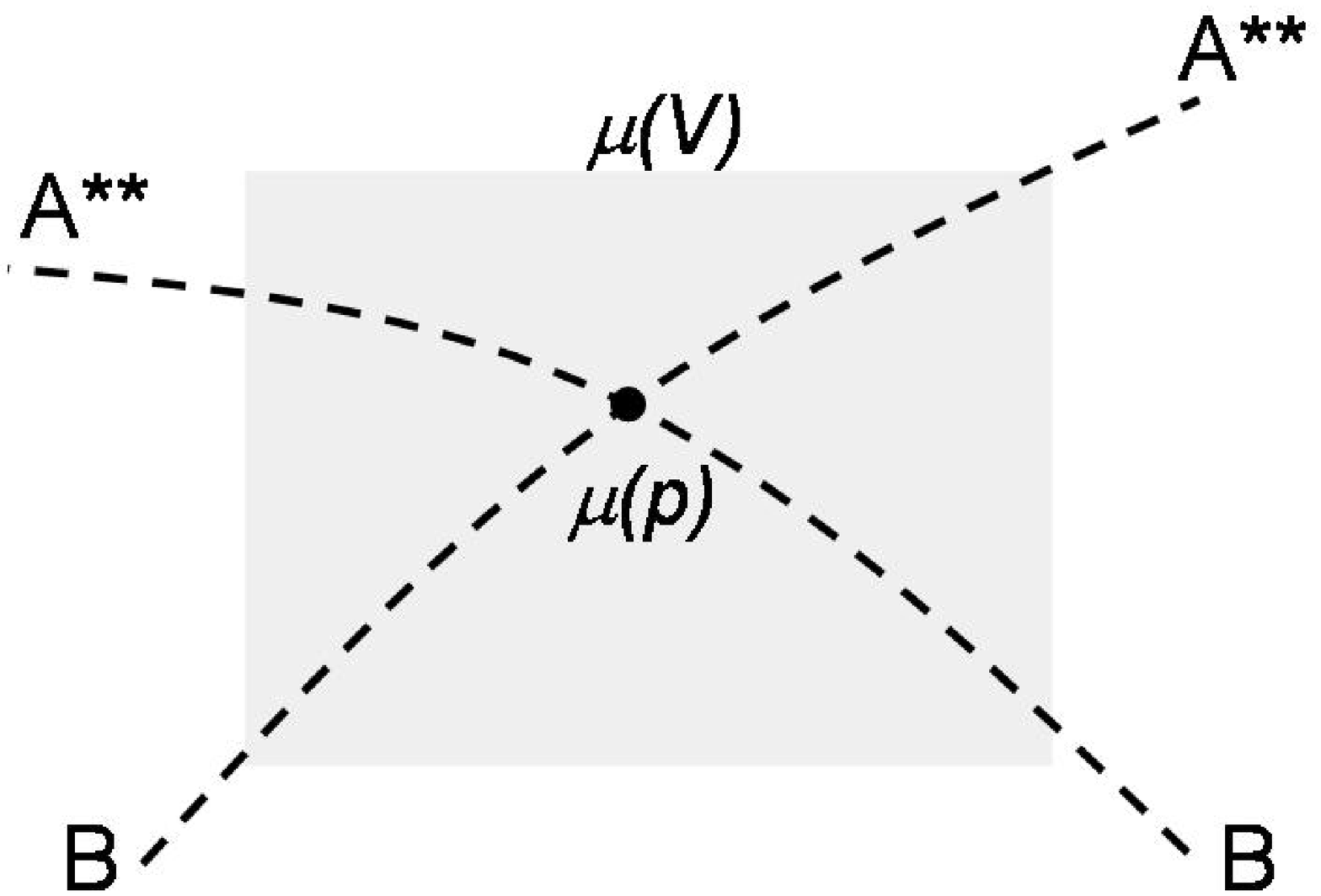}
\end{minipage}
\begin{minipage}[h]{0.44\textwidth}
\centering
\includegraphics[width=4.5cm,height=3.6cm]{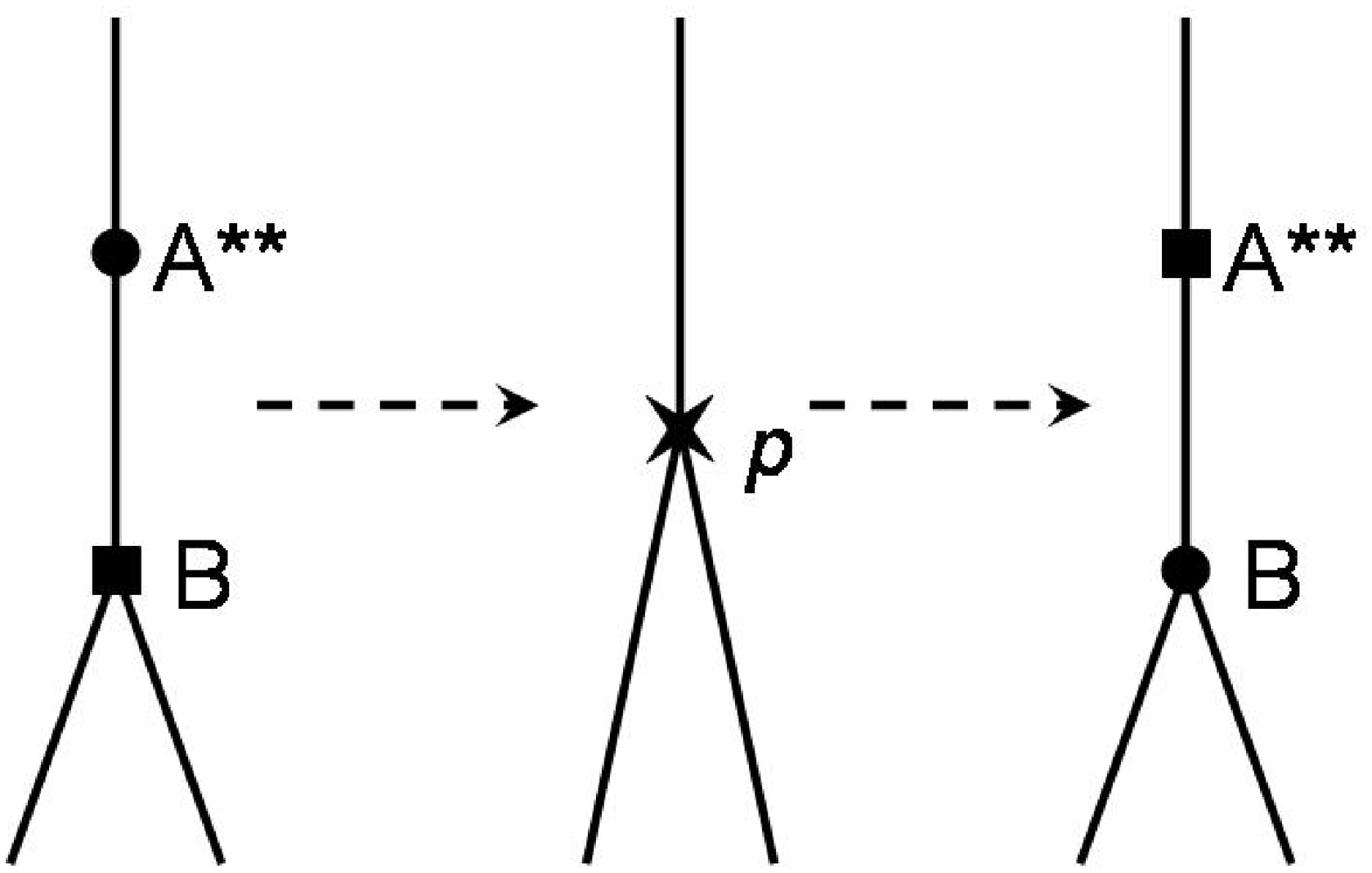}
\end{minipage}
\caption{The 4-non-orientable saddle-saddle fixed point. The two
possible bifurcation sets and Fomenko graphs near a fixed point of
the \emph{saddle-saddle type}, when all four one-dimensional orbits
in its leaf are non-orientable.}
\label{fig:saddle-saddle.4nonorientable}%
\end{figure}

\begin{figure}[h]
\centering
\begin{minipage}[h]{0.44\textwidth}
\centering
\includegraphics[width=4.5cm,height=3.6cm]{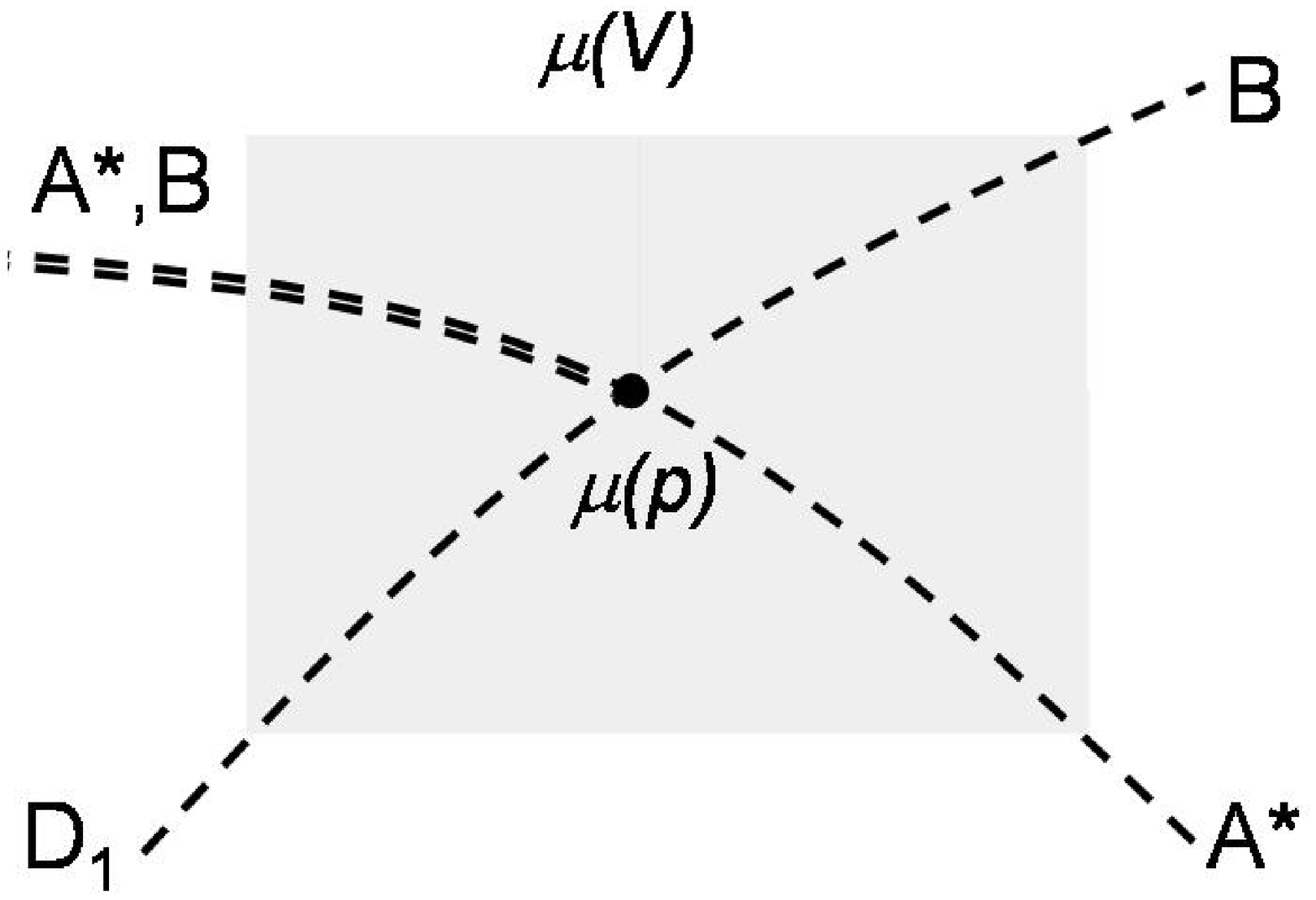}
\end{minipage}
\begin{minipage}[h]{0.44\textwidth}
\centering
\includegraphics[width=4.5cm,height=3.6cm]{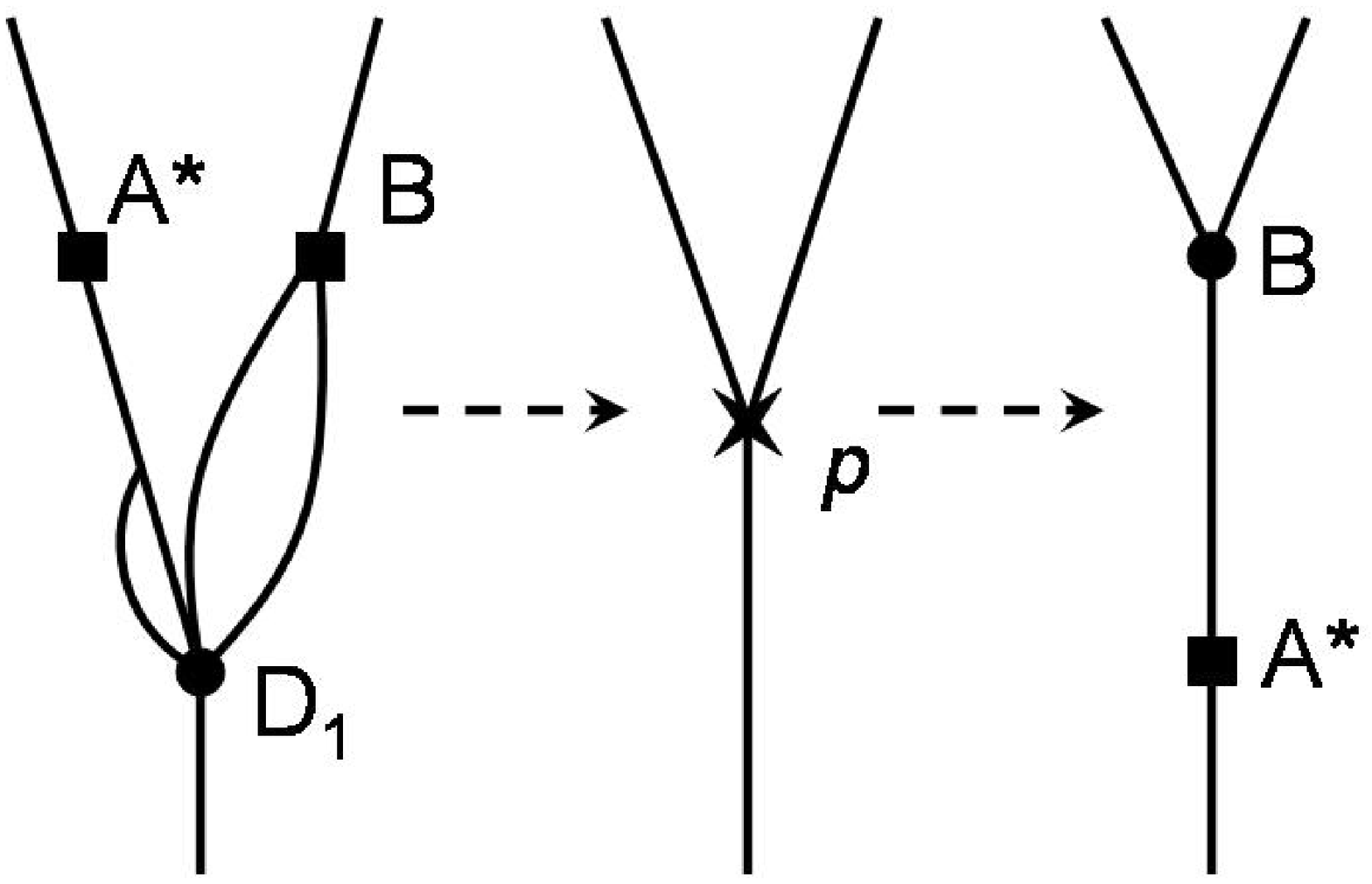}
\end{minipage}
\begin{minipage}[h]{0.44\textwidth}
\centering
\includegraphics[width=4.5cm,height=3.6cm]{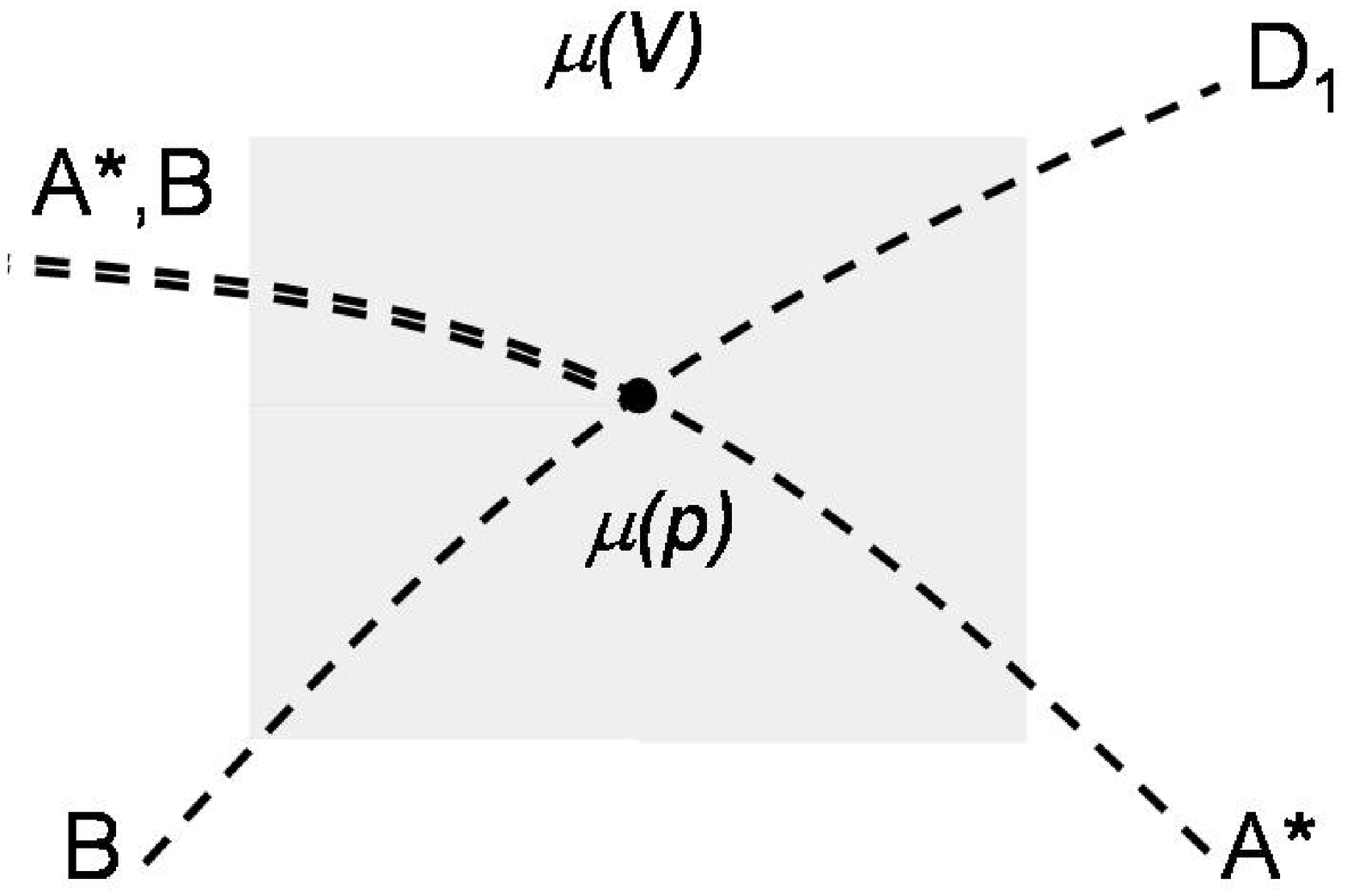}
\end{minipage}
\begin{minipage}[h]{0.44\textwidth}
\centering
\includegraphics[width=4.5cm,height=3.6cm]{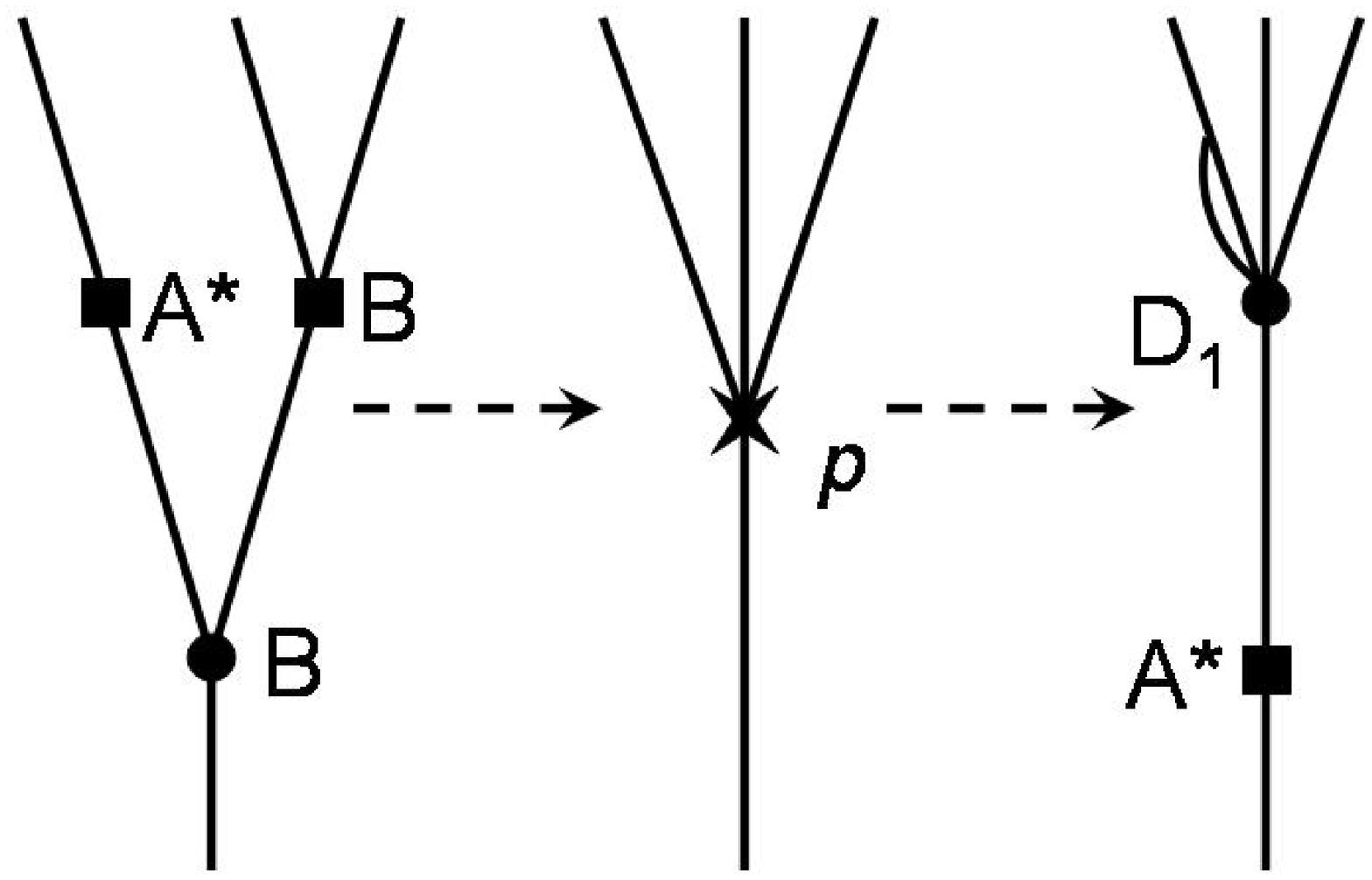}
\end{minipage}
\caption{The 3-non-orientable saddle-saddle fixed point. The two
possible bifurcation sets and Fomenko graphs near a fixed point of
the \emph{saddle-saddle type}, when three one-dimensional orbits
in its leaf are orientable and one is non-orientable.}
\label{fig:saddle-saddle.3orientable}%
\end{figure}

\begin{figure}[h]
\centering
\begin{minipage}[h]{0.44\textwidth}
\centering
\includegraphics[width=4.5cm,height=3.6cm]{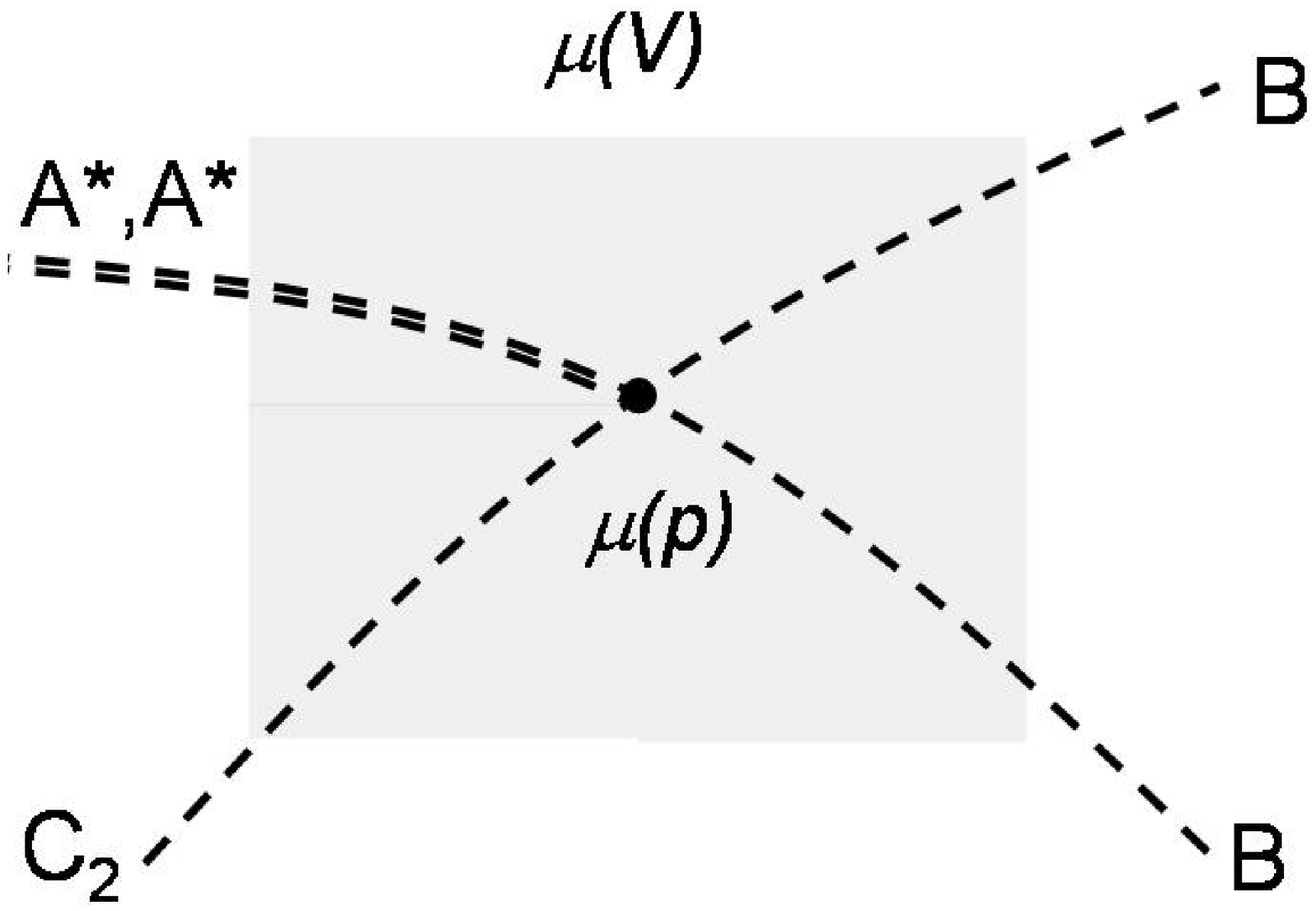}
\end{minipage}
\begin{minipage}[h]{0.44\textwidth}
\centering
\includegraphics[width=4.5cm,height=3.6cm]{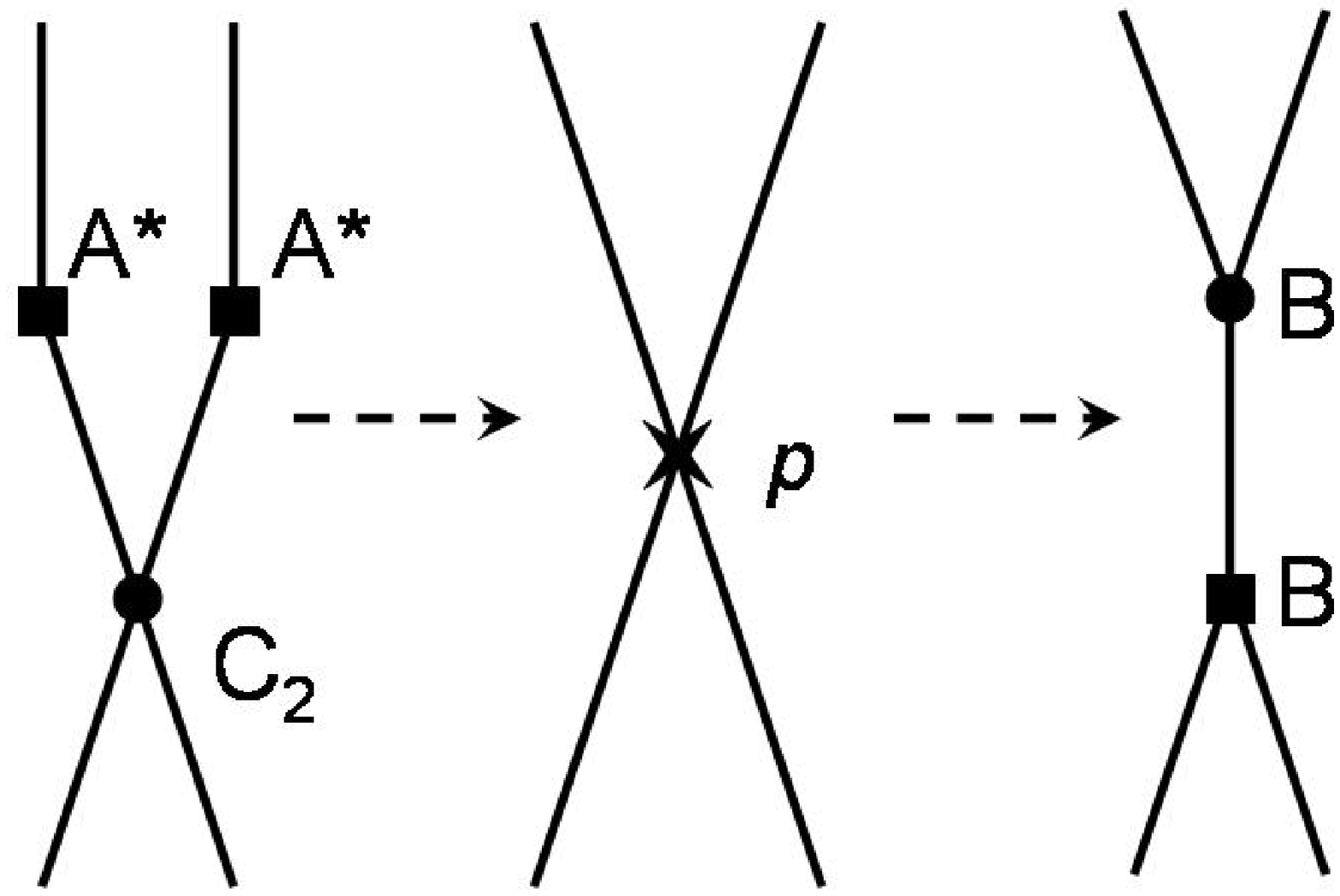}
\end{minipage}
\begin{minipage}[h]{0.44\textwidth}
\centering
\includegraphics[width=4.5cm,height=3.6cm]{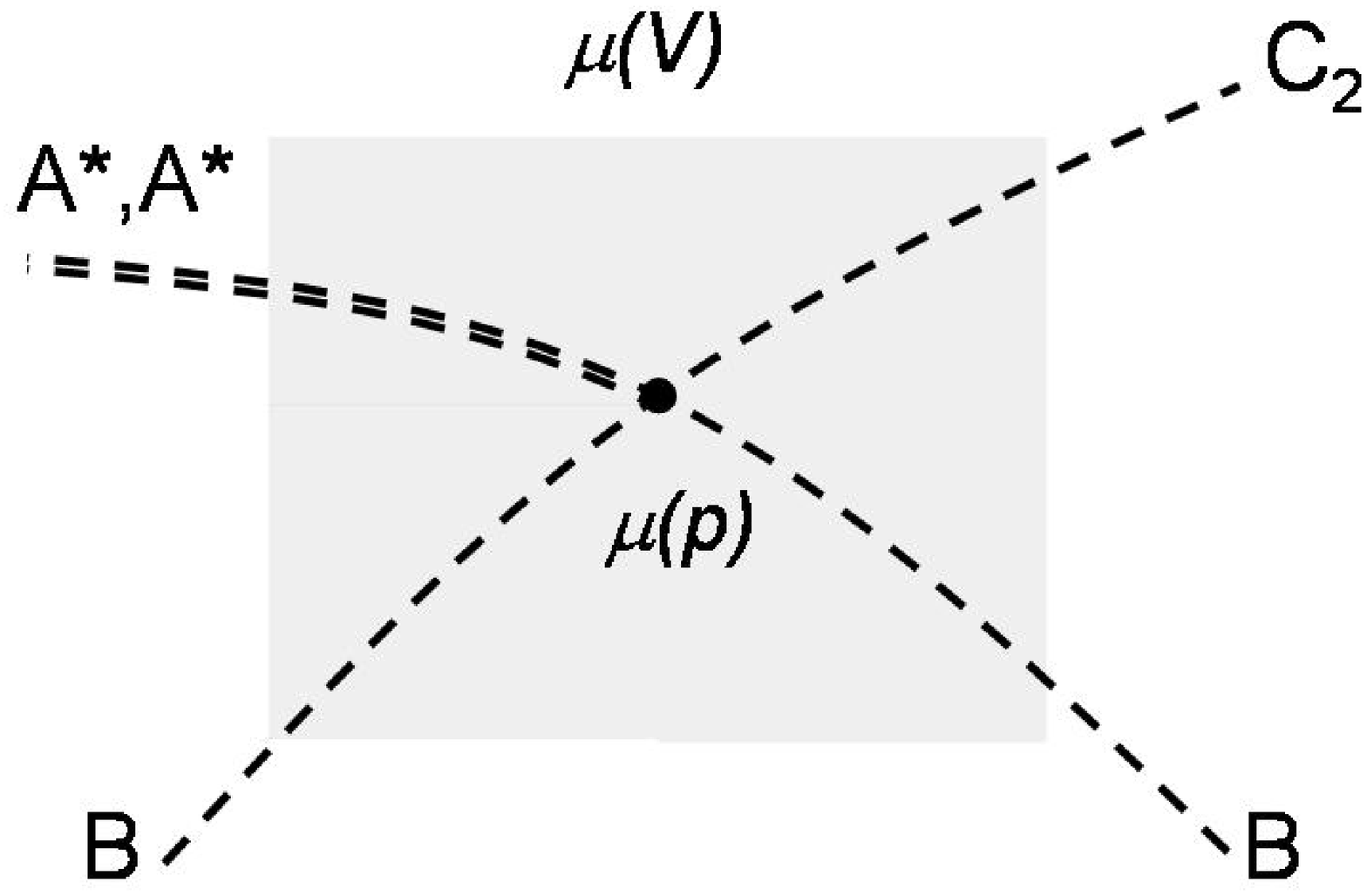}
\end{minipage}
\begin{minipage}[h]{0.44\textwidth}
\centering
\includegraphics[width=4.5cm,height=3.6cm]{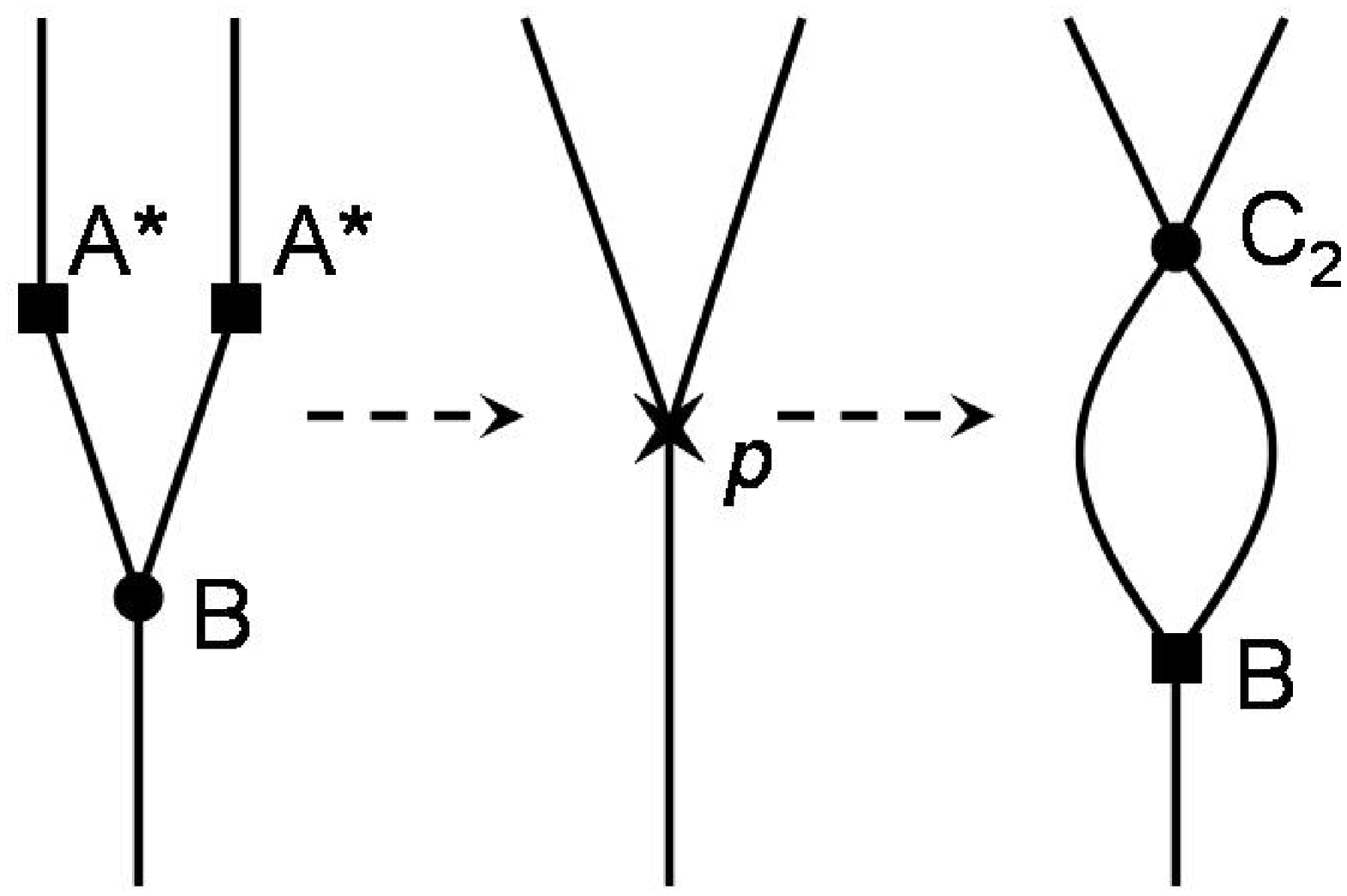}
\end{minipage}
\caption{The 2-non-orientable saddle-saddle fixed point. The two
possible bifurcation sets and Fomenko graphs near a fixed point of
the \emph{saddle-saddle type}, when two one-dimensional orbits in
its leaf are orientable and two are non-orientable.}
\label{fig:saddle-saddle.2orientable}%
\end{figure}

\begin{figure}[h]
\centering
\begin{minipage}[h]{0.44\textwidth}
\centering
\includegraphics[width=4.5cm,height=3.6cm]{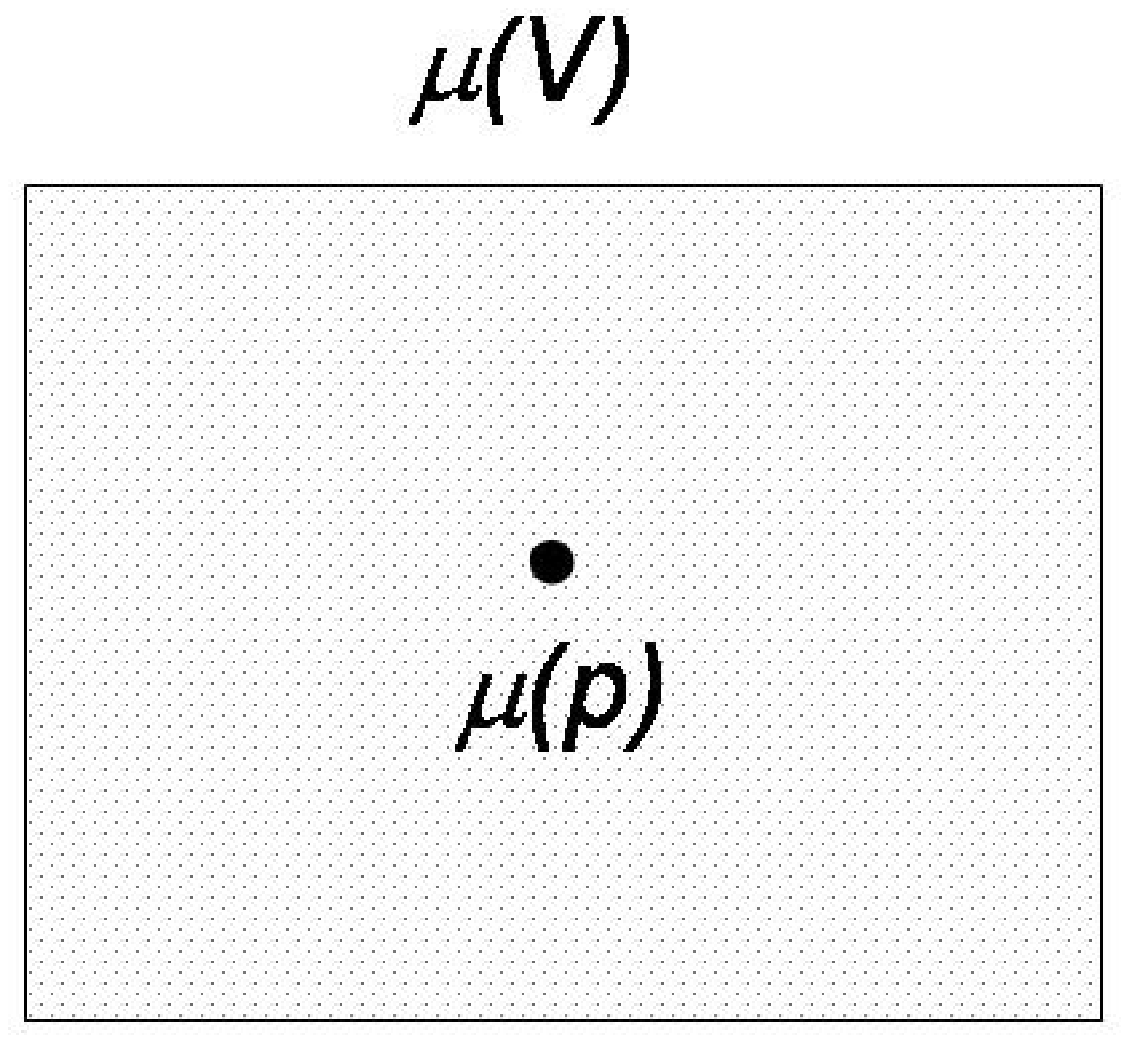}
\end{minipage}
\begin{minipage}[h]{0.44\textwidth}
\centering
\includegraphics[width=4.5cm,height=3.6cm]{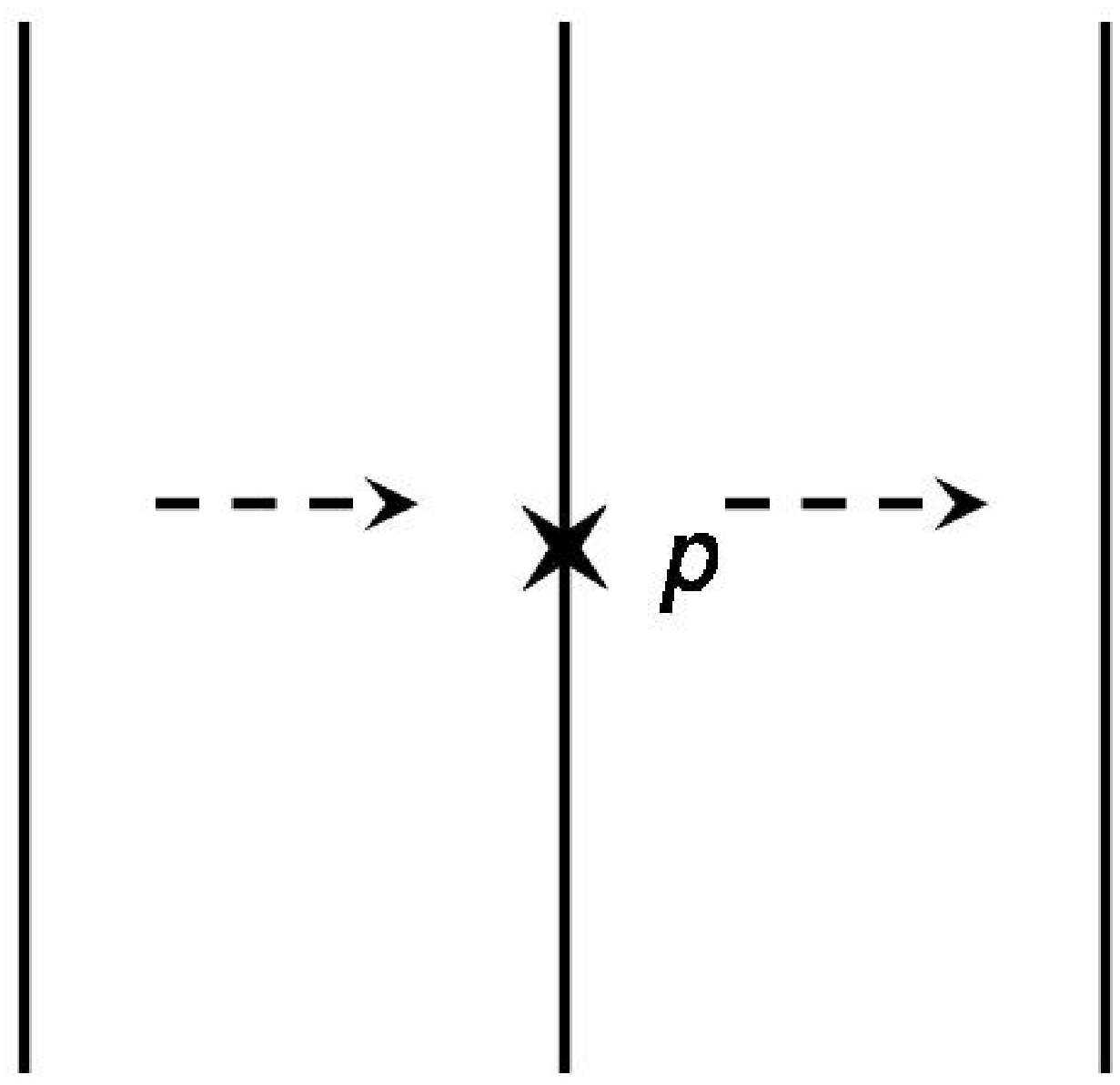}
\end{minipage}
\caption{The focus-focus fixed point. The bifurcation set and the
Fomenko graph near a fixed point of the \emph{focus-focus type}.
Notice that this simple diagram corresponds to a countable
infinity number of possible r-marked Fomenko graphs, see Remark
\ref{rem:monodromy}.}
\label{fig:focus}%
\end{figure}
\end{corollary}

\begin{remark}\label{rem:zamena}
Let us emphasize that, despite the different structures of the
isoenergy surfaces foliations, the extended neighborhoods $V$
corresponding to the pairs of subcases shown on Figures
\ref{fig:center-center}--\ref{fig:saddle-saddle.2orientable} are
isomorphic to each other. In other words, if we consider the Poisson
action $\Phi$ without distinguishing $H$ as the Hamiltonian
function, these cases do not split into subcases. Therefore, in
\cite{Bol} these cases are identified.
\end{remark}

\begin{remark}\label{rem:monodromy}
Fixed points of focus-focus type have interesting monodromy
properties, see \cite{CB,DS,Zung.focus}. In particular, it can be
shown that the twisting associated with a focus-focus point can
change the $r$-marks of the edge of the graph across this point in a
countable infinity number of ways, depending on the way the extended
neighborhood of the focus point is glued in the phase space. These
infinite number of possible changes in the foliation across the
focus point are considered as one type of change in The
Singularities and Foliations Theorem.
\end{remark}

Next, we describe the changes in the Liouville foliations of the
isoenergy surfaces near parabolic circles. Thus we formulate the
following consequence of the work \cite{BRF}.

\begin{corollary}\label{cor:parabolic}
Consider a system $(\mathcal{M},\omega,H)$ satisfying Assumptions
\ref{as:smooth}--\ref{as:sing.leaves} and let $\gamma$ be its
parabolic circle. Then, for a sufficiently small extended
neighborhood $V$ of $\gamma$, the bifurcation set $\Sigma(V)$ and
the corresponding Fomenko graphs are listed by Figures
\ref{fig:parabolic1}, \ref{fig:parabolic2}, and
\ref{fig:parabolic3}.

\begin{figure}[h]
\centering
\begin{minipage}[h]{0.44\textwidth}
\centering
\includegraphics[width=5cm,height=4cm]{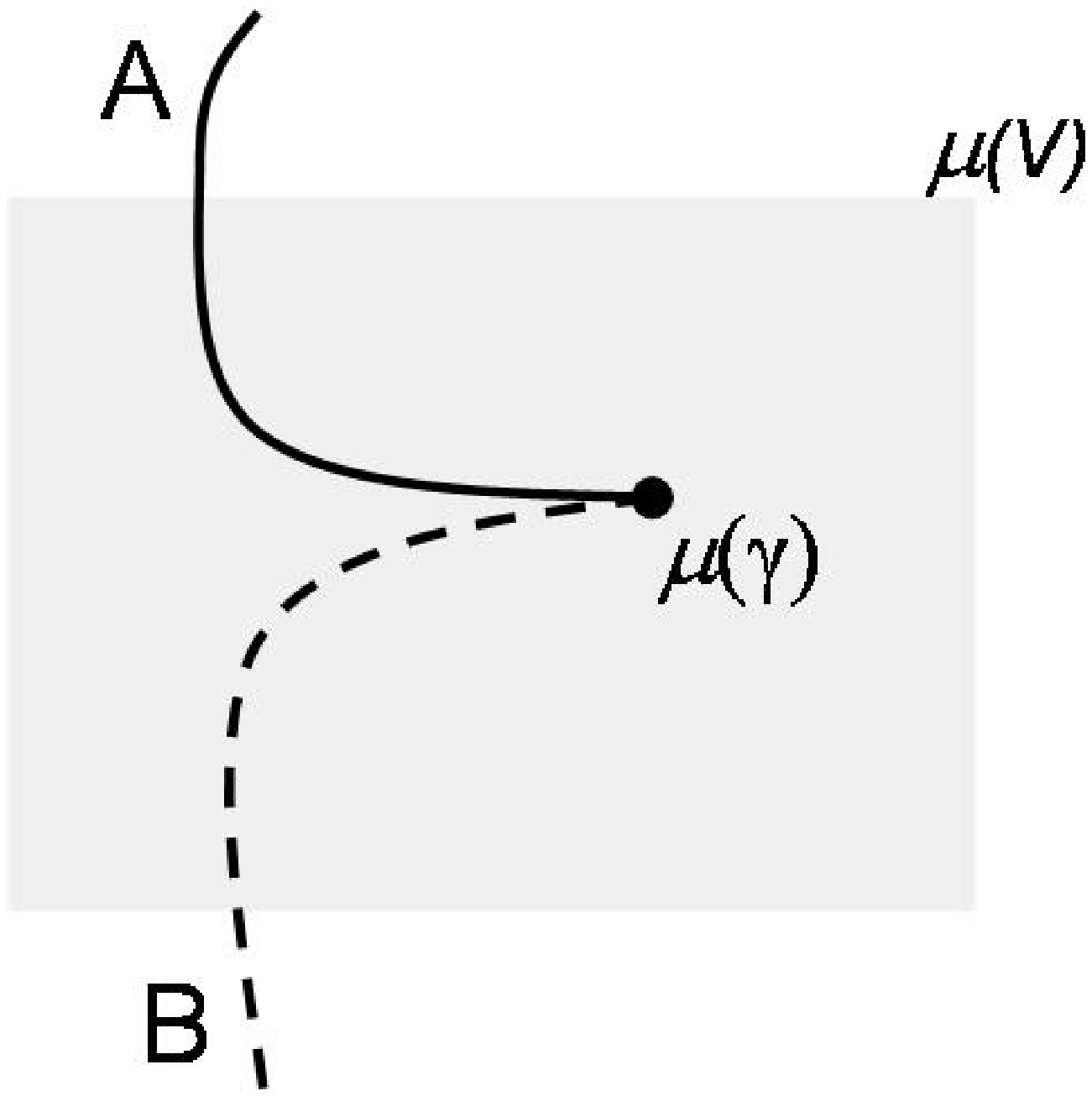}
\end{minipage}
\begin{minipage}[h]{0.44\textwidth}
\centering
\includegraphics[width=5cm,height=4cm]{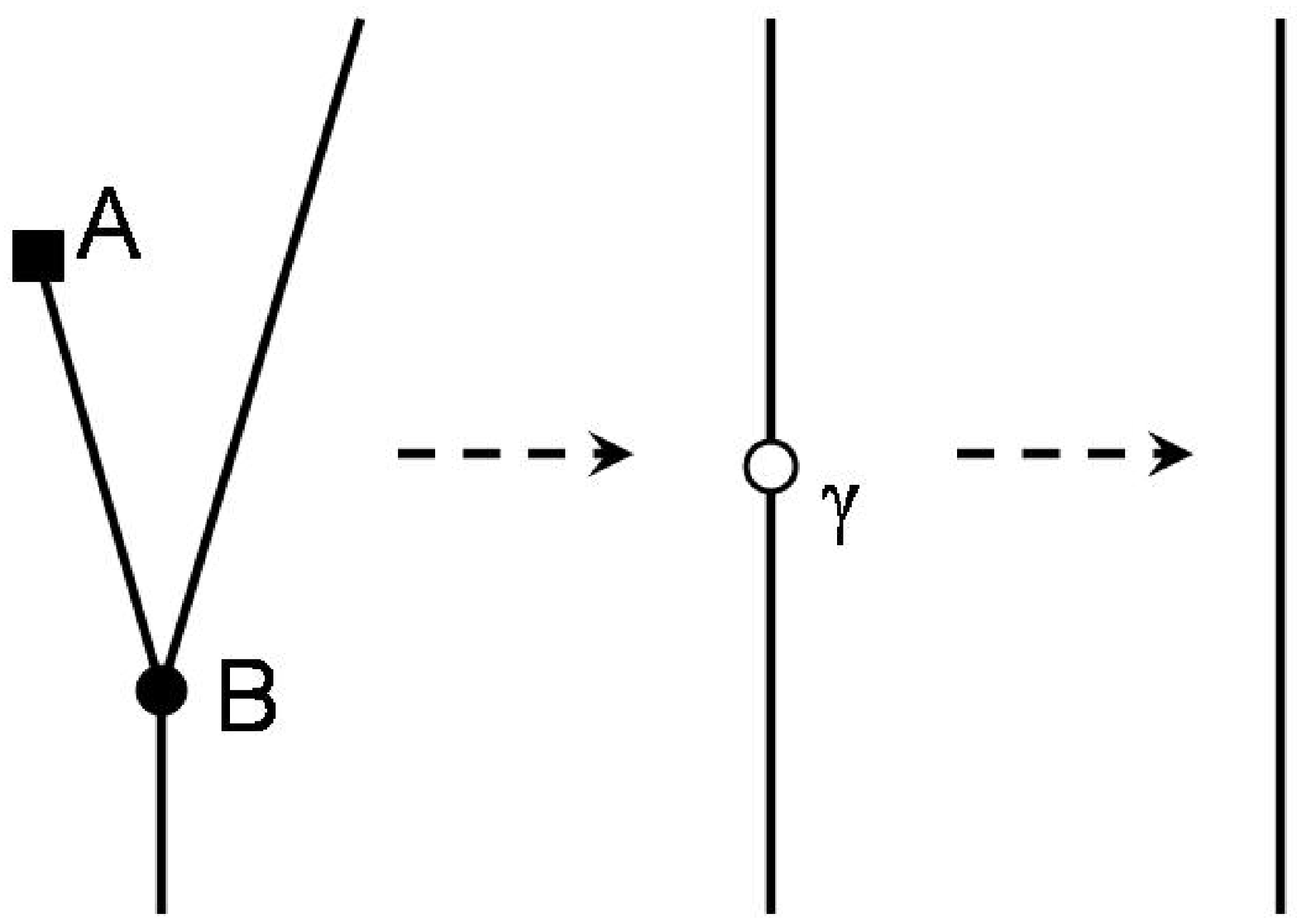}
\end{minipage}
\begin{minipage}[h]{0.44\textwidth}
\centering
\includegraphics[width=5cm,height=4cm]{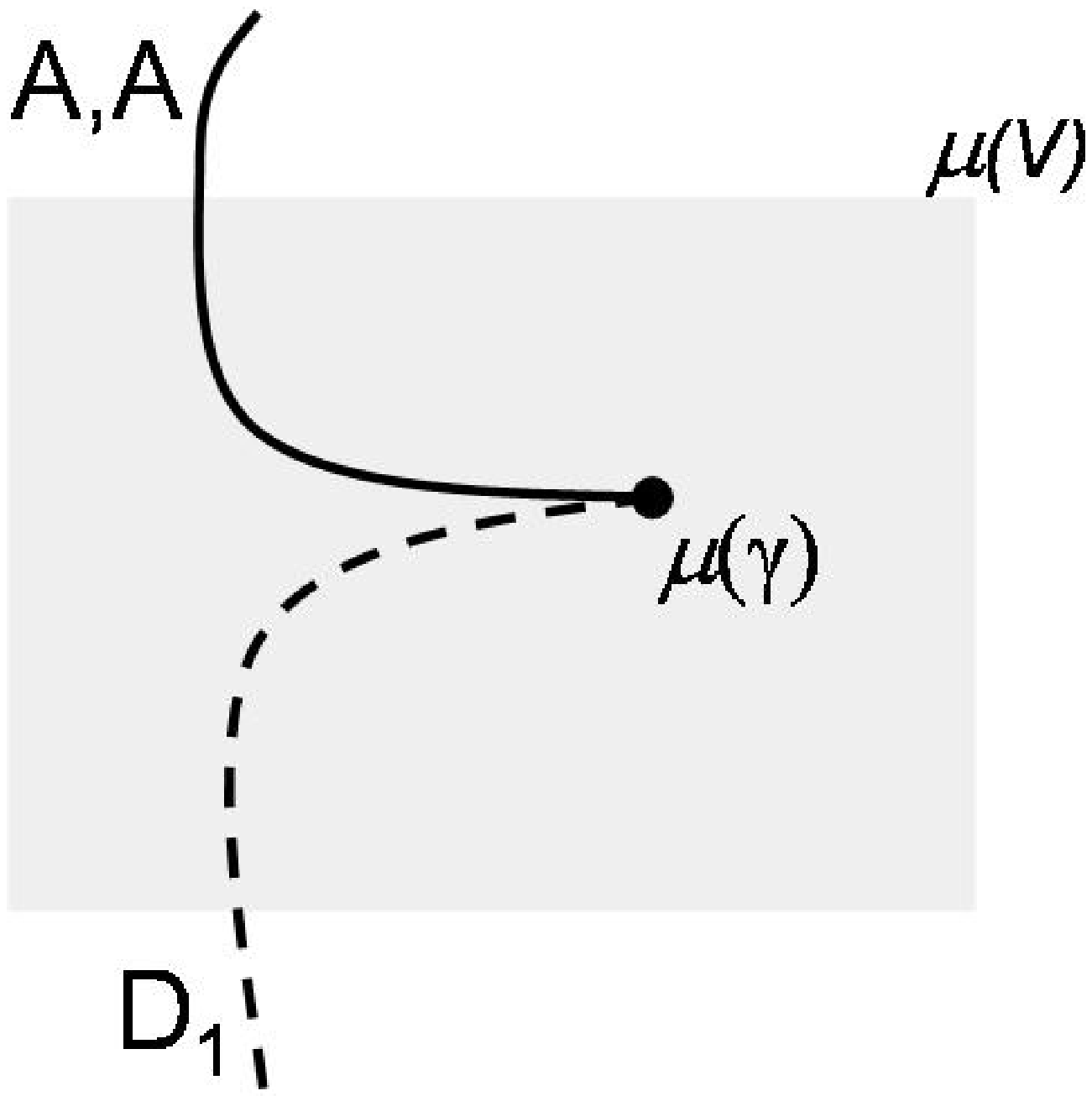}
\end{minipage}
\begin{minipage}[h]{0.44\textwidth}
\centering
\includegraphics[width=5cm,height=4cm]{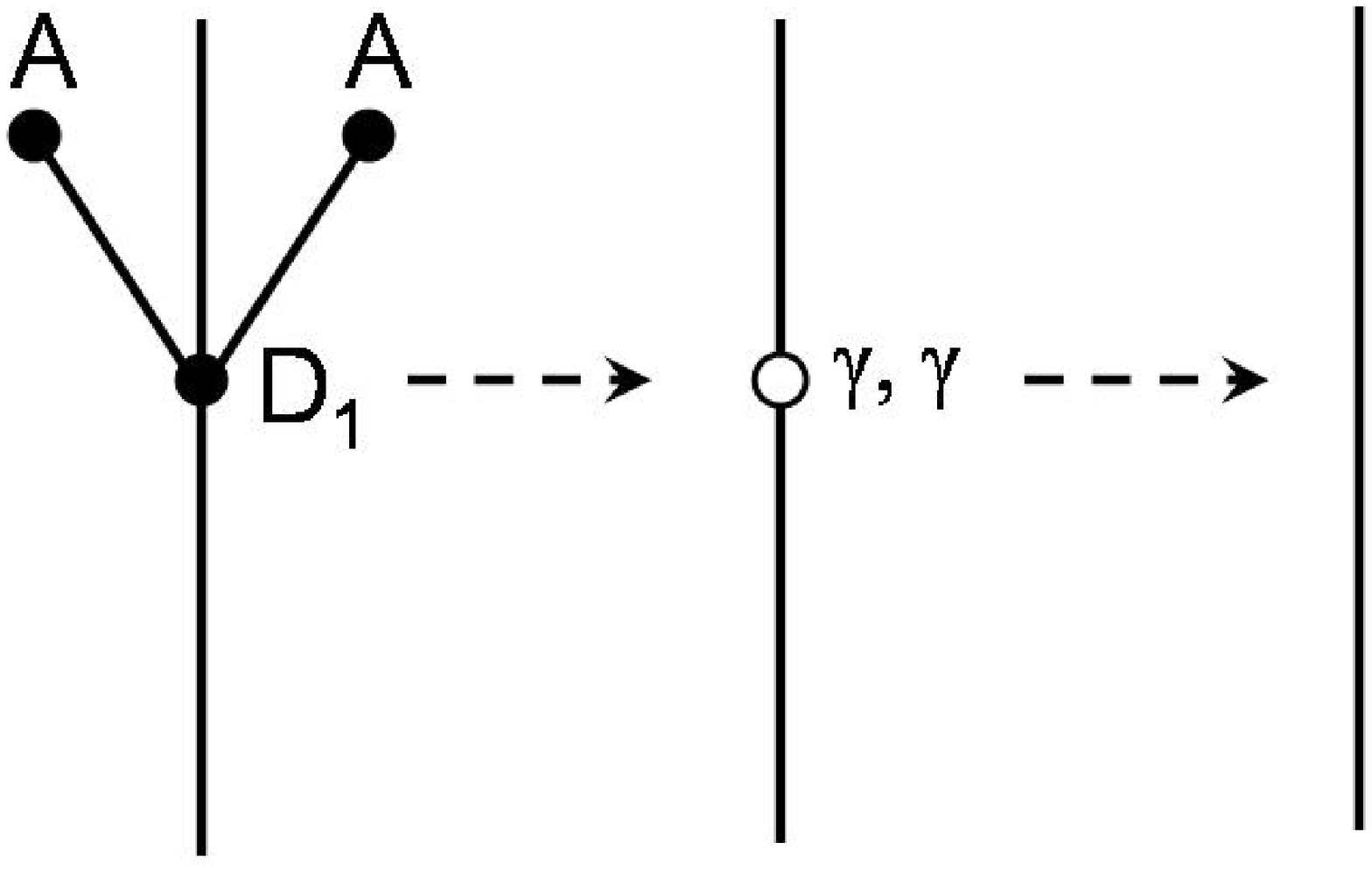}
\end{minipage}
\caption{Parabolic circles: the normal form of the family $K_h$ is
$hx+x^{3}+y^{2}$. Upper line: the bifurcation set and the Fomenko
graphs near the saddle-center bifurcation. Lower line: the
bifurcation set and the Fomenko graph when two parabolic circles
appear simultaneously on the same leaf. Notice the appearance of a
curve corresponding to leaves of complexity
$2$.}\label{fig:parabolic1}
\end{figure}

\begin{figure}[h]
\centering
\begin{minipage}[h]{0.44\textwidth}
\centering
\includegraphics[width=5cm,height=4cm]{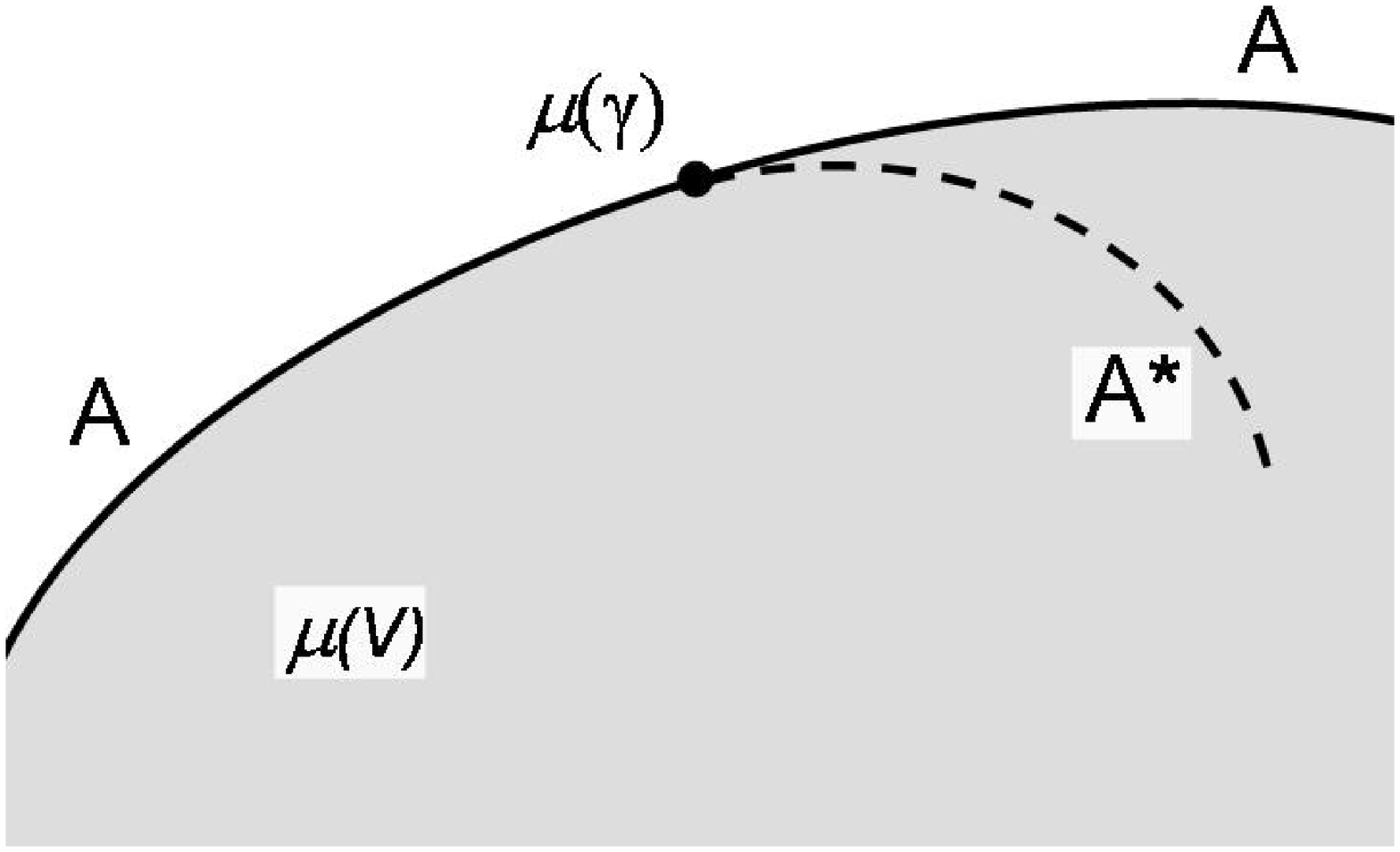}
\end{minipage}
\begin{minipage}[h]{0.44\textwidth}
\centering
\includegraphics[width=5cm,height=4cm]{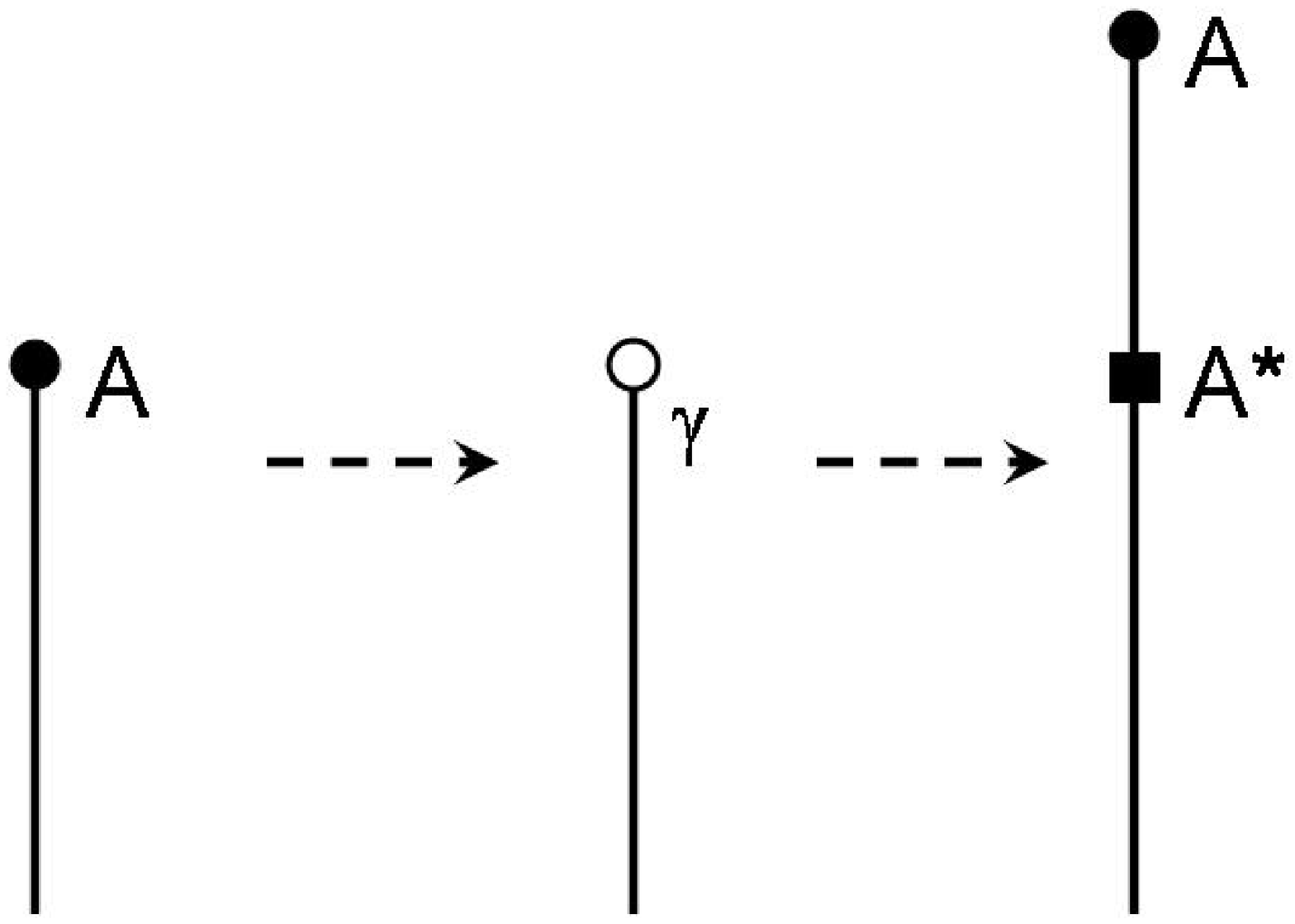}
\end{minipage}
\begin{minipage}[h]{0.44\textwidth}
\centering
\includegraphics[width=5cm,height=4cm]{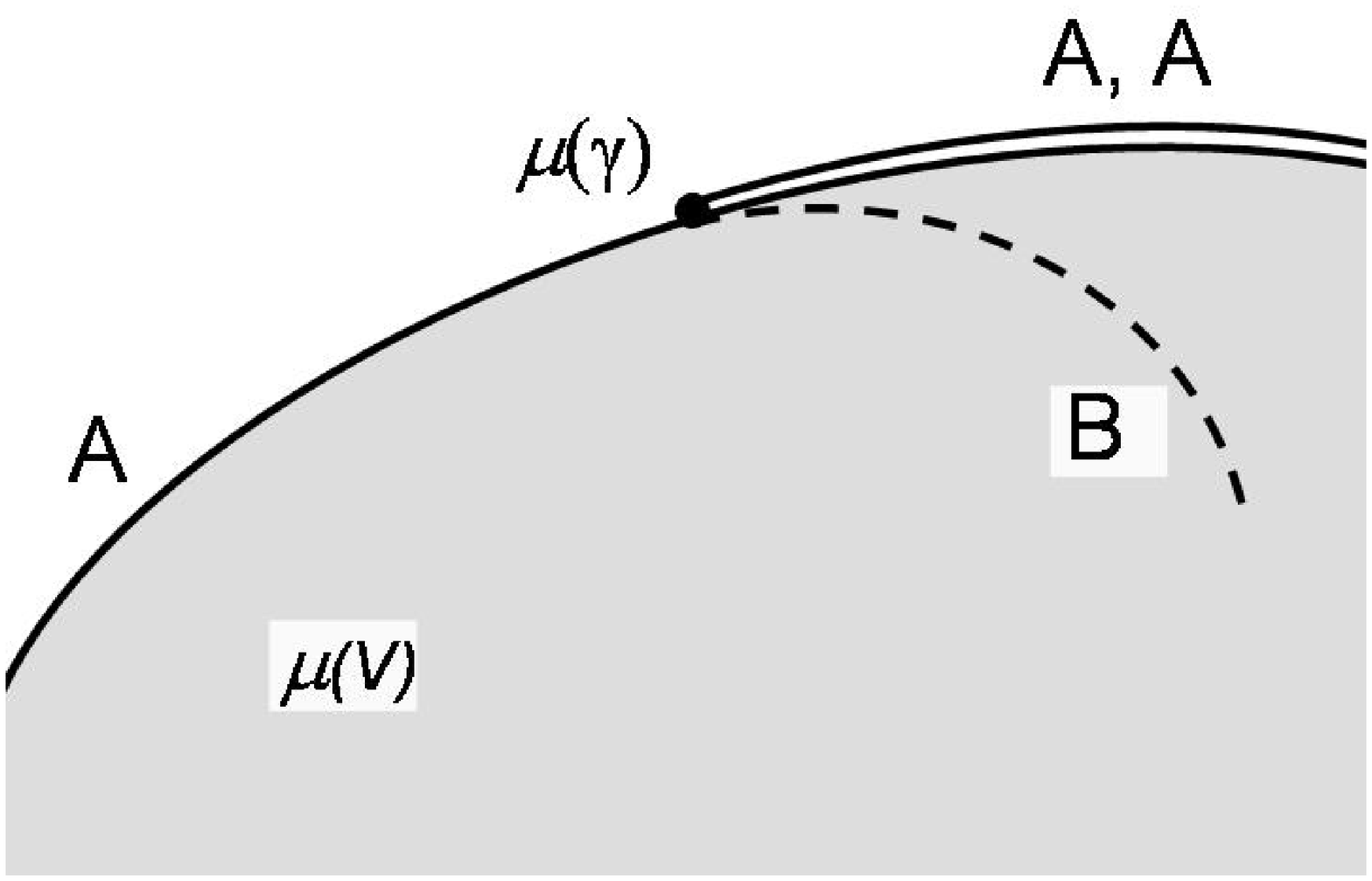}
\end{minipage}
\begin{minipage}[h]{0.44\textwidth}
\centering
\includegraphics[width=5cm,height=4cm]{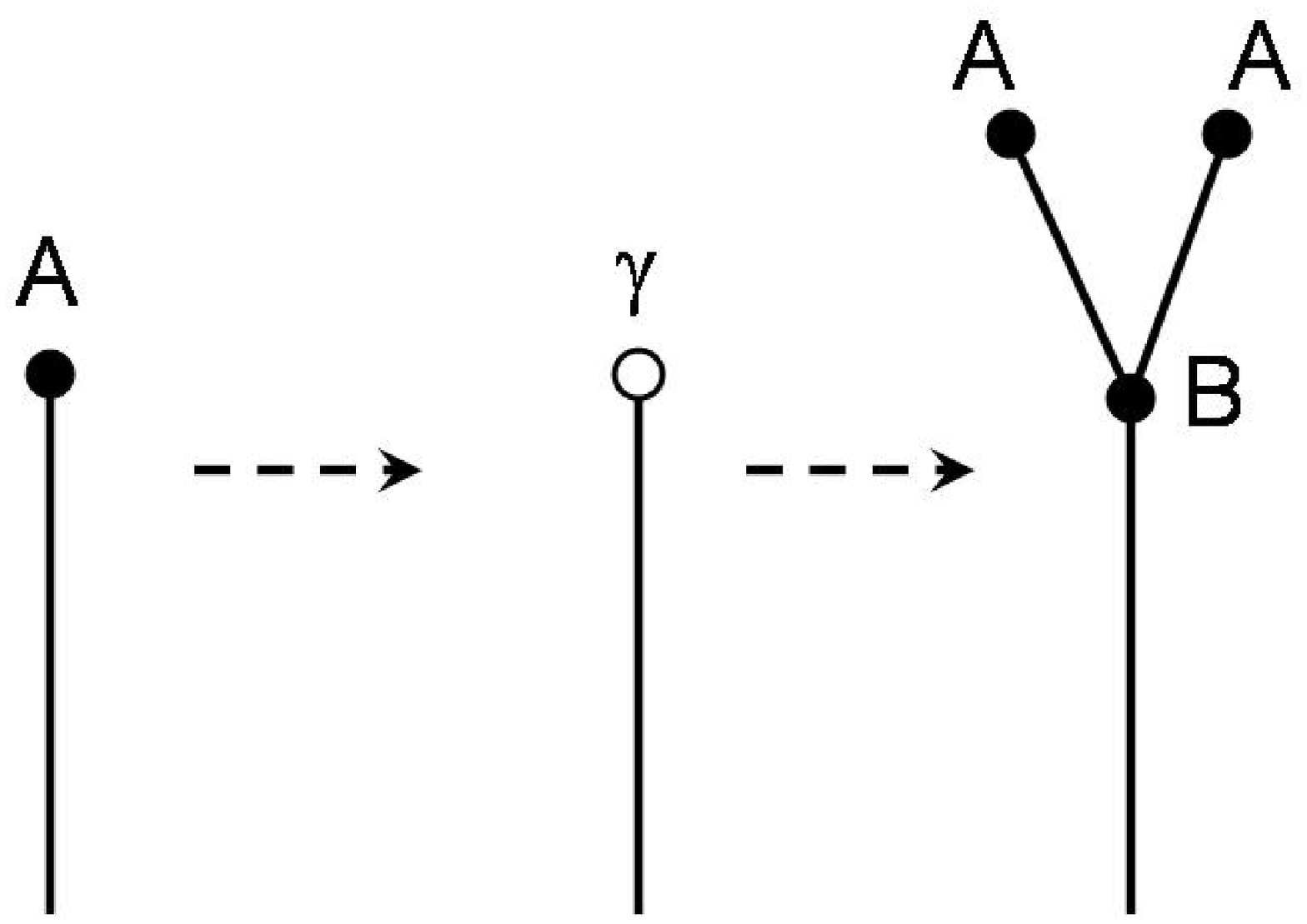}
\end{minipage}
\caption{Parabolic circles: the normal form of the family $K_h$ is
$hx^2+x^4+y^2$. Upper line: the bifurcation set and the Fomenko
graphs near elliptic period doubling bifurcation. Lower line: the
bifurcation set and the Fomenko graphs near elliptic pitchfork
bifurcation.}\label{fig:parabolic2}
\end{figure}

\begin{figure}[h]
\centering
\begin{minipage}[h]{0.44\textwidth}
\centering
\includegraphics[width=5cm,height=4cm]{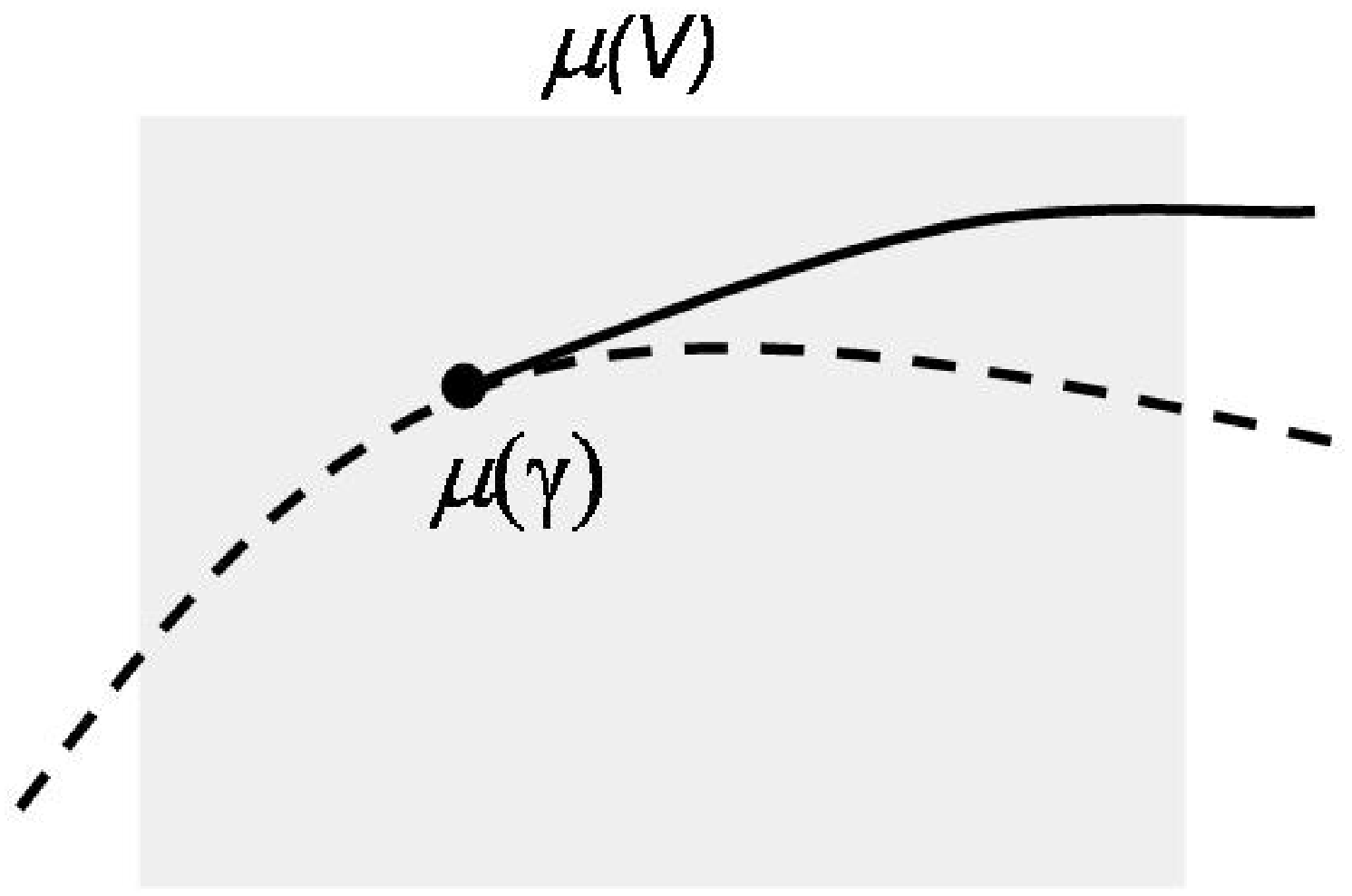}
\end{minipage}
\begin{minipage}[h]{0.44\textwidth}
\centering
\includegraphics[width=5cm,height=4cm]{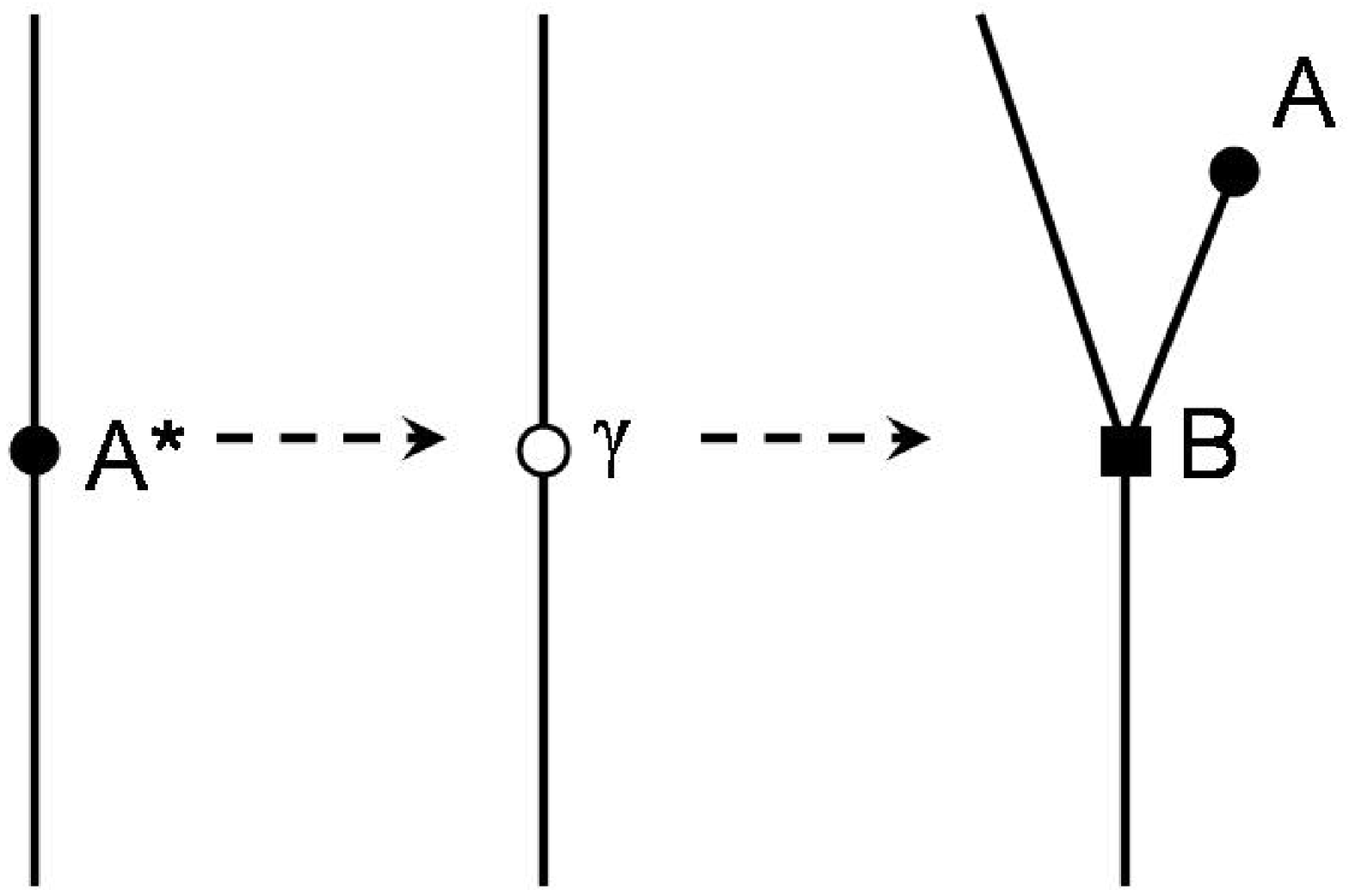}
\end{minipage}
\begin{minipage}[h]{0.44\textwidth}
\centering
\includegraphics[width=5cm,height=4cm]{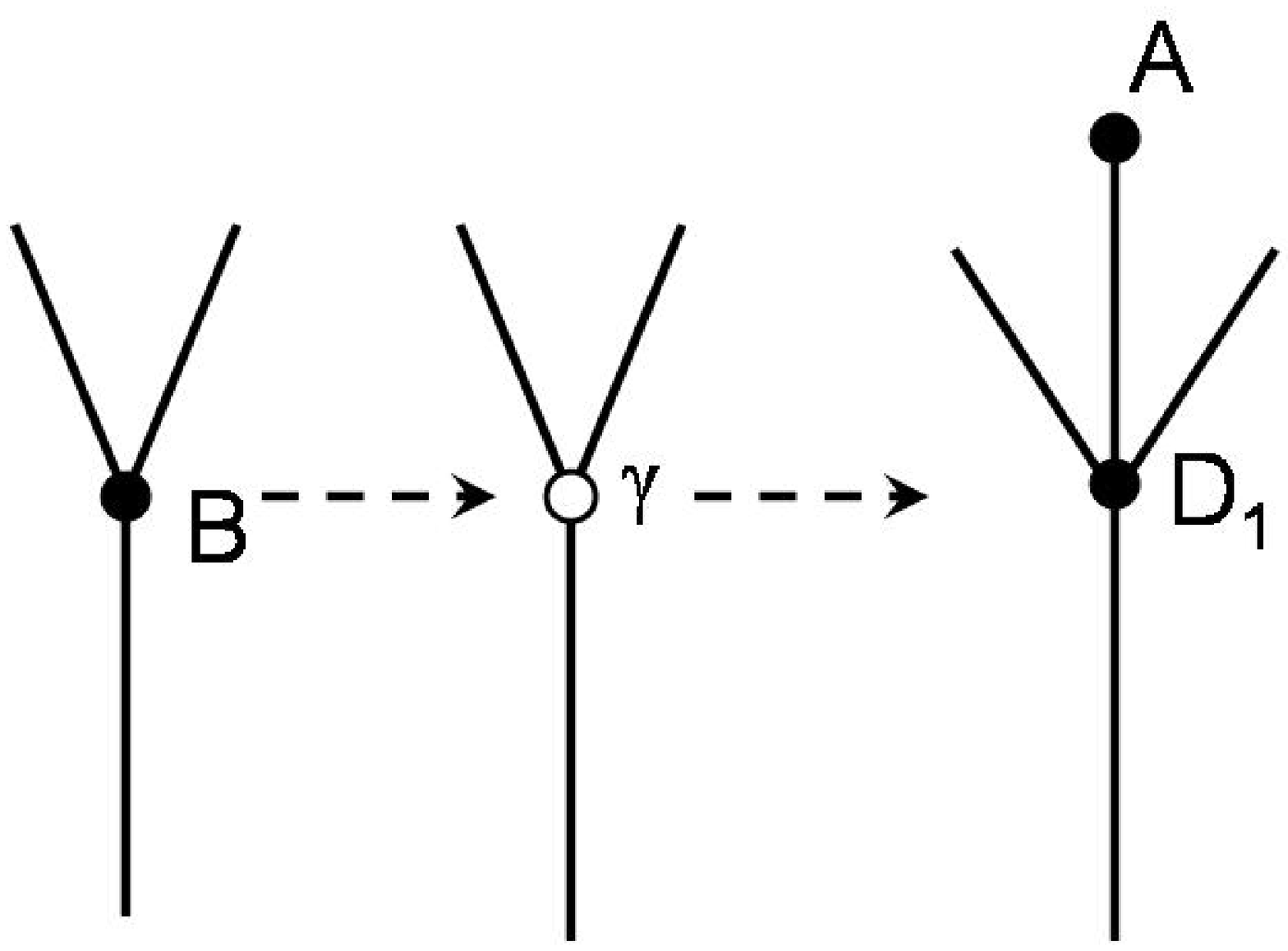}
\end{minipage}
\begin{minipage}[h]{0.44\textwidth}
\centering
\includegraphics[width=5cm,height=4cm]{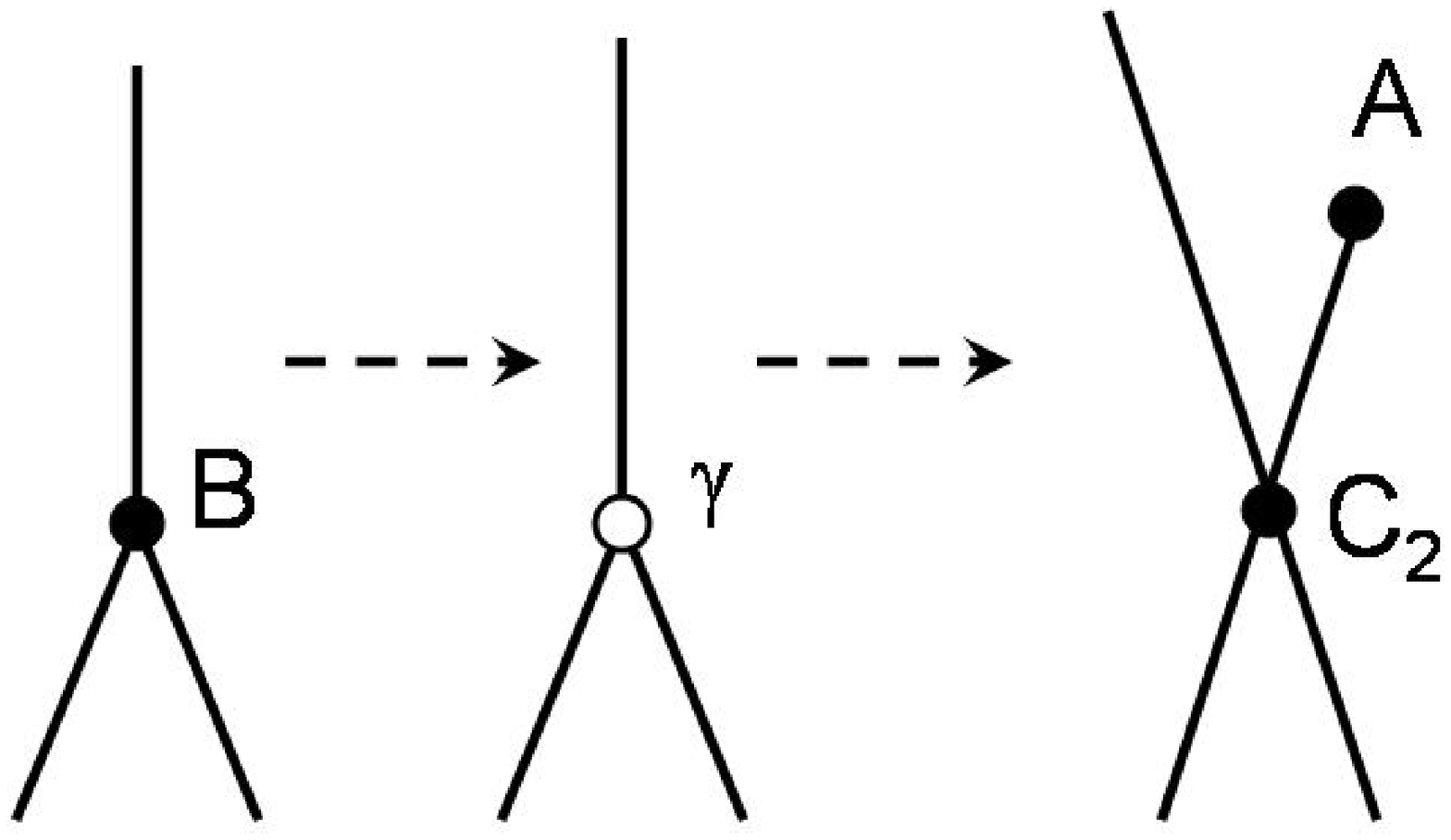}
\end{minipage}
\caption{Parabolic circles: the normal form of the family $K_h$ is
$hx^2+x^4-y^2$. Upper line: the bifurcation set and Fomenko graphs
near hyperbolic period doubling bifurcation. Lower line: Fomenko
graphs near hyperbolic pitchfork and modified hyperbolic pitchfork
bifurcations.}\label{fig:parabolic3}
\end{figure}
\end{corollary}

\end{document}